%% file: sample62.tex
\documentclass[twocolumn]{aastex62}

\usepackage{color}
\newcommand{\kawashima}[1]{{#1}}
\newcommand{\kawa}[1]{{#1}}

\newcommand{\ikomat}{\textcolor{black}}

\received{}
\revised{}
\accepted{}
\submitjournal{ApJ}

\shorttitle{Sample article}
\shortauthors{Kawashima \& Ikoma}

\begin{document}

\title{Theoretical Transmission Spectra of Exoplanet Atmospheres with Hydrocarbon Haze: \\
Effect of Creation, Growth, and Settling of Haze Particles. II. \\ Dependence on UV Irradiation Intensity, Metallicity, C/O Ratio, Eddy Diffusion Coefficient, and Temperature}

\correspondingauthor{Yui Kawashima}
\email{y.kawashima@sron.nl}

\author[0000-0003-3800-7518]{Yui Kawashima}
\affiliation{SRON Netherlands Institute for Space Research, Sorbonnelaan 2, 3584 CA Utrecht, The Netherlands}
\affiliation{Earth-Life Science Institute, Tokyo Institute of Technology 2-12-1-IE-1 Ookayama, Meguro-ku, Tokyo 152-8550, Japan}
\affiliation{Department of Earth and Planetary Science, Graduate School of Science, The University of Tokyo, 7-3-1 Hongo, Bunkyo-ku, Tokyo 113-0033, Japan}

\author{Masahiro Ikoma}
\affiliation{Department of Earth and Planetary Science, Graduate School of Science, The University of Tokyo, 7-3-1 Hongo, Bunkyo-ku, Tokyo 113-0033, Japan}
\affiliation{Research Center for the Early Universe (RESCEU), Graduate School of Science, The University of Tokyo, 7-3-1 Hongo, Bunkyo-ku, Tokyo 113-0033, Japan}



\begin{abstract}
Recent transmission spectroscopy has revealed that clouds and hazes are common in the atmospheres of close-in exoplanets.
In this study, using the photochemical, microphysical, and transmission spectrum models for close-in warm ($\lesssim$~1000~K) exoplanet atmospheres that we newly developed in our preceding paper \citep{2018ApJ...853....7K}, we investigate the vertical distribution{s} of haze particles and gaseous species and the resultant transmission spectra over wide ranges of the model parameters including UV irradiation intensity, metallicity, carbon-to-oxygen ratio (C/O), eddy diffusion coefficient, and temperature. 
The sensitivity to metallicity is of particular interest. 
We find that a rise in metallicity leads basically to reducing the photodissociation rates of the hydrocarbons and therefore the haze monomer production rates.
This is due to an enhanced photon-shielding effect by the major photon absorbers such as $\mathrm{H_2O}$, $\mathrm{CO}$, $\mathrm{CO_2}$, and $\mathrm{O_2}$, existing at higher altitudes than the hydrocarbons.
We also find that at relatively short wavelengths ($\lesssim$~ 2-3~$\mu$m), the absorption features in transmission spectra are most pronounced for moderate metallicities such as 100 times the solar metallicity, whereas the lower the metallicity the stronger the absorption features at relatively long wavelengths ($\gtrsim$~2-3~$\mu$m), where the contribution of haze is small.
These are because of the two competing effects of reduced haze production rate and atmospheric scale height for higher metallicities.
For the other model parameters, we show that stronger absorption features appear {in transmission spectra of} the atmospheres with lower UV irradiation, lower C/O ratio, higher eddy diffusion coefficient, and higher temperature.
\end{abstract}



\input{introduction}

\input{method}

\input{result}

\input{result_uv}

\input{result_metallicity}
\input{result_co}

\input{result_eddy}

\input{result_temp}

\input{discussion}

\input{summary}

\acknowledgments
We would like to express special thanks to the following people.
N. Narita and A. Fukui motivated us to work on this study and gave fruitful suggestions through observational collaboration. Advice and comments from Y. Sekine and S. Okuzumi were great help in modeling the properties of haze particles.
\kawashima{We thank the anonymous referee for his/her careful reading and constructive comments, which helped us improve this paper greatly.}
Y.K. is supported by the Grant-in-Aid for JSPS Fellow (JSPS KAKENHI No.15J08463), Leading Graduate Course for Frontiers of Mathematical Sciences and Physics, Grant-in-Aid for Scientific Research (A) (JSPS KAKENHI No.15H02065), and the European Union's Horizon 2020 Research and Innovation Programme under Grant Agreement 776403.
M.~I. is also supported by the Astrobiology Center Program of National Institutes of Natural Sciences (NINS) (No.~AB291004) and JSPS Core-to-Core Program “International Network of Planetary Sciences”.
This work has made use of the MUSCLES Treasury Survey High-Level Science Products (doi:10.17909/T9DG6F).

\bibliography{ads,ykawashima-RefList,plus}



\end{document}

%% file: introduction.tex
\section{Introduction} \label{intro}
Multi-wavelength simultaneous observations have been done for transits of several exoplanets. 
A set of transit depths thus observed is termed a transmission spectrum. 
It provides information of radiative absorption and scattering by gaseous molecules and particles such as haze and clouds in the planetary atmosphere\footnote{In this study, we refer to thermochemical condensates as ``clouds'' and photochemical products as ``haze''.}.
Thus, comparison between observational and theoretical transmission spectra can constrain the properties of the planetary atmosphere.
Recent multi-wavelength transit observations have revealed that many of those observed planets show spectra with steep slope features in the visible and/or featureless spectra in the near-infrared \citep[e.g.,][]{2014Natur.505...69K, Sing:2016hi}, inferring the existence of haze and/or clouds in the atmospheres, which prevents us from probing the atmospheric molecular composition.

As for planets with hydrogen-rich atmospheres orbiting M stars, which will be the main targets for near-future exoplanet characterization, hydrocarbon haze has attracted particular interest from the exoplanetary science community as a likely candidate for particles in relatively cool atmospheres.
\kawashima{(In this study, we also refer to organic haze as “hydrocarbon haze”.)}
While previous theoretical modeling of transmission spectra of 
such atmospheres \citep[e.g.,][]{2012ApJ...756..176H, 2013ApJ...775...33M, 2014A&A...570A..89E}
made ad hoc assumptions about the properties of the haze layer (namely, the size and number density of haze particles and the altitude and thickness of the haze layer),
recent studies \citep{2017ApJ...847...32L, 2018ApJ...853....7K, 2019arXiv190210151K} considered 
the microphysics of particles, namely, collisional growth, sedimentation, and transport by eddy diffusion 
and thereby determined the distribution of the size and number density of haze particles directly.
Such microphysical modeling has been also applied to condensation clouds to derive their distributions in exoplanet atmospheres recently \citep{2018ApJ...859...34O, 2018ApJ...860...18P, 2018ApJ...863..165G, 2019A&A...622A.121O}.

In our preceding paper \citep[][hereafter Paper~I]{2018ApJ...853....7K}, we developed new photochemical and microphysical models of the creation, growth, and settling of haze particles for deriving their size and number-density distributions in 
atmospheres of close-in warm ($\lesssim$~1000~K) exoplanets. 
We also developed radiative extinction models for generating theoretical transmission spectra of the atmospheres, combined with obtained properties of haze.
In Paper~I, we focused on describing the methodology and demonstrating the sensitivity of transmission spectra to the production rate of haze monomers.
We found that the haze was distributed in the atmosphere much more broadly than previously assumed and consisted of particles of various sizes. 
We also demonstrated that {difference\ikomat{s} in production rate of haze monomers, which is related to UV irradiation intensity from host stars, could explain} the observed diversity of transmission spectra{; completely flat spectrum, spectrum with only extinction features of hazes (i.e., \ikomat{a} spectral slope due to Rayleigh scattering and absorption features of hazes), spectrum with \ikomat{a} slope due to Rayleigh scattering and some molecular absorption features, and spectrum with only molecular absorption features.}

In this paper, we make a detailed investigation of the dependence of transmission spectra on model parameters, namely, UV irradiation intensity, metallicity, carbon-to-oxygen ratio (C/O), eddy diffusion coefficient, and \ikomat{atmospheric} temperature \ikomat{for} close-in warm ($\lesssim$~1000~K) exoplanets, adopting a more realistic assumption for \ikomat{haze} monomer production rate than in Paper~I. 
{From this investigation}, we explore {\ikomat{possible} combinations} of the model parameters that result in {larger or smaller absorption features in the transmission spectra\ikomat{, aiming to provide a strategy for future observations}.} 
\ikomat{Such an exploration has been done for hot Jupiters by \cite{2017ApJ...847...32L}, but not yet for warm super-Earths.
}
In our forthcoming papers, we explore in detail the composition of the atmospheres of known warm exoplanets by comparing the observed spectra with our theoretical ones.

{The rest of this paper is organized as follows: In \S\ref{method}, we briefly describe our photochemical, particle growth, and transmission spectrum models. In \S\ref{results}, we show the dependence of the distributions of gaseous species and haze particles and transmission spectra on the above model parameters. Then, we discuss implications for observations, those from experiments, comparison with previous studies, and caveats in \S\ref{discussion} and finally conclude this paper in \S\ref{summary}.}

%% file: method.tex
\section{Method} \label{method}
We outline our modeling of transmission spectra of an atmosphere with hydrocarbon haze as follows: 
First, we perform photochemical calculations to derive the steady-state vertical distribution of gaseous species for a given temperature distribution in the atmosphere (\S~\ref{subsec: pc}). 
Then, 
we define the production rate of {the smallest-size haze particles} \ikomat{(called haze monomers, hereafter)} as the sum of the photodissociation rates of the major hydrocarbons included in our photochemical calculations such as $\mathrm{CH_4}$, $\mathrm{HCN}$, and $\mathrm{C_2H_2}$.
After that, using the profile of the haze-monomer production rate,
we derive the steady-state distribution of haze particles by the particle growth calculations (\S~\ref{subsec: pg}).
Finally, we model transmission spectra of the atmospheres with the obtained distributions of haze particles and gaseous species (\S~\ref{subsec: ts}).

Same as in Paper~I, we model the transmission spectra, assuming the properties of GJ~1214~b, the atmosphere of which has been probed most by transmission spectroscopy observations among super-Earths detected so far.

The photochemical, particle growth, and transmission spectrum models used in this study are basically the same as those developed and described in Paper~I.
Below, we briefly describe our models and explain the assumptions and parameters we adopt, with focus on differences from Paper~I.

\subsection{Photochemical Model} \label{subsec: pc}
Our photochemical model includes 
five elements\ikomat{,} C, H, O, N, and He, \ikomat{29 chemical species, }$\mathrm{O}$, $\mathrm{O_2}$, $\mathrm{H_2O}$, $\mathrm{H}$, $\mathrm{OH}$, $\mathrm{CO_2}$, $\mathrm{CO}$, $\mathrm{HCO}$, $\mathrm{CH_4}$, $\mathrm{CH_3}$, $\mathrm{CH_3O}$, $\mathrm{CH_3OH}$, $\mathrm{CH}$, $\mathrm{CH_2}$, $\mathrm{C}$, $\mathrm{C_2}$, $\mathrm{C_2H}$, $\mathrm{C_2H_2}$, $\mathrm{N}$, $\mathrm{N_2}$, $\mathrm{NH}$, $\mathrm{NH_2}$, $\mathrm{NH_3}$, $\mathrm{CN}$, $\mathrm{HCN}$, $\mathrm{H_2}$, $\mathrm{He}$, $\mathrm{O(^1D)}$, and $\mathrm{^1CH_2}$.
It contains 154 thermochemical reactions (and their reverse reactions) and 16 photochemical reactions, which are listed in Tables~1 and 2 of Paper~I, respectively.

For the boundary conditions, we assume the diffusion fluxes of all the species to be zero at the upper boundary and the volume mixing ratios of gaseous species to be the thermochemical equilibrium values at the lower boundary.

We prepare 165 atmospheric layers with same thickness $\Delta z$.
We place the lower boundary at 1000~bar and set the thickness $\Delta z$ to 45~km except for the cases of 10, 100, and 1000 times the Solar metallicity, for which $\Delta z$ = 40, 18, and 4~km, respectively, (see \S~\ref{metallicity}), and the case with the irradiation temperature $T_\mathrm{irr}$ = 1290~K, for which $\Delta z$ = 100~km \ikomat{(see \S~\ref{temp})}.

\subsection{Particle Growth Model} \label{subsec: pg}
We derive the steady-state number density of haze particles of various sizes at each altitude, considering the collisional growth, sedimentation, transport by eddy diffusion, and monomer production.
As for collisional growth, we consider two processes, the Brownian diffusion and gravitational collection. 
The latter is the collisional process that occurs as a result of the difference in sedimentation velocity between different-size particles.
We neglect the effect of thermal decomposition, which \cite{2017ApJ...847...32L} considered, since it does not occur in the pressure and temperature ranges considered in this study.
Same as in Paper~I, we prepare 40 volume bins, setting the volume ratio of two adjacent bins to be 3.

In Paper~I, as for monomer production, we assumed that the {integrated} mass of the monomers produced per unit time throughout the atmosphere was equal to the production rate observed in Titan's atmosphere multiplied by the ratio of the incident stellar Ly$\alpha$ flux at the planet's orbital distance to that currently received by Titan.
Then, we distributed the production rate according to the distribution of the sum of the number densities of haze precursors, which were assumed to be HCN and $\mathrm{C_2H_2}$, at each altitude.
{Those are similar \ikomat{assumptions} used in the previous studies such as \cite{2006PNAS..10318035T} and \cite{2013ApJ...775...33M}.}
Instead, in this study, we define the monomer production rate at each altitude as the sum of the photodissociation rate of $\mathrm{CH_4}$, $\mathrm{HCN}$, and $\mathrm{C_2H_2}$ {as is \ikomat{a} more realistic assumption}.
Thus, the production rate of monomers at altitude $z$, $p(v_1, z)$, is given 
as a function of the monomer volume $v_1$ and mass $m_{p, 1}$, 
\begin{equation}
p(v_1, z) = \frac{\sum_i^{\mathrm{CH_4}, \mathrm{HCN}, \mathrm{C_2H_2}} m_i J_i \left(z \right) n_i \left(z \right)}{m_{p, 1}},
\end{equation}
where $m_i$, $J_i$, and $n_i$ are the mass, photodissociation rate (the number of molecules dissociated per unit time), and number density of molecule $i$, respectively.

As for the boundary conditions, we consider that all the particles are lost at the lower boundary at a rate corresponding to the larger of the sedimentation velocity and the downward velocity imposed by the atmospheric mixing, following \cite{2010Icar..210..832L}.
As the upper boundary conditions, we set zero fluxes for all the particle sizes.
We set the lower and upper boundaries at the pressure levels of 10~bar and $1 \times 10^{-10}$~bar, respectively, and divide the atmosphere in that pressure range into 200 layers \ikomat{of same thickness}.

\subsection{Transmission Spectrum Model} \label{subsec: ts}
We use exactly the same method to model the transmission spectra as the one we used in Paper~I. 
This modeling had been already applied to WASP-80b in \cite{2014ApJ...790..108F} and for HAT-P-14b in \cite{Fukui:2016ky}.

We assume that all the parts of the sphere inside the 10~bar level are optically thick enough to block the incident stellar light completely. 
This assumption is valid, since we have confirmed that the chord optical depth at the pressure level of 10~bar is sufficiently larger than unity in the atmosphere considered in this study.
We calculate the opacity and transit depth over a wavenumber grid with a width of 0.1~$\mathrm{cm}^{-1}$.

\begin{figure}
\plotone{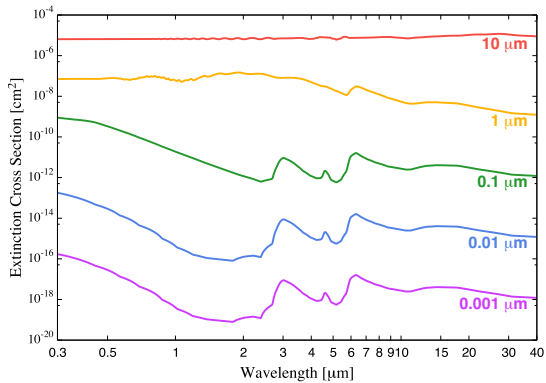}
\caption{{Extinction cross sections of the tholin-like haze particles of five different particle \kawashima{radii} of 0.001, 0.01, 0.1, 1, and 10~$\mu$m, which are smoothed with the resolution of $R = 100$.}
\kawashima{Note that this figure is the same as Figure~1 of Paper~I, but extended to longer wavelengths.}
\label{fig-opacity_haze}}
\end{figure}

We use the bhmie code \citep{RefWorks:9} to calculate the extinction opacity of haze particles.
As for the complex refractive indices of haze, we use the experimental values of tholin from \cite{1984Icar...60..127K}.
\kawashima{Note that since these are the laboratory-experimental values in a simulated Titan's atmospheric composition of 0.9$\mathrm{N_2}$/0.1$\mathrm{CH_4}$ gas mixture, the complex refractive indices of haze in atmospheres of different composition such as ones of interest in this study can be different.}
{In Figure~\ref{fig-opacity_haze}, we show the extinction cross sections of the
haze particles of five different particle \kawashima{radii}, namely, 0.001,
0.01, 0.1, 1, and 10~$\mu$m, which are smoothed with the resolution of $R = 100$. Note that the bumps found around 3.0 and 4.6~$\mu$m come from the vibrational transitions of the C-H bond and C$\equiv$N bond of the tholin-like haze particles, respectively \citep{1984Icar...60..127K}.}

For the opacities of gaseous molecules, we consider line absorption by $\mathrm{H_2O}$, $\mathrm{CO_2}$, $\mathrm{CO}$, $\mathrm{CH_4}$, $\mathrm{O_2}$, $\mathrm{NH_3}$, $\mathrm{OH}$, $\mathrm{N_2}$, $\mathrm{HCN}$, $\mathrm{C_2H_2}$, and $\mathrm{H_2}$, taking their line data from HITRAN2012 \citep{2013JQSRT.130....4R} in addition to their Rayleigh scattering and the collision-induced absorption by $\mathrm{H_2}$-$\mathrm{H_2}$ and $\mathrm{H_2}$-$\mathrm{He}$. 
We ignore the Rayleigh scattering by OH, because of its low abundance in the atmosphere and thus negligible effect.

\subsection{Model Setting and Input Parameters} \label{subsec: setting}
As described in Introduction, this study is aimed at exploring the sensitivity of the haze particle distribution and transmission spectrum especially to {UV irradiation intensity,} metallicity, carbon-to-oxygen ratio (C/O), eddy diffusion {coefficient}, and temperature in the atmosphere.
Here, we explain the model setting and input parameters.

We use observed properties of GJ1214~b and its host star: 
The stellar radius is $0.201$~$R_\sun$ \citep[][]{2013AA...551A..48A}.  
The planetary mass is $6.26$~$M_\earth$ \citep[][]{2013AA...551A..48A}.
We choose 2.07~$R_\earth$ as the planetocentric distance at 1000~bar so as to roughly match the observed transit radii of GJ~1214~b with the assumption of a clear solar-composition atmosphere {and use this value for all the atmospheric scenarios presented in this paper}.
{Note that when we investigate atmospheric compositions from observed transit depths, we often suffer from degeneracy among the reference radius and those inferred properties \citep[e.g.,][]{2017MNRAS.470.2972H}. Relative \ikomat{values of} transit depths at different wavelengths rather than the absolute \ikomat{ones} are \ikomat{useful for constraining} the atmospheric properties.}

As for the stellar UV spectrum, same as in Paper~I, we use the GJ~1214's spectrum constructed by the MUSCLES Treasury Survey \citep{2016ApJ...820...89F, Youngblood:2016ib, 2016ApJ...824..102L}. 
We adopt its version 1.1 {of} the panchromatic SED binned to a constant 1~$\mbox{\AA}$ resolution and downsample{d} in low signal-to-noise regions to avoid \ikomat{any} negative flux.
Unlike in Paper~I, however, in this study, we adopt 8~$\mbox{\AA}$  resolution since we have confirmed that this resolution is sufficient for our calculations.
We derive the 8~$\mbox{\AA}$ resolution spectrum by averaging the original data points within the wavelength range of the resolution.
{When we explore the sensitivity to UV irradiation intensity in \S~\ref{uv}, we vary the intensities of the incoming stellar UV flux at all the wavelengths.}

The elemental abundance ratios for the fiducial case are assumed to be equal to the solar system abundance ratios, which we take from Table 2 of \citet{2003ApJ...591.1220L}. 
That corresponds to C/O, O/H, and N/H of $5.010 \times 10^{-1}$, $5.812 \times 10^{-4}$, and $8.021 \times 10^{-5}$, respectively. 
When we consider high metallicity cases including 10, 100, and 1000 times the Solar metallicity {in \S~\ref{metallicity}}, we enhance the abundances of all the elements except H and He by a factor of 10, 100, and 1000, respectively. 
{The mass proportions of all the elements except H and He for the cases of 1, 10, 100, and 1000 times the Solar metallicity are $Z = 9.89 \times 10^{-3}$, $9.08 \times 10^{-2}$, 0.500, and 0.909, respectively.}
When we explore the sensitivity to the C/O ratio (i.e., 1, 10, and 1000 times the Solar C/O ratio) in \S~\ref{co}, we change its value, keeping the ratio of the sum of C and O abundances to the other element abundances unchanged.

As for the eddy diffusion coefficient $K_{zz}$, we set its fiducial value to be $1 \times 10^7$~$\mathrm{cm^2}$~$\mathrm{s^{-1}}$. 
In addition, 
we explore the cases of $K_{zz} = 1 \times 10^9$ and $1 \times 10^5$~$\mathrm{cm^2}$~$\mathrm{s^{-1}}$ in \S~\ref{eddy}.

\begin{figure}
\plotone{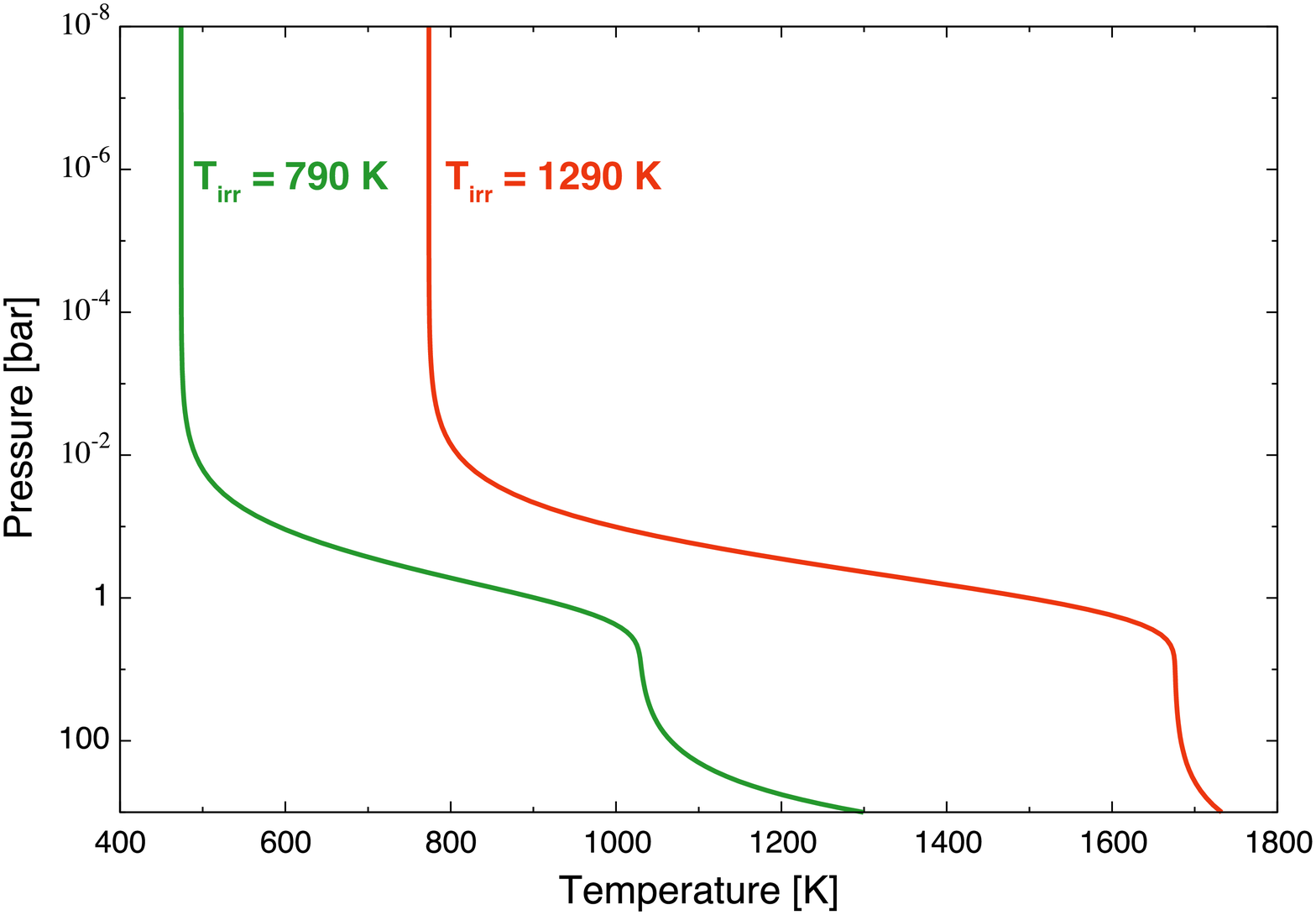}
\caption{Temperature-pressure profiles of the atmosphere that we use to explore the dependence on temperature in \S~\ref{temp}. 
The green line represents the fiducial profile, which is calculated by the use of the analytical formula of \citet{2010A&A...520A..27G}. Also shown is the profile for the irradiation temperature $T_\mathrm{irr}$ higher by 500~K than the fiducial value of 790~K, namely 1290~K, (red line) with the other parameters unchanged from the fiducial values.
\label{fig-temp}}
\end{figure}

For the temperature-pressure profile, same as in Paper~I, we use the analytical formula of Eq.~(29) of \citet{2010A&A...520A..27G}.
We choose the fiducial values of the parameters, namely, the intrinsic temperature $T_\mathrm{int}$, irradiation temperature $T_\mathrm{irr}$, averaged opacity in the optical $k_\mathrm{v}$, and that in the infrared $k_\mathrm{th}$ so as to match the temperature-pressure profile of GJ~1214b that \cite{2010ApJ...716L..74M} derived for a solar composition atmosphere under the assumption of efficient heat redistribution 
from the day and night sides.
This yields $T_\mathrm{int} = 120$~K, $T_\mathrm{irr} = 790$~K, 
$k_\mathrm{v} = 10^{-4.1}$~\ikomat{$\mathrm{cm^2}$~$\mathrm{g^{-1}}$}, and 
$k_\mathrm{th} = 10^{-2.7}$~\ikomat{$\mathrm{cm^2}$~$\mathrm{g^{-1}}$}
\footnote{Note that the values of $k_\mathrm{v}$ and $k_\mathrm{th}$ are slightly different from those used in Paper~I, in which 
$k_\mathrm{v} = 10^{-4.0}$~$\mathrm{cm^2}$~$\mathrm{g^{-1}}$ and 
$k_\mathrm{th} = 10^{-2.6}$~$\mathrm{cm^2}$~$\mathrm{g^{-1}}$.
This difference comes from the method used to solve the equation of hydrostatic equilibrium.
We used a first-order (Euler) integration method in Paper~I, but use the fourth-order (Runge-Kutta) method in this study.
We have confirmed that the difference coming from the different choice of $k_\mathrm{v}$ and $k_\mathrm{th}$ has little effect on our results and conclusions of Paper~I, as realized by comparison between Figure~2 of Paper~I and Figure~\ref{fig-photo}(a) of this paper (the differences being in the value of $k_\mathrm{v}$ and $k_\mathrm{th}$, the integration method for the equation of hydrostatic equilibrium, and the resolution of the input stellar spectrum only.)}.
Even when we consider the atmosphere with elemental abundance ratios other than the solar ones, we use the same parameters, since the focus of this study is on the effect of elemental abundance ratios.
In \S~\ref{temp}, we explore the case where $T_\mathrm{irr}$ is higher by 500~K than the fiducial value of 790~K, namely 1290~K, keeping the other parameters unchanged from the fiducial values to investigate the dependence on temperature.
Figure~\ref{fig-temp} shows the temperature-pressure profiles of the atmosphere that we use as the fiducial (green line) and higher-$T_\mathrm{irr}$ cases, which we use in \S~\ref{temp}.

Finally, as for the monomer \kawashima{radius}, we adopt 1~nm.
Thus, \kawashima{radius} bins range from 1~nm to 1600~$\mu$m.
{We have confirmed it has a little effect on transmission spectra. A brief discussion is made in \S~\ref{monomers}}.
\kawashima{For the value of the material density of haze particles, we adopt 1.0~g~$\mathrm{cm^{-3}}$ as the widely used value for the microphysical models of Titan's haze \citep[e.g.,][]{1992Icar...95...24T, 2011ApJ...728...80L}. We also make a brief discussion on this effect in \S~\ref{density}.}

%% file: result.tex
\section{Results} \label{results}
In this section, we present the results of our numerical simulations. 
First, we overview the distributions of gaseous species and haze particles and transmission spectra that we obtain in the fiducial case in \S~\ref{fiducial}.
Then, we explore the dependence of the transmission spectra on {UV irradiation intensity in \S~\ref{uv}}, metallicity in \S~\ref{metallicity}, C/O ratio in \S~\ref{co}, eddy diffusion coefficient in \S~\ref{eddy}, and temperature in \S~\ref{temp}.

\subsection{Fiducial Case} \label{fiducial}
\subsubsection{\ikomat{Photochemistry and Haze Precursor Production}}
\begin{figure*}
\gridline{\fig{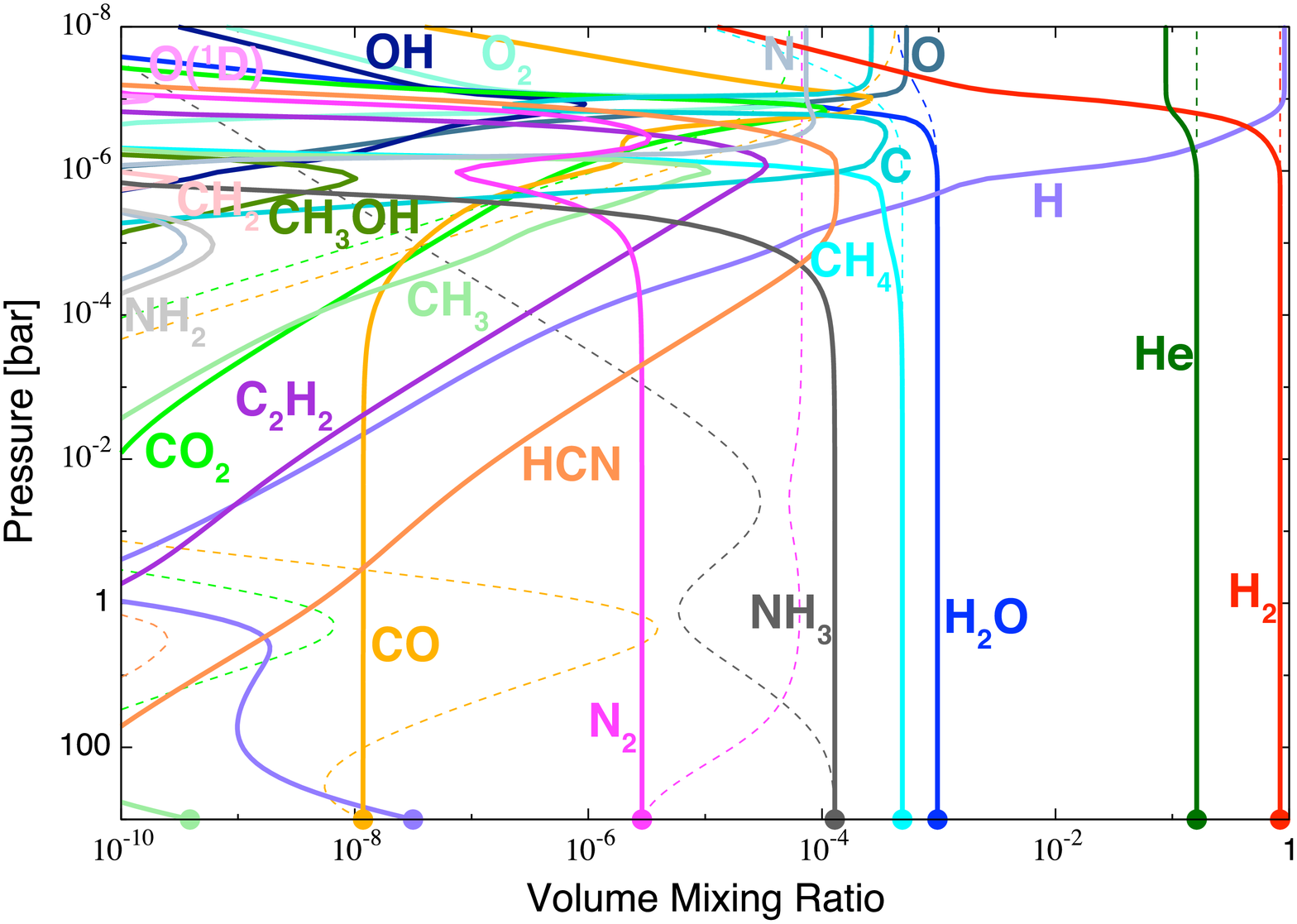}{0.5\textwidth}{(a)}
\fig{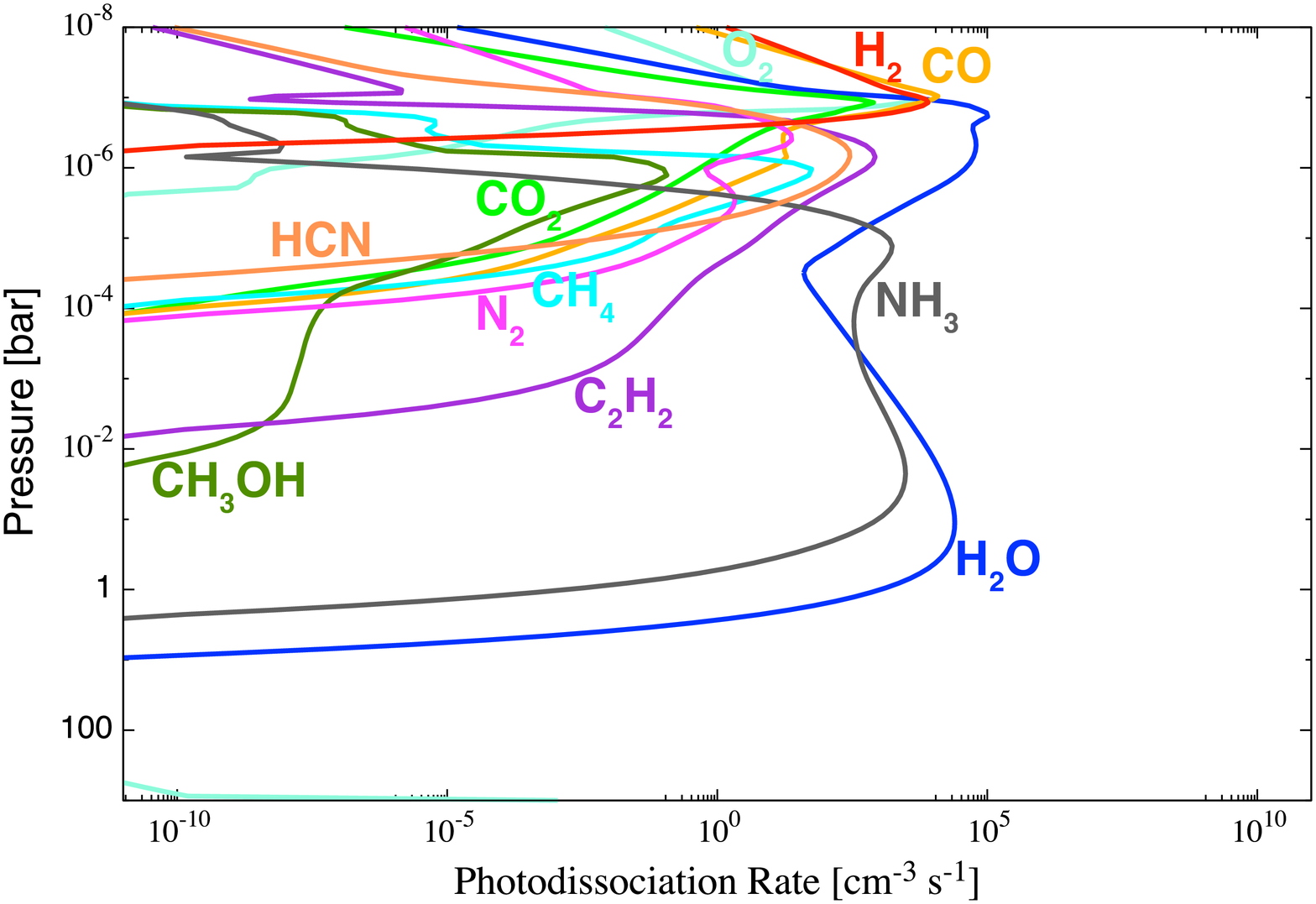}{0.5\textwidth}{(b)}
}
\caption{
Photochemical properties of the atmosphere for the fiducial case.
(a)~Vertical distributions of gaseous species in photochemical equilibrium (solid line). 
The filled circles represent the thermochemical equilibrium values at the lower boundary. 
For reference, shown by the dashed lines are the abundances in thermochemical equilibrium, where the eddy diffusion transport is also ignored.
(b)~Vertical profile of the photodissociation rate of each species, namely, the number of molecules that photodissociate per unit volume and unit time.
\kawashima{Note that Figure~(a) is basically the same as Figure~2 of Paper~I with the differences being in the value of $k_\mathrm{v}$ and $k_\mathrm{th}$, the integration method for the equation of hydrostatic equilibrium, and the resolution of the input stellar spectrum only.}
\label{fig-photo}}
\end{figure*}

In Figure~\ref{fig-photo}, we show the results of our photochemical calculations for the fiducial case.
Although these results are basically the same as those of \S~3.1 of Paper~I, we show them because they are helpful in interpreting our later results.
Figure~\ref{fig-photo}~(a) shows the calculated vertical distributions of gaseous species in the photochemical equilibrium state (solid lines). 
For reference, we also present the distributions obtained by thermochemical equilibrium calculations (dashed lines) that ignore photochemical processes and eddy diffusion transport. 
In the lower atmosphere ($P \gtrsim 10^{-4}$~bar), $\mathrm{NH_3}$, $\mathrm{N_2}$, and CO are found to have constant values of the volume mixing ratio equal to their lower boundary values because of the eddy diffusion mixing.
In the upper atmosphere ($P \lesssim 10^{-4}$~bar), many species that hardly exist in thermochemical equilibrium (i.e., H, O, C, HCN, N, $\mathrm{O_2}$, $\mathrm{C_2H_2}$, $\mathrm{CH_3}$, OH, $\mathrm{CH_3OH}$, $\mathrm{NH_2}$, $\mathrm{CH_2}$, and $\mathrm{O(^1D)}$) are produced due to photochemical reactions, and H is the most abundant species.

\begin{deluxetable*}{llcccc}
\tablecaption{Integrated photodissociation rates of haze precursors \label{tab-photo}}
\tablehead{
\colhead{} & \colhead{Figure} & \colhead{$\mathrm{CH_4}$} & \colhead{HCN} & \colhead{$\mathrm{C_2H_2}$} & \colhead{Total}
}
\startdata
Fiducial & Fig.~\ref{fig-photo}~(b) & $1.99 \times 10^{-14}$ & $4.07 \times 10^{-13}$ & $9.49 \times 10^{-13}$ & $1.38 \times 10^{-12}$ \\
UV $\times 10^5$ & Fig.~\ref{fig-photo_uv}~(b) & $6.05 \times 10^{-16}$ & $1.02 \times 10^{-12}$ & $5.33 \times 10^{-9}$ & $5.33 \times 10^{-9}$ \\
UV $\times 10^{2.5}$ & Fig.~\ref{fig-photo_uv}~(d) & $1.78 \times 10^{-13}$ & $1.25 \times 10^{-11}$ & $7.69 \times 10^{-11}$ & $8.96 \times 10^{-11}$ \\
UV $\times 10^{-2.5}$ & Fig.~\ref{fig-photo_uv}~(f) & $1.25 \times 10^{-15}$ & $1.86 \times 10^{-15}$ & $1.01 \times 10^{-15}$ & $4.12 \times 10^{-15}$ \\
UV $\times 10^{-5}$ & Fig.~\ref{fig-photo_uv}~(h) & $1.63 \times 10^{-16}$ & $5.34 \times 10^{-19}$ & $3.01 \times 10^{-23}$ & $1.64 \times 10^{-16}$ \\
10 $\times$ Solar & Fig.~\ref{fig-photo_metallicity}~(b) & $3.03 \times 10^{-16}$ & $1.17 \times 10^{-13}$ & $3.94 \times 10^{-13}$ & $5.12 \times 10^{-13}$ \\
100 $\times$ Solar & Fig.~\ref{fig-photo_metallicity}~(d) & $9.42 \times 10^{-17}$ & $2.37 \times 10^{-14}$ & $1.62 \times 10^{-13}$ & $1.86 \times 10^{-13}$ \\
1000 $\times$ Solar & Fig.~\ref{fig-photo_metallicity}~(f) & $1.80 \times 10^{-13}$ & $9.04 \times 10^{-15}$ & $1.00 \times 10^{-20}$ & $1.89 \times 10^{-13}$ \\
$\mathrm{C/O} = 1$ & Fig.~\ref{fig-photo_co}~(b) & $4.88 \times 10^{-14}$ & $4.59 \times 10^{-13}$ & $2.47 \times 10^{-12}$ & $2.98 \times 10^{-12}$ \\
$\mathrm{C/O} = 10$ & Fig.~\ref{fig-photo_co}~(d) & $1.90 \times 10^{-13}$ & $5.98 \times 10^{-13}$ & $1.81 \times 10^{-11}$ & $1.89 \times 10^{-11}$ \\
$\mathrm{C/O} = 1000$ & Fig.~\ref{fig-photo_co}~(f) & $1.64 \times 10^{-13}$ & $5.61 \times 10^{-13}$ & $7.73 \times 10^{-11}$ & $7.80 \times 10^{-11}$ \\
$K_{zz} = 1 \times 10^9$~$\mathrm{cm}^2$~$\mathrm{s}^{-1}$ & Fig.~\ref{fig-photo_eddy}~(b) & $4.12 \times 10^{-13}$ & $5.68 \times 10^{-13}$ & $5.84 \times 10^{-15}$ & $9.86 \times 10^{-13}$ \\
$K_{zz} = 1 \times 10^5$~$\mathrm{cm}^2$~$\mathrm{s}^{-1}$ & Fig.~\ref{fig-photo_eddy}~(d) & $1.71 \times 10^{-14}$ & $3.66 \times 10^{-13}$ & $1.08 \times 10^{-12}$ & $1.46 \times 10^{-12}$ \\
$T_\mathrm{irr} = 1290$~K & Fig.~\ref{fig-photo_temp}~(b) & $7.75 \times 10^{-15}$ & $2.23 \times 10^{-13}$ & $2.06 \times 10^{-14}$ & $2.51 \times 10^{-13}$ \\
\enddata
\tablecomments{
Photodissociation rates (in g~$\mathrm{cm^{-2}}$~$\mathrm{s^{-1}}$) of \ikomat{the} haze precursors, $\mathrm{CH_4}$, HCN, and $\mathrm{C_2H_2}$, integrated over the calculation range of \ikomat{pressure in our} particle growth simulations, namely, between 10$^{-10}$ and 10~bar. 
The sum of the photodissociation rates of these haze precursors (the 6th column) \ikomat{is} assumed as the monomer production rate in the particle growth simulations. Note the unit of photodissociation rate is different from that used in the figures for photodissociation rate profiles. 
Here {"UV $\times n$" and "$m \times$ Solar" mean the UV irradiation intensity and atmospheric metallicity are $n$ and $m$ times higher than \ikomat{their} fiducial \ikomat{values}, respectively}, C/O is the carbon-to-oxygen abundance ratio, $K_\mathrm{zz}$ is \ikomat{the} eddy diffusion coefficient, $T_\mathrm{irr}$ is the irradiation temperature. The fiducial values are {$m = 1$}, $n = 1$, C/O = 0.5, $K_\mathrm{zz} = 1 \times 10^7$~cm$^2$ s$^{-1}$, and $T_\mathrm{irr}$ = 790~K.
}
\end{deluxetable*}

In Figure~\ref{fig-photo}~(b), we plot the vertical profile of the photodissociation rate of each species, namely, the number of molecules that photodissociate per unit volume and unit time. 
Note that the photodissociation rate of $\mathrm{O_2}$ (light blue) just above the lower boundary (1000~bar) is calculated from its volume mixing ratio whose value is out of our computational precision, so that its value is physically meaningless.
One finds that $\mathrm{H_2}$, CO, $\mathrm{H_2O}$, and $\mathrm{NH_3}$ make major contribution to absorption of stellar photons via photodissociation. 
Among the haze precursors ($\mathrm{CH_4}$, HCN, and $\mathrm{C_2H_2}$), $\mathrm{C_2H_2}$ photodissociates most in spite of the lowest abundance. 
This is because $\mathrm{C_2H_2}$ can use photons of low-energy, namely, incoming stellar flux at long wavelength\ikomat{s} {(\ikomat{up to} {$\sim$}220~nm)}, for its photodissociation, unlike the other two molecules. 
Therefore, in our simulations, $\mathrm{C_2H_2}$ makes the greatest contribution to the haze monomer production. 
Also, in Table~\ref{tab-photo}, we tabulate the photodissociation rate\ikomat{s} of \ikomat{the} haze precursors integrated over the calculation range of \ikomat{pressure in our} particle growth simulations, between 10$^{-10}$ and 10~bar. 
The sum of the photodissociation rate\ikomat{s} of these haze precursors (the 6th column of the table) is assumed as the monomer production rate in the particle growth simulations. 

\subsubsection{Particle Growth}
\begin{figure}
\plotone{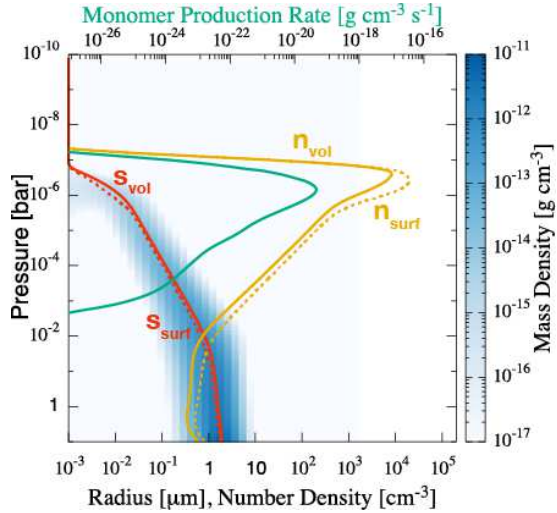}
\caption{Vertical profiles of the volume average radius $s_\mathrm{vol}$ (red solid line) and number density $n_\mathrm{vol}$ (orange solid line), and the surface average radius $s_\mathrm{surf}$ (red dashed line) and number density $n_\mathrm{surf}$ (orange dashed line), along with that of the monomer mass production rate (green solid line)\ikomat{, for the fiducial case}. See the text for the definition of each quantity.
The mass densities for all the size bins at each pressure level are also plotted with the blue color contour.
\kawashima{Note that the mass density here refers not to material density of haze particles, but to their spatial density.}
\label{fig-growth}}
\end{figure}

Figure~\ref{fig-growth} shows the calculated vertical profiles of haze properties. Here, we define the volume average radius $s_\mathrm{vol}$ (red solid line) 
and surface average radius $s_\mathrm{surf}$ (red dashed line), respectively, as 
\begin{equation}
s_\mathrm{vol} = \frac{\sum_{i = 1}^{\mathcal{N}} n \left(s_i \right) s_i^4}{\sum_{i = 1}^{\mathcal{N}} n \left(s_i \right) s_i^3}, 
\hspace{1ex}
s_\mathrm{surf} = \frac{\sum_{i = 1}^{\mathcal{N}} n \left(s_i \right) s_i^3}{\sum_{i = 1}^{\mathcal{N}} n \left(s_i \right) s_i^2},
\end{equation}
where $n \left(s_i \right)$ is the number density of particles with radius $s_i$ and $\mathcal{N}$ is the total number of volume bins used in the calculation\ikomat{s}.
If the two average sizes \ikomat{coincide} with each other at a certain altitude, the size distribution is unimodal at the altitude.
The volume average number density $n_\mathrm{vol}$ (orange solid line) and the surface average number density $n_\mathrm{surf}$ (orange dashed line) are calculated, respectively, as
\begin{equation}
n_\mathrm{vol} = \frac{\sum_{i = 1}^{\mathcal{N}} n \left(s_i \right) s_i^3}{s_\mathrm{vol}^3},
\hspace{1ex}
n_\mathrm{surf} = \frac{\sum_{i = 1}^{\mathcal{N}} n \left(s_i \right) s_i^3}{s_\mathrm{surf}^3}.
\end{equation}
Additionally, the mass densities for all the size bins at each pressure level are plotted with the blue color contour and the vertical profile of the monomer mass production rate is plotted with the green solid line.

In Figure~\ref{fig-growth}, one finds that both average radii increase from $1 \times 10^{-3}$ to $\sim 2$~$\mu$m via collisional growth.
The number densities take their peak values at $\sim$ 2--4 $\times 10^{-7}$~bar. 
This is because, below this altitude, the atmospheric pressure is high enough that the collisional growth occurs via the Brownian diffusion rapidly compared to the monomer production.
Change of the trend found at $P \sim 10^{-2}$~bar results from the transition in gas drag law from the slip-flow to Stokes-flow regimes (see Paper~I for the details), increasing the sedimentation velocity and thereby inhibiting collision between particles.
The slight difference between $s_\mathrm{vol}$ and $s_\mathrm{surf}$ means that the haze contains different size particles at each altitude. The color contour indicates that particles in some narrow range of size are abundant at each altitude.

{Here we note that compared to the fiducial case of Paper~I, the {integrated} monomer production rate \ikomat{adopted in Fig.~\ref{fig-growth}} is smaller \ikomat{by a factor $\sim 40$} and \ikomat{is distributed only} in the upper atmosphere. As a result, both the mass density and average radii of the haze particles are slightly smaller and result in the spectrum with more prominent molecular-absorption features as shown in next section (see also Figs.~6 and 9 of Paper~I).}

\subsubsection{Transmission Spectrum Models}
\begin{figure}
\plotone{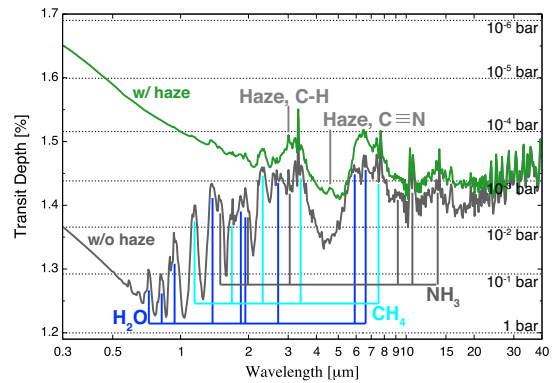}
\caption{Transmission spectrum models for the atmosphere with haze (green line) and without haze (black line) \ikomat{for the fiducial case}. The horizontal dotted lines represent the transit depths corresponding to the pressure levels from $1 \times 10^{-6}$~bar to 1~bar. Note that the transmission spectrum models are smoothed for clarity 
{with the resolution \ikomat{$R$} of 100}.
\label{fig-spectra}}
\end{figure}

Figure~\ref{fig-spectra} shows the transmission spectrum models for the atmosphere with haze (green line) and without haze (black line).
The relative cross section of the planetary disk with a radius corresponding to a certain pressure level is presented by horizontal dotted lines from $P = 1 \times 10^{-6}$~bar to 1~bar. 
Note that the transmission spectrum models are smoothed for clarity {with the resolution \ikomat{$R$ of 100}}. 
We use the same smoothing method for \ikomat{all the spectrum models shown below}.

In the spectrum model for the atmosphere without haze (black line), the major spectral features come from absorption features of $\mathrm{H_2O}$, $\mathrm{CH_4}$, and $\mathrm{NH_3}$.
Distinct features of $\mathrm{H_2O}$ are found around 0.7, 0.8, 0.9, 1.4, 1.8, 1.9, 2.5--2.9, 5.9, and 6.5~$\mu$m, those of $\mathrm{CH_4}$ around 1.2, 1.7, 2.3, 3.3, and 7.2--8.4~$\mu$m, and those of $\mathrm{NH_3}$ around 1.5, 2.0, 3.0, and 8.4--16.4~$\mu$m. The Rayleigh scattering feature mainly due to $\mathrm{H_2}$ can be seen in the optical wavelength region.

The spectrum for the atmosphere with haze (green line) is relatively featureless, compared to that for the atmosphere without haze (black line). 
This is because the haze particles in the upper atmosphere ($P \lesssim 10^{-3}$~bar) make the atmosphere optically thick and prevent the molecules in the lower atmosphere ($P \gtrsim 10^{-3}$~bar) from showing their absorption features. 
One finds that the absorption features at shorter wavelengths are more obscured by haze particles, since small haze particles of $\lesssim 1$~$\mu$m floating in the upper atmosphere ($P \lesssim 10^{-3}$~bar) have larger opacity at shorter wavelengths{, producing \ikomat{the so-called} Rayleigh-scattering slope,} and roughly speaking, the optical depth of the clear atmosphere is smaller at shorter wavelengths.
The small features of $\mathrm{CH_4}$ above $10^{-3}$~bar can be seen at 2.3, 3.3, and 7.2--8.4~$\mu$m, those of $\mathrm{H_2O}$ at {2.5--2.9 and }5.9~$\mu$m, and those of $\mathrm{NH_3}$ at 8.4--16.4~$\mu$m.
Also, the spectral features of the haze particles appear at 3.0, 4.6{, and 6.3}~$\mu$m, respectively.
%

%% file: result_uv.tex
\subsection{Dependence on UV Irradiation Intensity} \label{uv}
{
\ikomat{When exploring} the dependence of the transmission spectra on UV irradiation intensity in Paper~I, we defined the monomer production rate \ikomat{on} the assumption similar to those \ikomat{adopted} in \cite{2006PNAS..10318035T} and \cite{2013ApJ...775...33M}. 
In this paper, \ikomat{more realistically,} we \ikomat{define} the monomer production rate as the sum of the photodissociation rate of the hydrocarbons $\mathrm{CH_4}$, $\mathrm{HCN}$, and $\mathrm{C_2H_2}$ (see \S~\ref{subsec: pg}).
}

\begin{figure}
\plotone{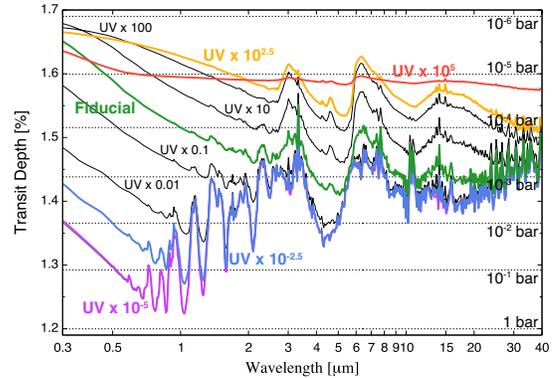}
\caption{{
Transmission spectrum models for the atmosphere with haze for \ikomat{different nine choices of UV irradiation intensity: $10^5$ (red line), $10^{2.5}$ (yellow line), 100, 10, 0.1, 0.01 (all represented by black thin lines), $10^{-2.5}$ (blue line), and $10^{-5}$ (purple line) times the fiducial value}
(green line, same as the green line in Figure~\ref{fig-spectra}). 
\ikomat{All the lines except the fiducial one are labeled ``UV $\times$ $n$'' ($n$ = {$10^{5}$, $10^{2.5}$, 100, 10, 0.1, 0.01, $10^{-2.5}$, $10^{-5}$}}).
The haze-free transmission spectrum for the atmosphere without haze for the fiducial \ikomat{UV intensity} is also plotted (black line, same as the black line in Fig.~\ref{fig-spectra}), but can be hardly seen because it overlaps with that for the atmosphere with haze in the case of UV~$\times 10^{-5}$ (purple line).
As in Fig.~\ref{fig-spectra}, horizontal dotted lines represent the transit depths corresponding to the pressure levels from $1 \times 10^{-6}$~bar to 1~bar for the atmosphere in the fiducial UV case.
Note that the transmission spectrum models are smoothed for clarity.
}
\label{fig-spectra_uv}}
\end{figure}

{Figure~\ref{fig-spectra_uv} shows the transmission spectrum models with haze for \ikomat{different nine choices of} the UV irradiation intensity.
From this figure, we realize that the transmission spectrum varies significantly with the UV irradiation intensity. 
\ikomat{This is} because the monomer production rate becomes higher with increasing UV irradiation intensity \ikomat{(see \S~\ref{uv_photo})}. 
In the \ikomat{UV$\times 10^{5}$ case} (red line), the spectrum is \ikomat{characterized by} flat\ikomat{ness} \ikomat{broadly in the infrared and by a Rayleigh-scattering slope in the optical.} 
\ikomat{The former is due to the relatively} large ($\sim 0.1$~$\mu$m) haze particles floating at high altitudes ($P \sim 10^{-5}$~bar), 
{while the latter is} \ikomat{due to the smaller} haze particles 
\ikomat{at higher altitudes} 
\ikomat{(see \S~\ref{uv_growth})}.
In the \ikomat{UV$\times 10^{2.5}$ case} (yellow line), the absorption features of the haze can be seen at 3.0, 4.6, and 6.3~$\mu$m in addition to the Rayleigh scattering in the optical. 
It is notable that the transit depths in the optical and at 3.0 and 6.3~$\mu$m \ikomat{(i.e., haze absorption features)} are larger than \ikomat{those in the UV$\times 10^{5}$ case, in spite of the smaller monomer production rate}. This is because the monomer production occurs at higher altitudes for lower UV irradiation \ikomat{intensities} as shown \ikomat{below} and also, the opacities of haze particles at these wavelengths are large.
As the UV irradiation intensity decreases, the overall transit depth becomes \ikomat{smaller} and the molecular-absorption features become more prominent. 
In \ikomat{the UV$\times 10^{-5}$ case} (purple line), the spectrum is almost the same as that \ikomat{for} the atmosphere without haze (black line).
Note that since the monomer production rate assumed \ikomat{here} is slightly smaller than that in Paper~I, the transmission spectrum models \ikomat{for} the same UV irradiation intensity have more distinct Rayleigh scattering slope and/or absorption features\ikomat{, as found by comparison} with Figure~14 of Paper~I.
}

\subsubsection{\ikomat{Photochemistry} {and Haze Precursor Production}} \label{uv_photo}
\begin{figure*}
\gridline{
\fig{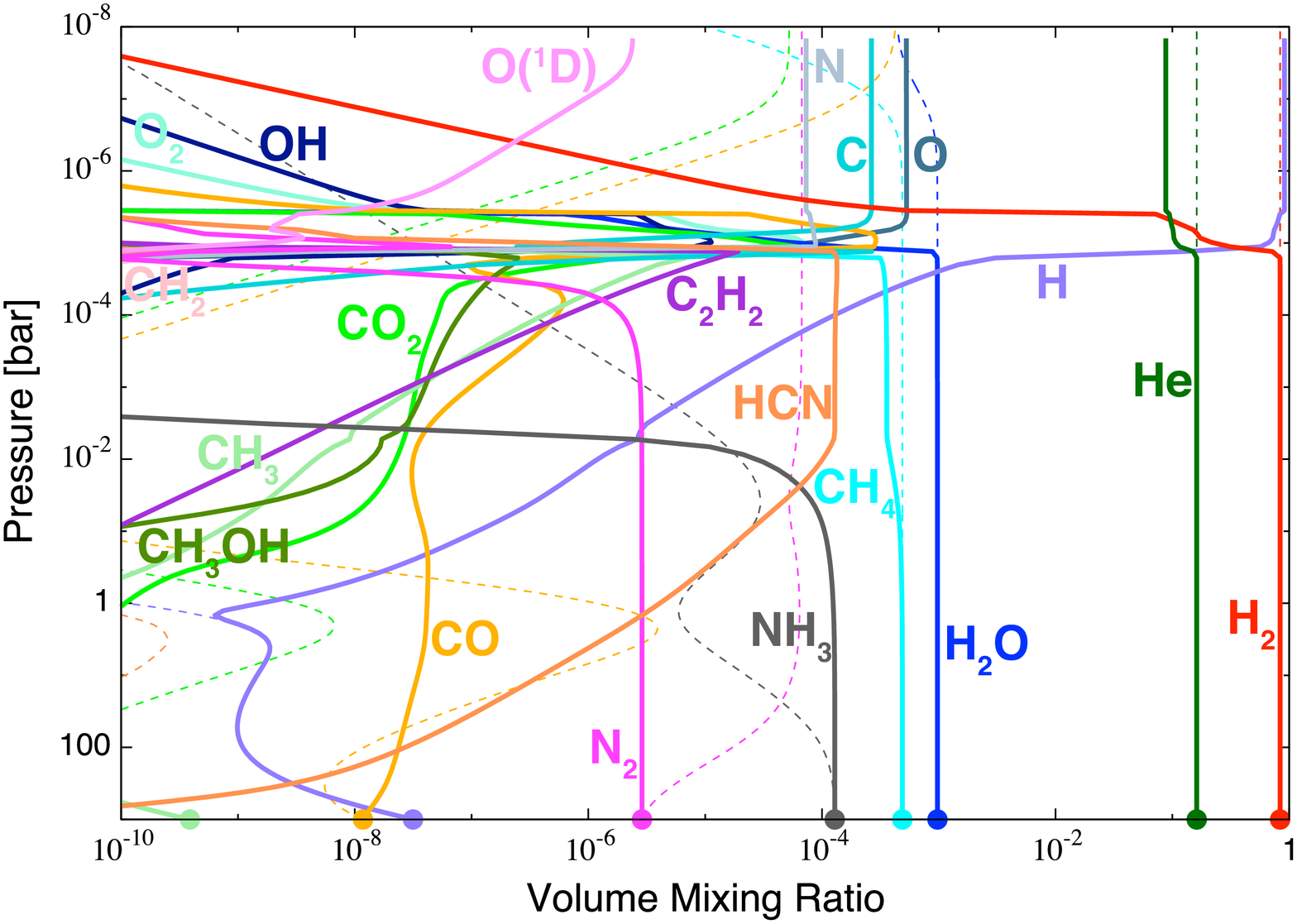}{0.4\textwidth}{(a) UV $\times 10^{5}$}
\fig{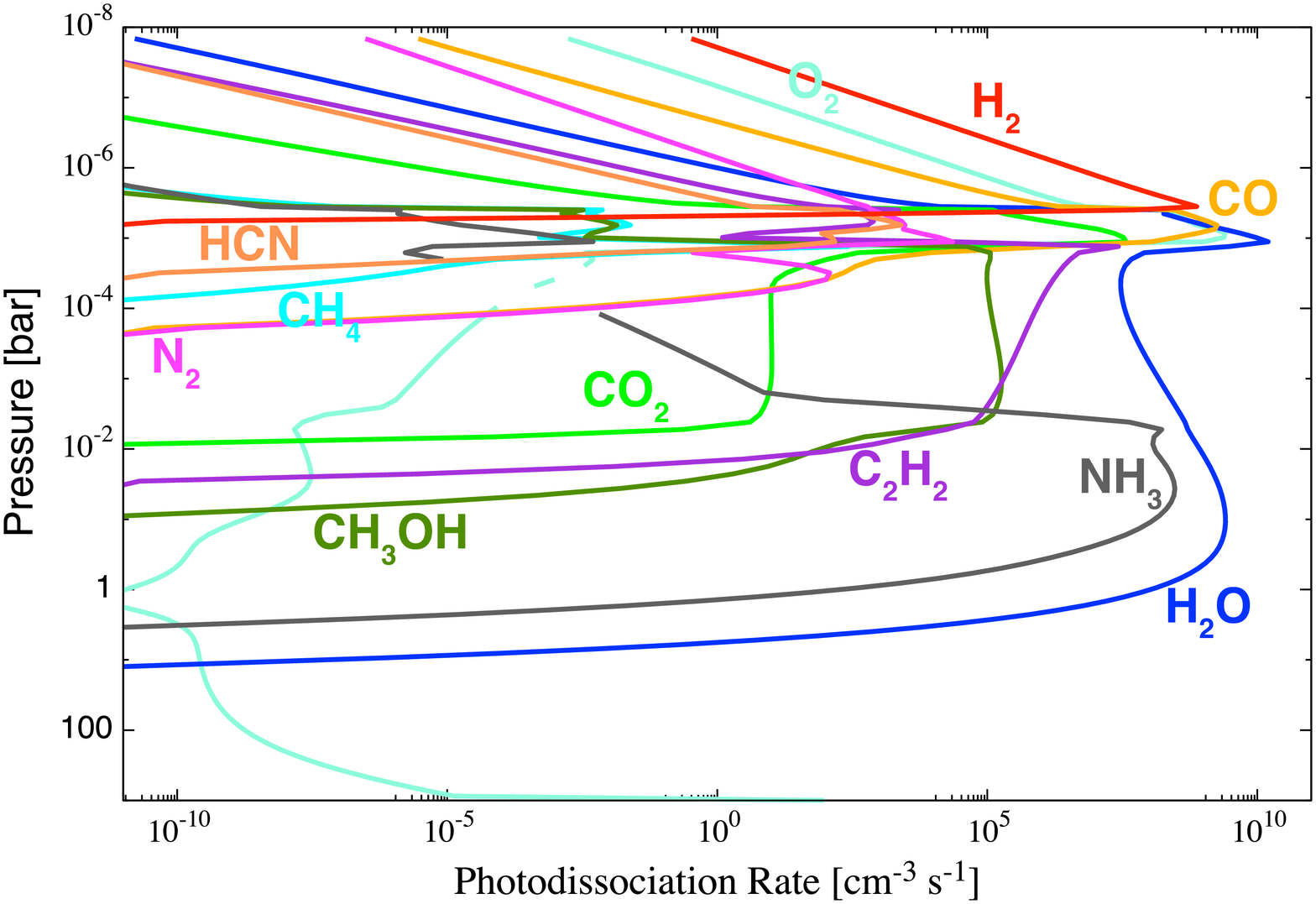}{0.4\textwidth}{(b) UV $\times 10^{5}$}
}
\gridline{
\fig{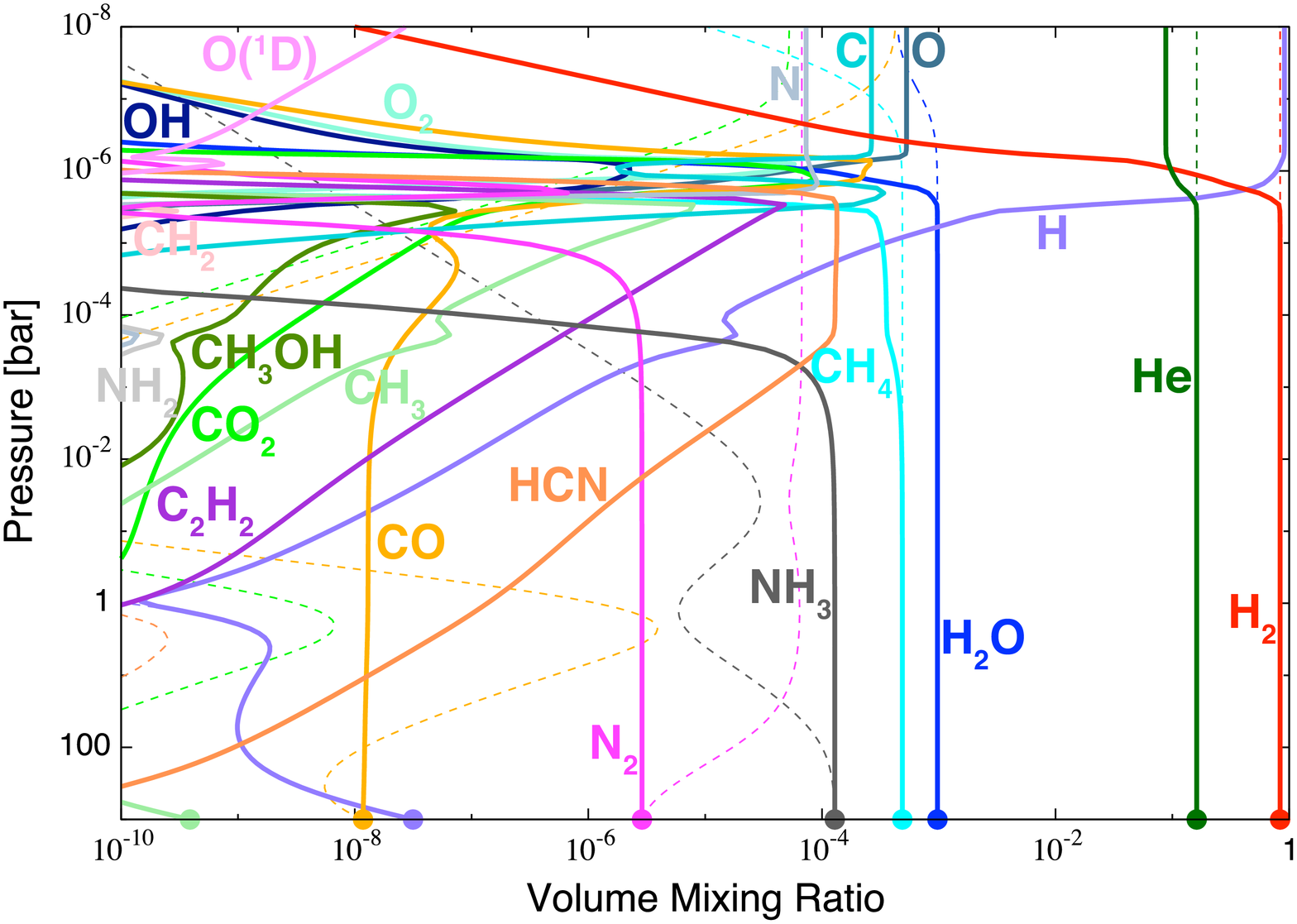}{0.4\textwidth}{(c) UV $\times 10^{2.5}$}
\fig{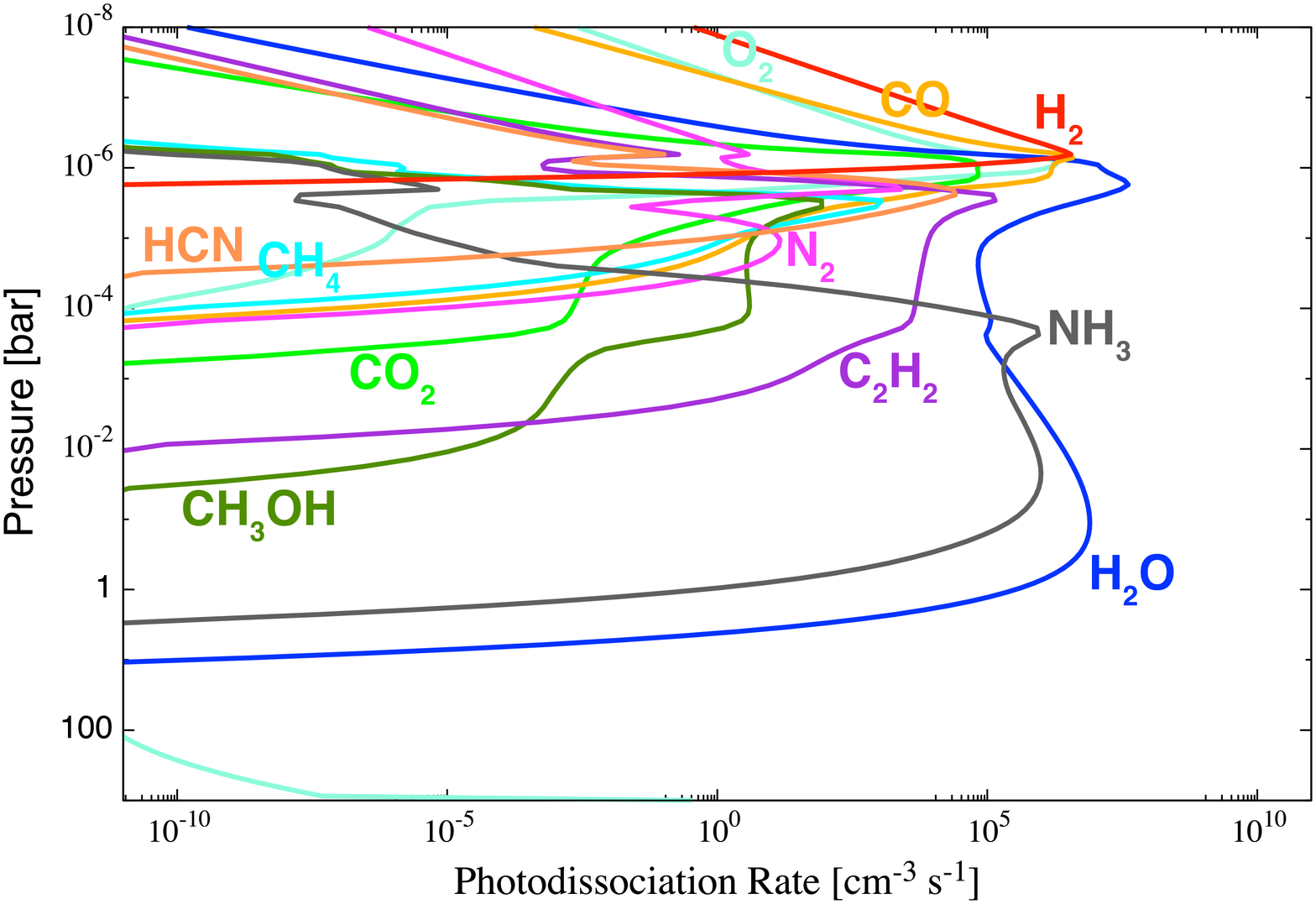}{0.4\textwidth}{(d) UV $\times 10^{2.5}$}
}
\gridline{
\fig{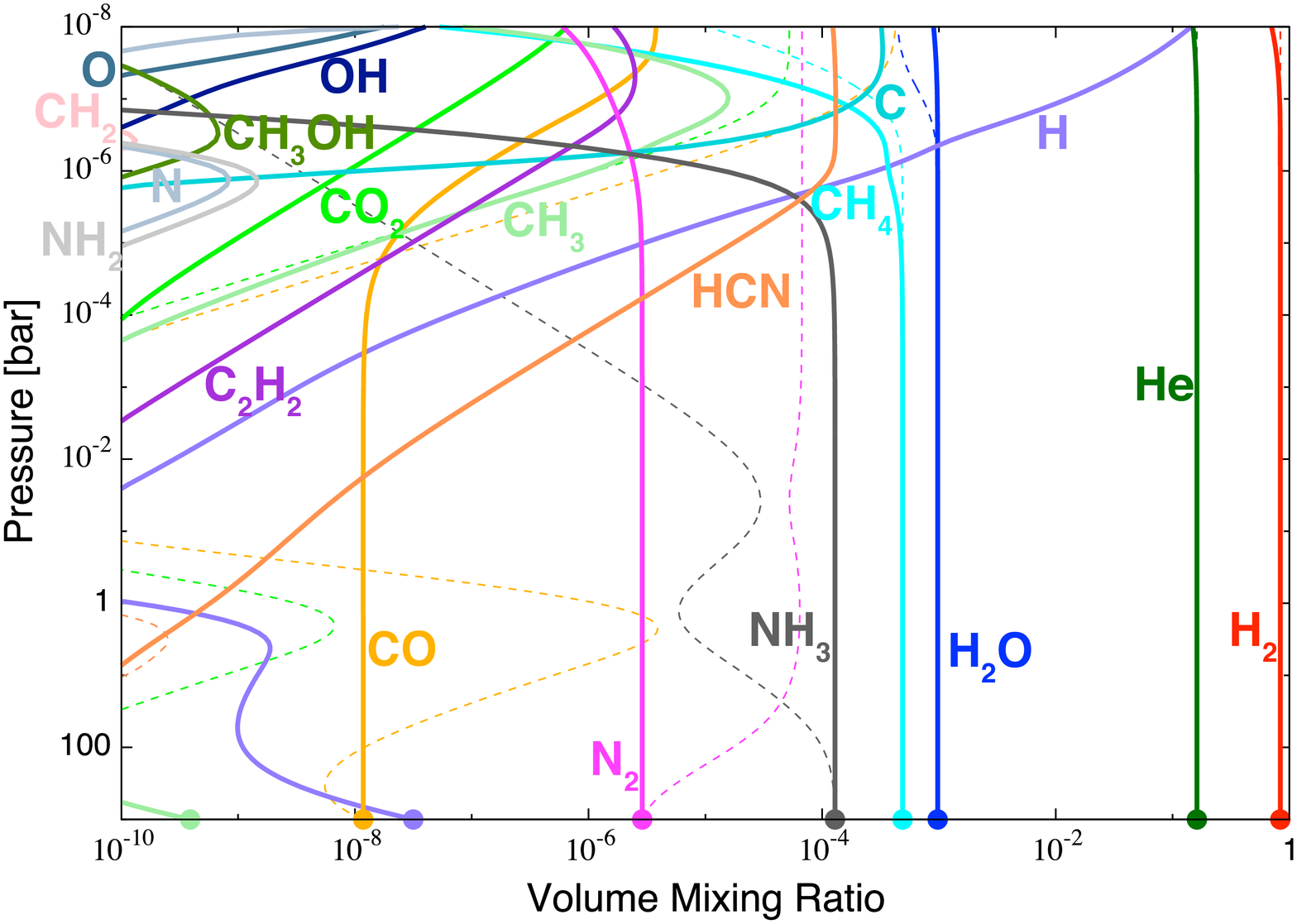}{0.4\textwidth}{(e) UV $\times 10^{-2.5}$}
\fig{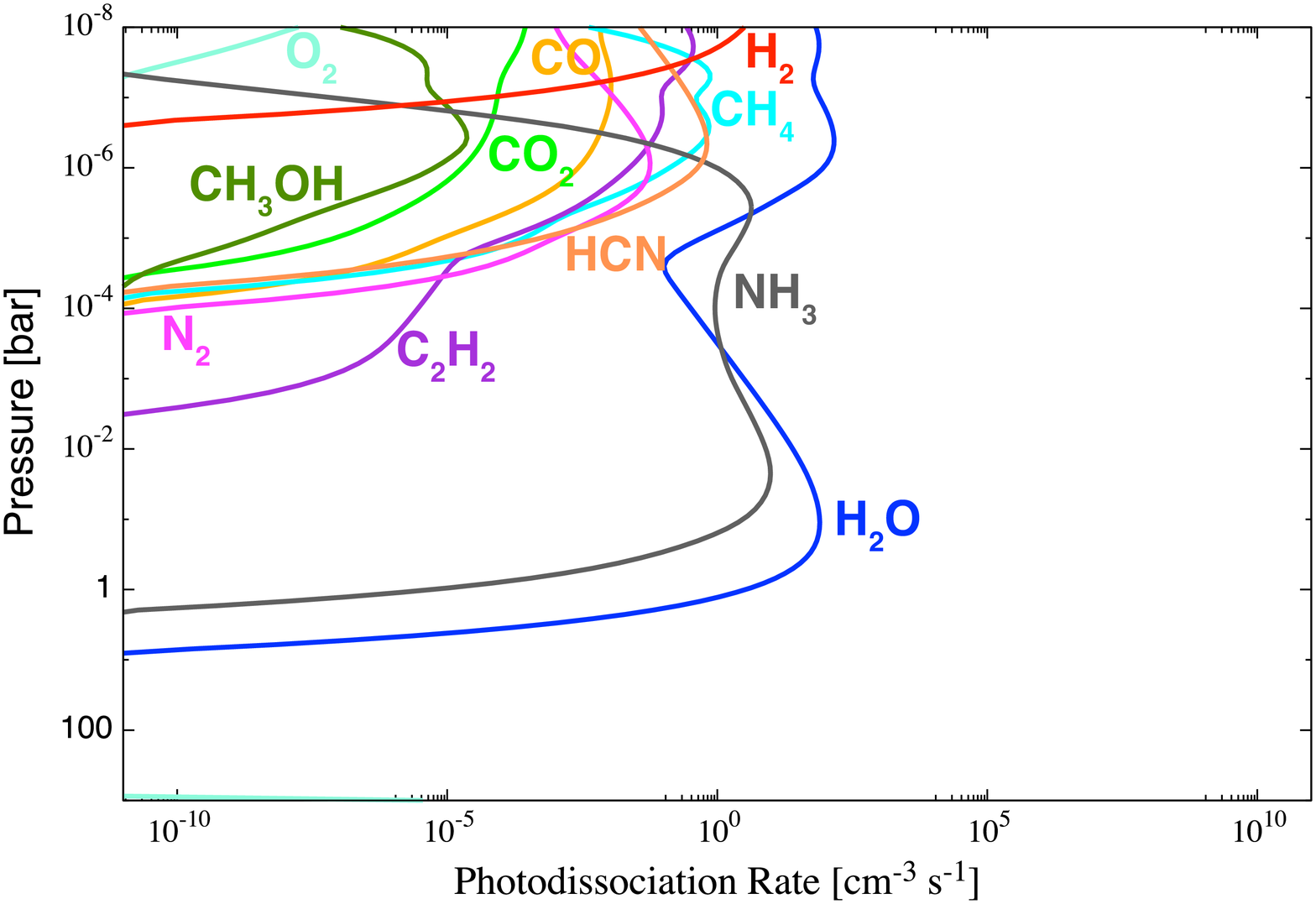}{0.4\textwidth}{(f) UV $\times 10^{-2.5}$}
}
\gridline{
\fig{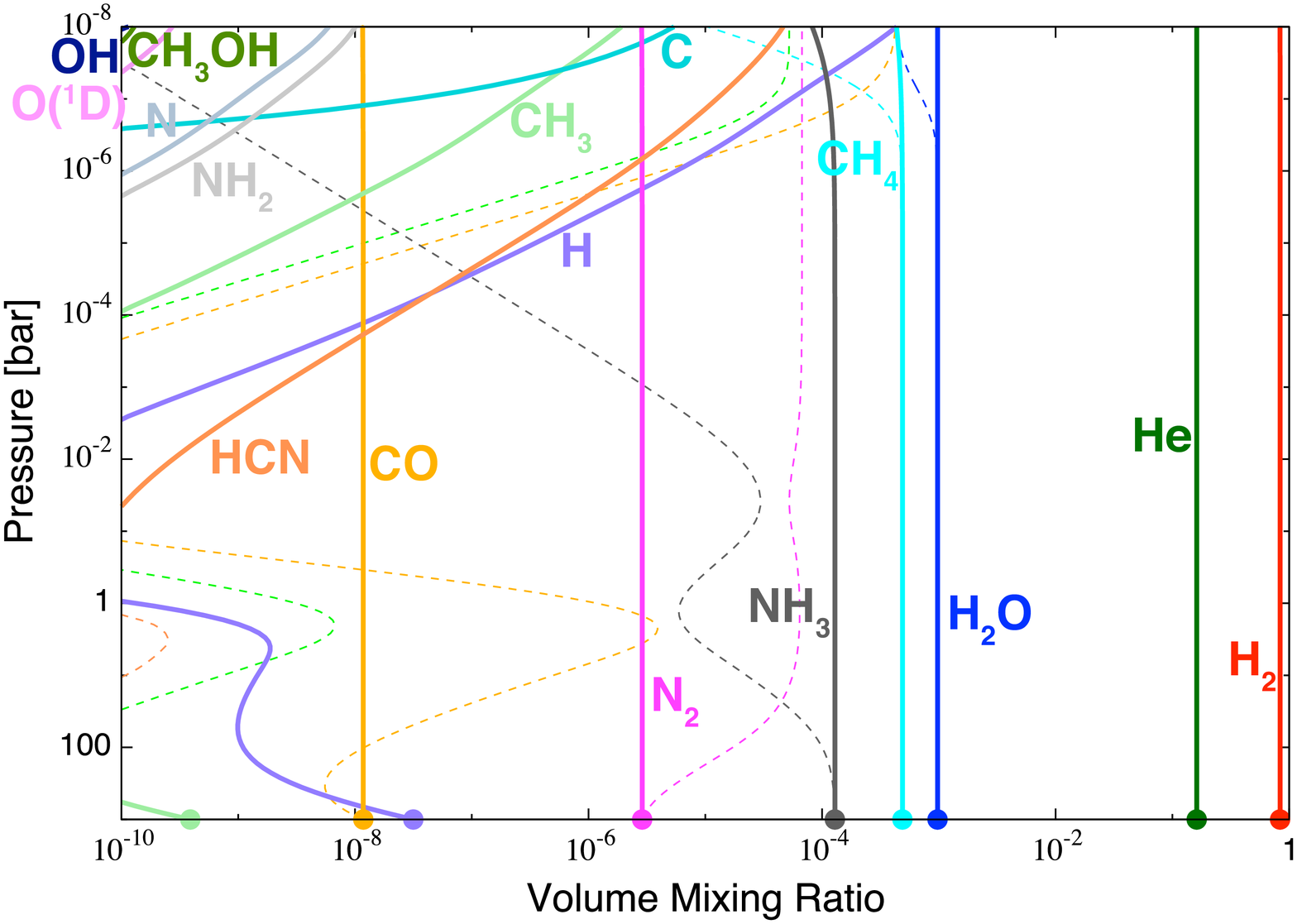}{0.4\textwidth}{(g) UV $\times 10^{-5}$}
\fig{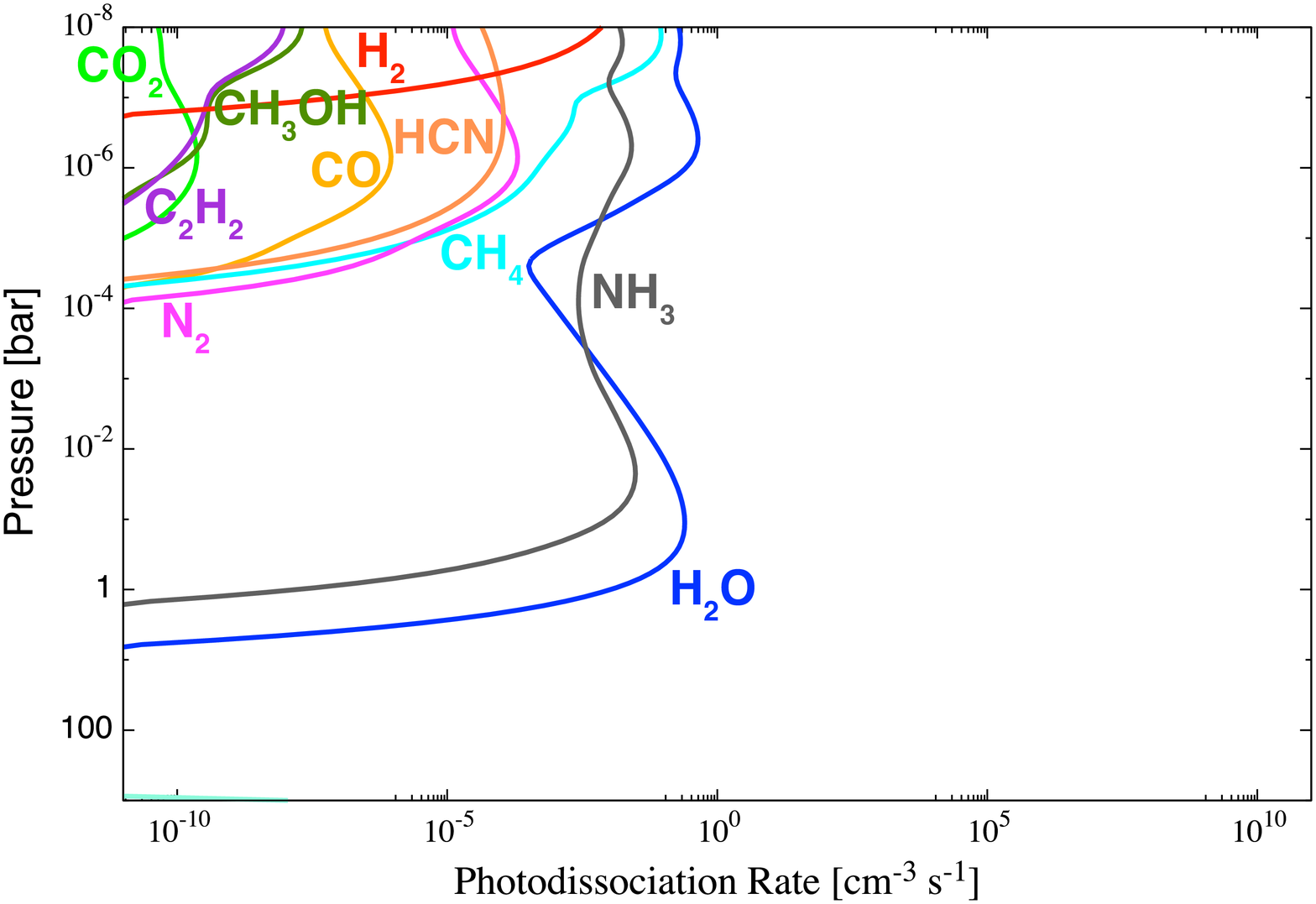}{0.4\textwidth}{(h) UV $\times 10^{-5}$}
}
\caption{{Same as Fig.~\ref{fig-photo} but for the cases of (a, b)~UV~$\times 10^{5}$, (c, d)~UV~$\times 10^{2.5}$, (e, f)~UV~$\times 10^{-2.5}$, and (g, h)~UV~$\times 10^{-5}$.}
\label{fig-photo_uv}}
\end{figure*}

{
In Figure~\ref{fig-photo_uv}, we show the calculated vertical distributions \ikomat{of volume mixing ratio{s} (left column) and photodissociation rate{s} (right column)} of gaseous species for the cases of UV $\times 10^{5}$, UV $\times 10^{2.5}$, UV $\times 10^{-2.5}$, and UV $\times 10^{-5}$.
{Note that the jumps of the photodissociation rate of $\mathrm{O_2}$ (light blue) found at $P \sim 10^{-4}$~bar for the case of UV $\times 10^{5}$~(b) come from its volume mixing ratio out of our computational precision, so that they are physically meaningless.}
In the high-UV cases (UV $\times 10^5$ and $10^{2.5}$), the photodissociation of molecules such as $\mathrm{H_2}$, $\mathrm{H_2O}$, $\mathrm{CH_4}$, and $\mathrm{NH_3}$ occurs deeper \ikomat{in the} atmosphere than in the fiducial case (Figure~\ref{fig-photo}).
\ikomat{This is because photochemistry works more effectively compared to eddy diffusion even at low altitudes because of the intense UV irradiation. Also, the higher the UV irradiation flux, the higher the photodissociation rates themselves.} 
On the other hand, in the low-UV cases (UV $\times 10^{-2.5}$ and $10^{-5}$), only at very high altitudes, photodissociation occurs effectively compared to eddy-diffusion transport, resulting in the constant abundance\ikomat{s} of molecules such as $\mathrm{H_2O}$, $\mathrm{CH_4}$, and $\mathrm{NH_3}$, up to higher altitudes.
From Table~\ref{tab-photo}, one finds that the dependence of the photodissociation rates of the haze precursors on the UV irradiation intensity is weaker than linear, {namely} \ikomat{the higher the UV flux,} the {smaller} proportion of the incoming photons used for the photodissociation of the haze precursors.}
{We have confirmed that this is mainly because larger proportion of the incoming photons are used for the photodissociation of CO and $\mathrm{O_2}$, which exist at higher altitudes than the hydrocarbons, for higher UV flux.}

\subsubsection{Particle Growth} \label{uv_growth}
\begin{figure*}
\plotone{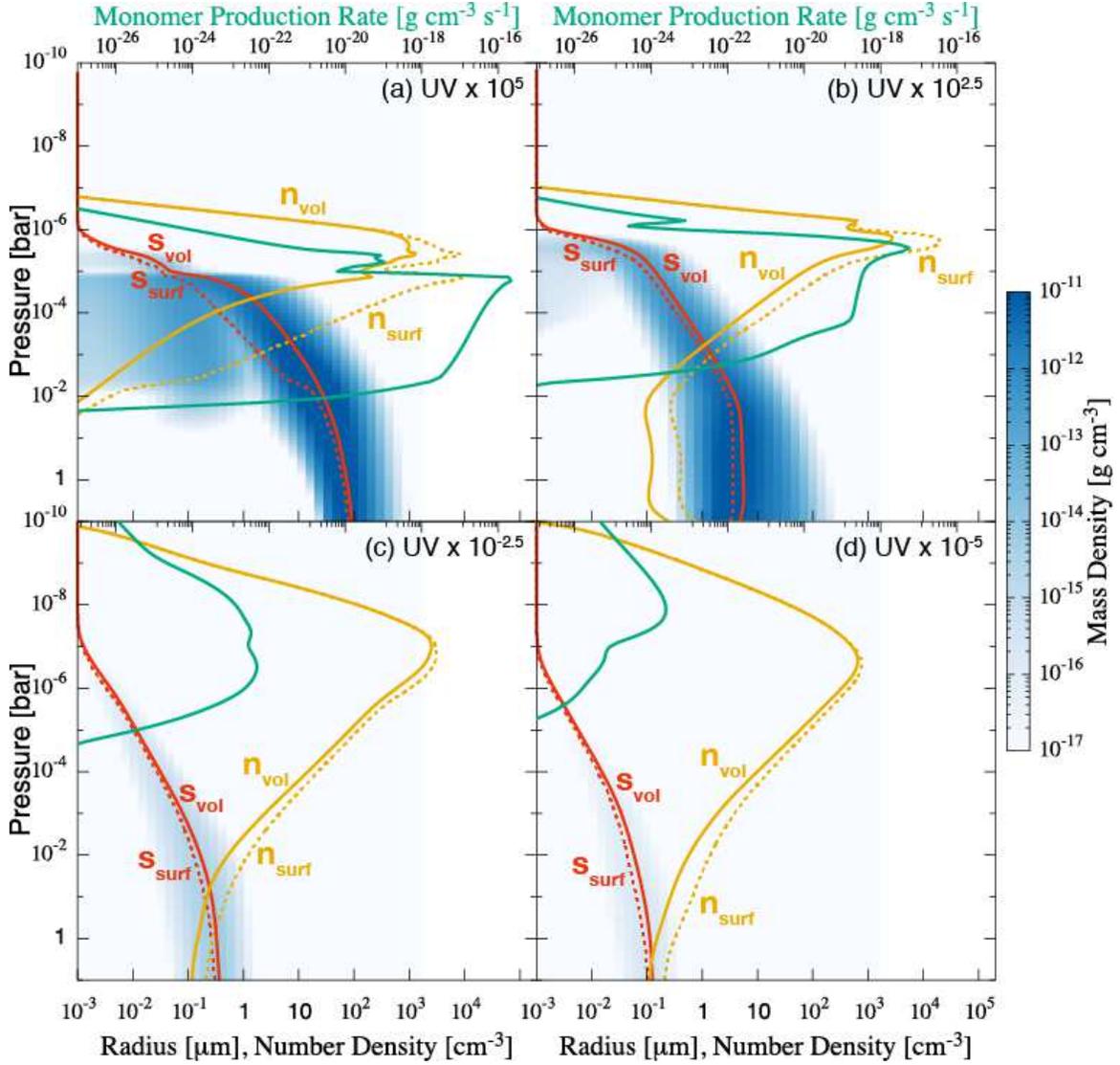}
\caption{{Same as Fig.~\ref{fig-growth} but for the cases of (a)~UV~$\times 10^{5}$, (b)~UV~$\times 10^{2.5}$, (c)~UV~$\times 10^{-2.5}$, and (d)~UV~$\times 10^{-5}$.}
\label{fig-growth_uv}}
\end{figure*}

{
Figure~\ref{fig-growth_uv} \ikomat{also} shows the vertical distribution of the haze particles for \ikomat{the four UV irradiation intensities}.
\ikomat{For higher UV intensities,} the mass density and average radii of haze particles are \ikomat{found to be} significantly larger \kawashima{because of the higher monomer production rates}, \kawashima{nevertheless} the particles start to grow at lower altitudes as explained in the previous section. 
\ikomat{For example,} 
$s_\mathrm{vol}$ becomes as large as 100~$\mu$m in the \ikomat{UV$\times 10^{5}$ case}, while it \ikomat{is} only less than 0.1~$\mu$m in the \ikomat{UV$\times 10^{-5}$ case} at the lower boundary of 10~bar.
For the UV$\times 10^{5}$ \ikomat{case}, because of the high monomer production rate, the disagreement between $s_\mathrm{vol}$ and $s_\mathrm{surf}$ is large in the middle atmosphere with the pressure range from $\sim10^{-5}$~bar to $\sim10^{-2}$~bar, indicating the broad distribution of the particle size in that region.
However, two average radii almost coincide with each other again below the pressure level of $\sim 10^{-2}$~bar because of the decrease \ikomat{in} the monomer production rate and the increase \ikomat{in} the sedimentation velocity \ikomat{after} the transition in gas drag law (from the slip-flow to Stokes-flow regimes), which hampers the particle growth at low altitudes.
}

%% file: result_metallicity.tex
\subsection{Dependence on Metallicity} \label{metallicity}

\begin{figure}
\plotone{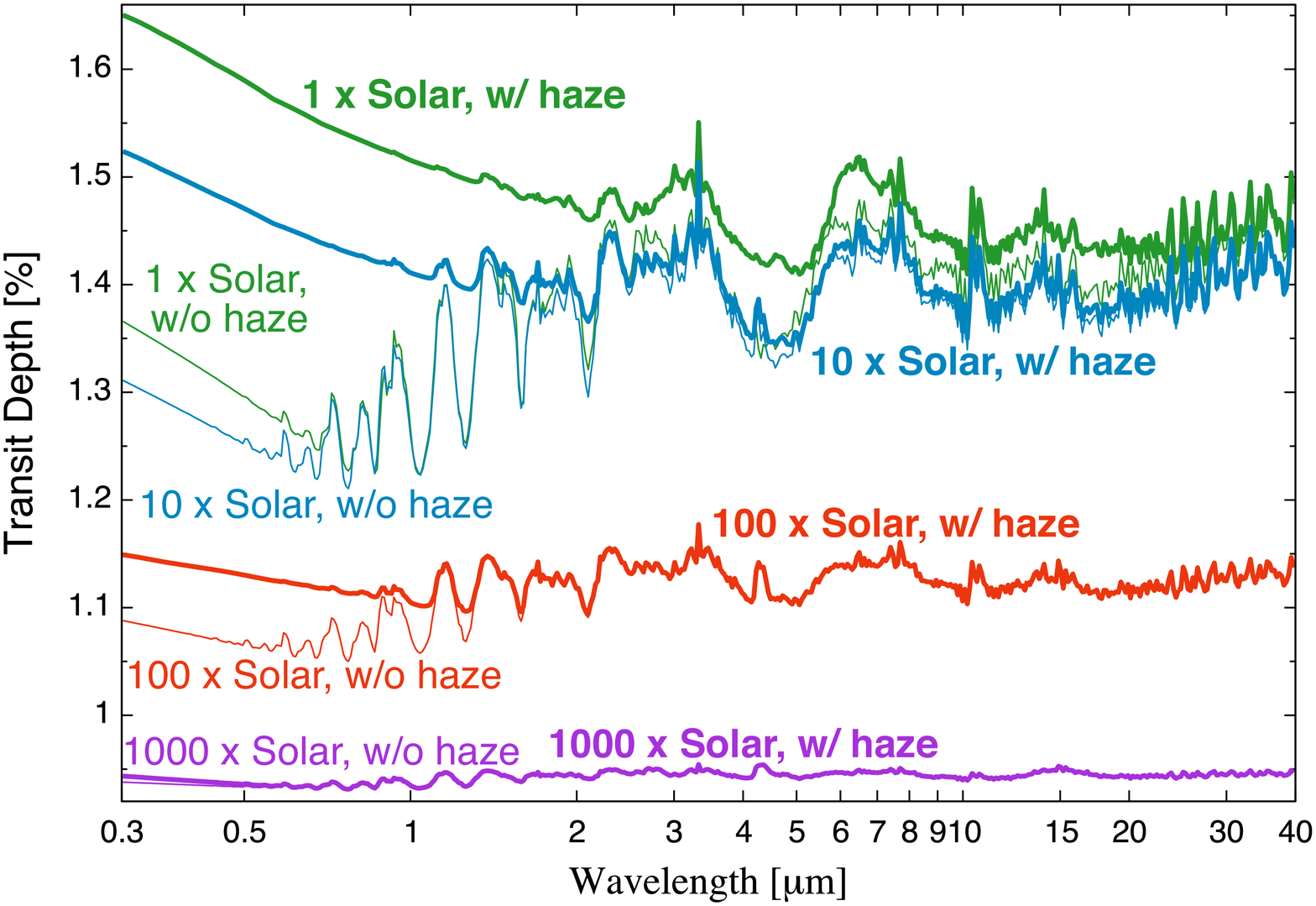}
\caption{Transmission spectrum models for four different atmospheric metallicities, solar (green, same as the green line in Fig.~\ref{fig-spectra}; "1 $\times$ Solar") and 10 times (blue; "10 $\times$ Solar"), 100 times (red; "100 $\times$ Solar"), and 1000 times (purple; "1000 $\times$ Solar") higher than the solar. 
The thick and thin lines represent the spectra of the haze-covered and haze-free atmospheres, respectively.
Note that the transmission spectrum models are smoothed for clarity.
\label{fig-spectra_metallicity}}
\end{figure}

Figure~\ref{fig-spectra_metallicity} shows the transmission spectrum models for 
the hazy atmosphere with the solar \ikomat{and super-solar metallicities (colored thick lines)}. 
\ikomat{Those} without haze are also plotted by thin lines with the same colors.

The main effect of enhanced metallicity is reduction in the scale height of the atmosphere \ikomat{and, thus, in the transit depth for both the clear and hazy atmospheres, a}s shown in Fig.~\ref{fig-spectra_metallicity}.
\ikomat{This makes} the effect of haze less pronounced in higher-metallicity cases. 
Also, haze affects \ikomat{the spectrum} only at shorter wavelengths with increasing metallicity.
{This is} because the mass density of the haze particles becomes smaller with increasing metallicity, as seen in \S~\ref{metallicity_growth}.
{
\ikomat{From an observational point of view, of particular importance are strengths of absorption features (i.e., transit depths relative to baselines).}
For the clear atmospheres, \ikomat{the strengths are found to} decrease with increasing metallicity because of the reduced scale height. 
On the other hand, \ikomat{in the case of} hazy atmospheres, \ikomat{its sensitivity to metallicity}  is \ikomat{somewhat} complicated because both the amount of haze and atmospheric scale height are smaller for higher metallicities.
Among the four different metallicity cases, absorption features at shorter wavelengths ($\lesssim$~2-3~$\mu$m) are \ikomat{strongest} for the $100 \times$ Solar case, while those at longer wavelengths ($\gtrsim$~2-3~$\mu$m), \ikomat{which are less affected by haze}, \ikomat{are strongest} for the smallest metallicity case of $1 \times$ Solar.}
{\ikomat{Note that} {we discuss} investigation of atmospheric metallicity from observations {in which we} suffer from degeneracy with the reference radius in \S~\ref{obs}.}

\subsubsection{Photochemistry {and Haze Precursor Production}}
\begin{figure*}
\gridline{
\fig{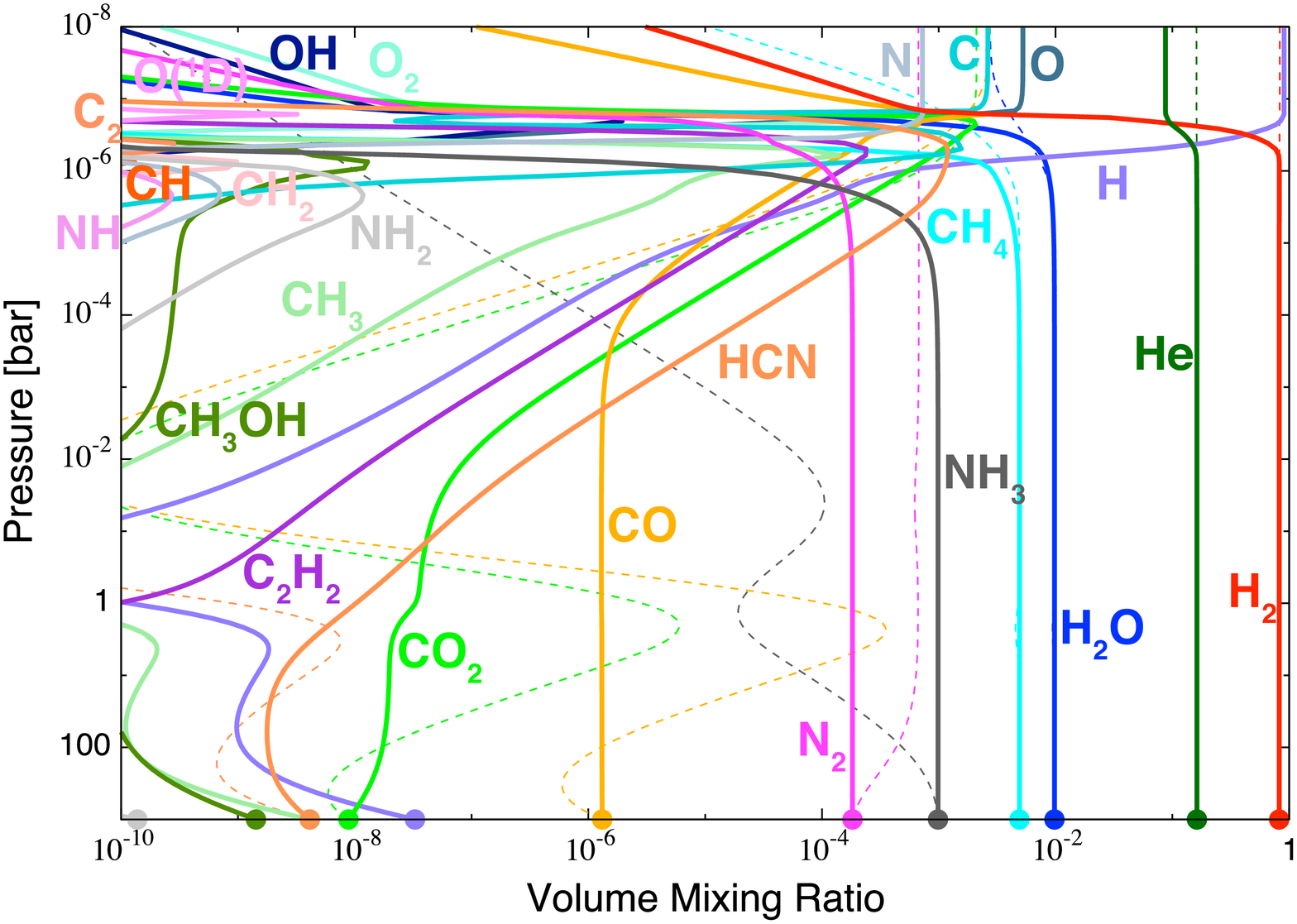}{0.5\textwidth}{(a)~10 $\times$ Solar}
\fig{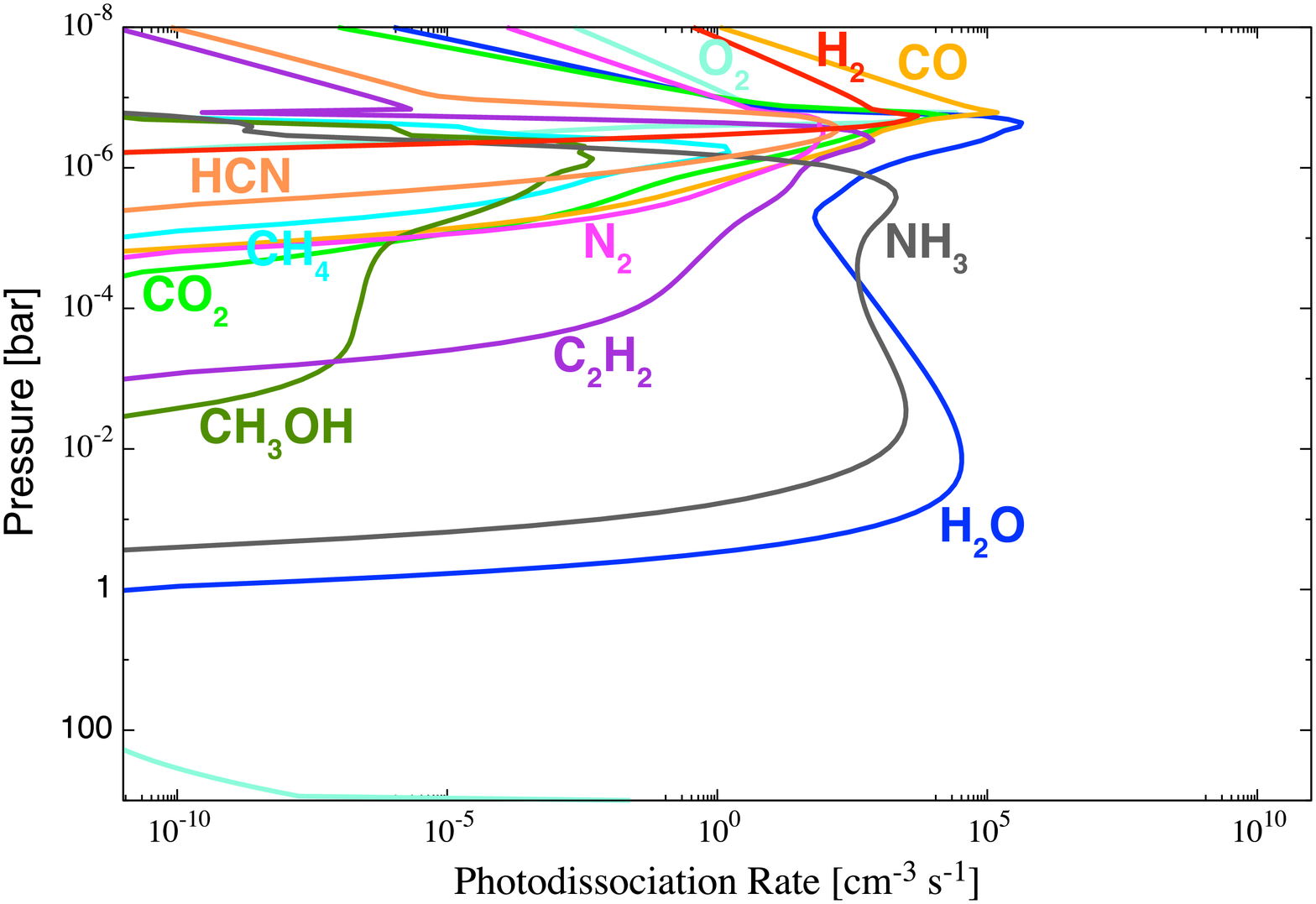}{0.5\textwidth}{(b)~10 $\times$ Solar}
}
\gridline{
\fig{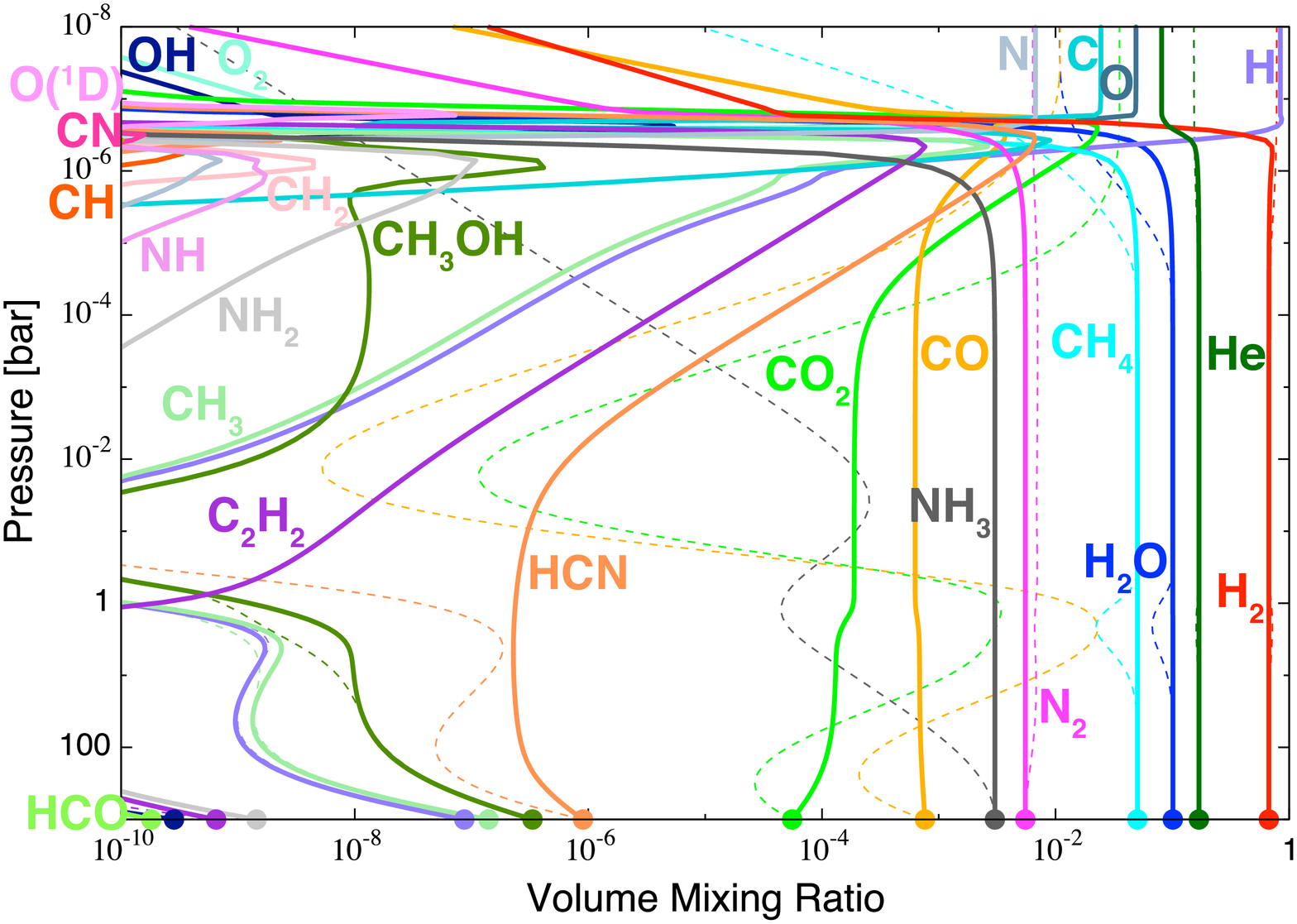}{0.5\textwidth}{(c)~100 $\times$ Solar}
\fig{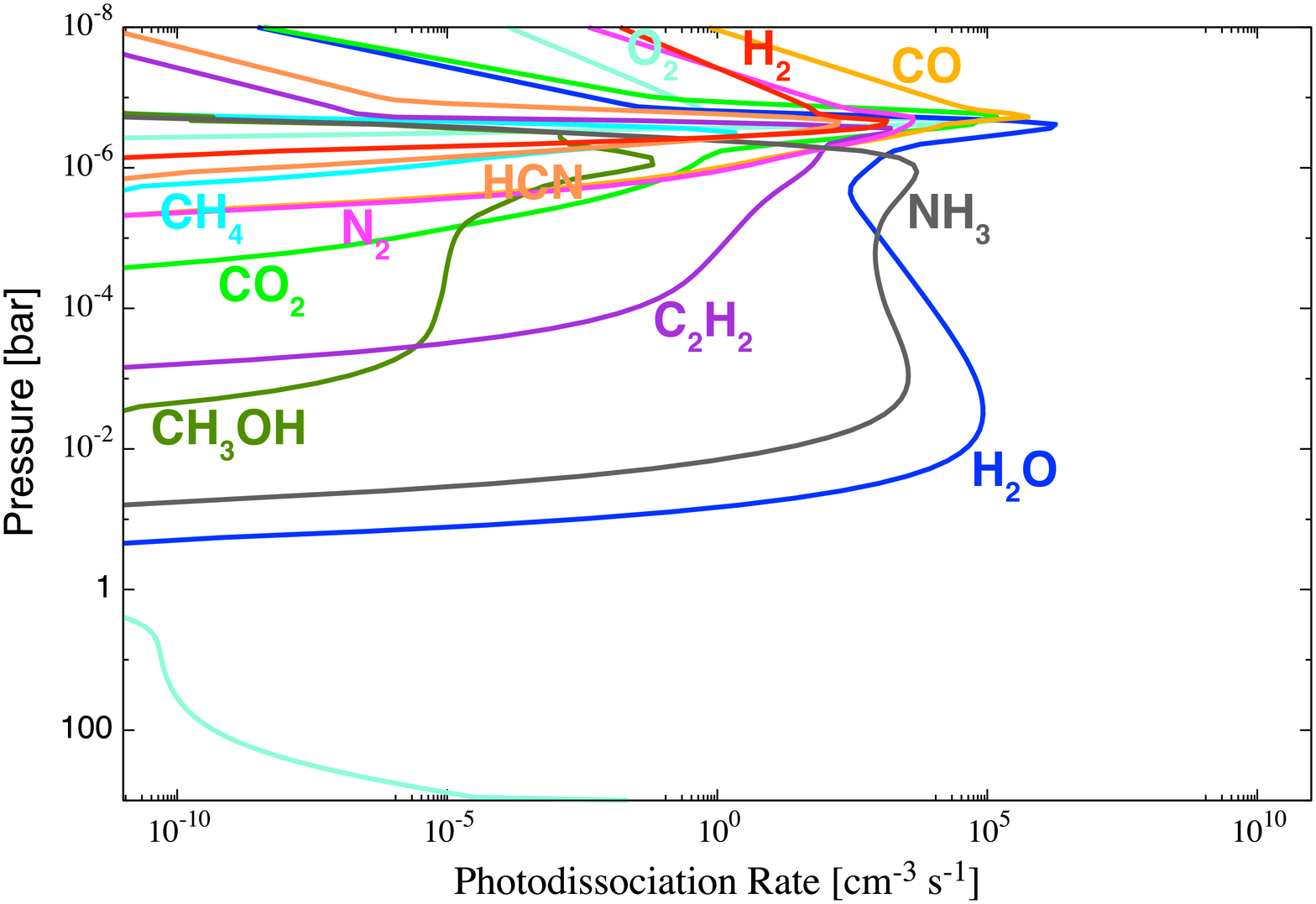}{0.5\textwidth}{(d)~100 $\times$ Solar}
}
\gridline{
\fig{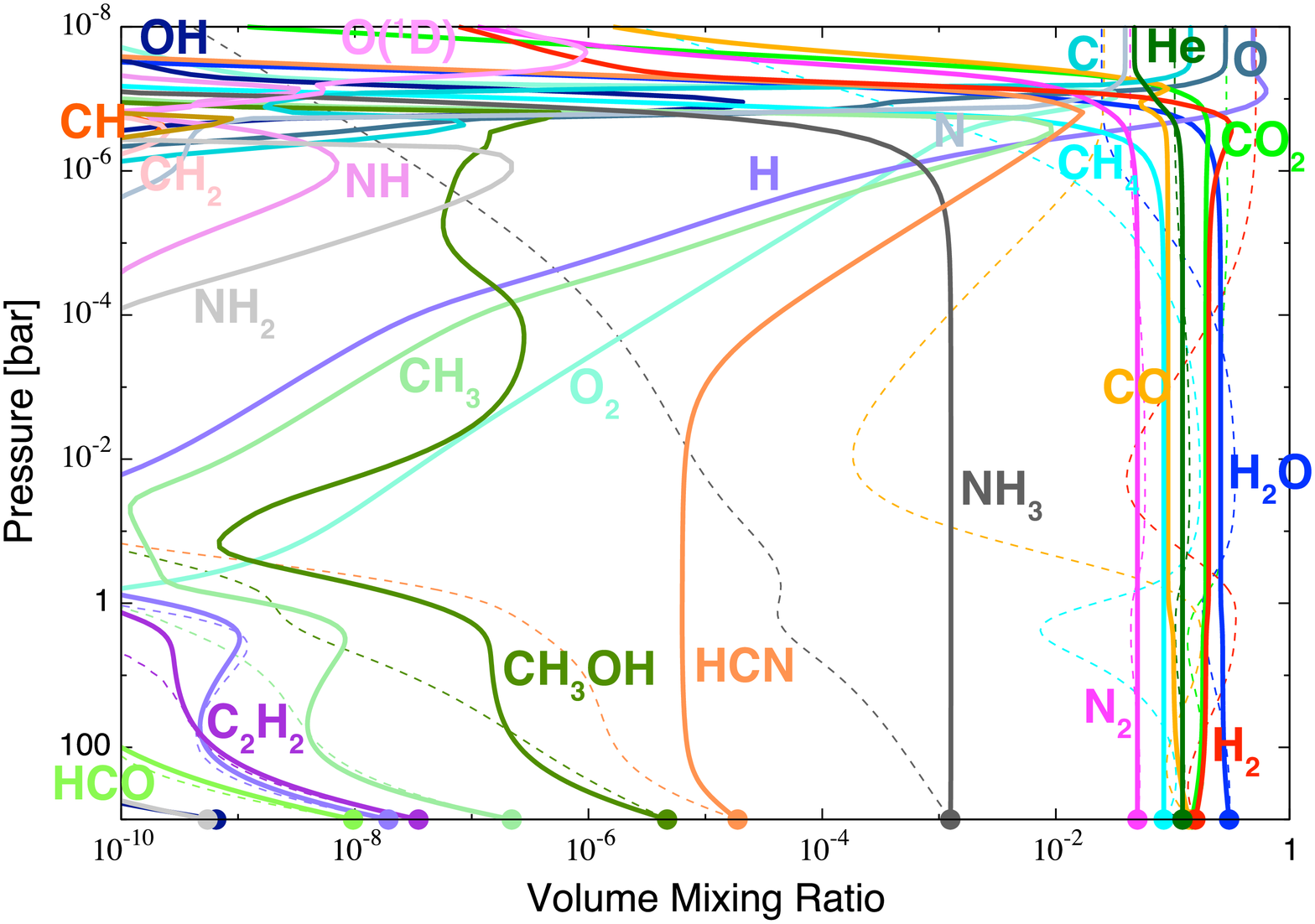}{0.5\textwidth}{(e)~1000 $\times$ Solar}
\fig{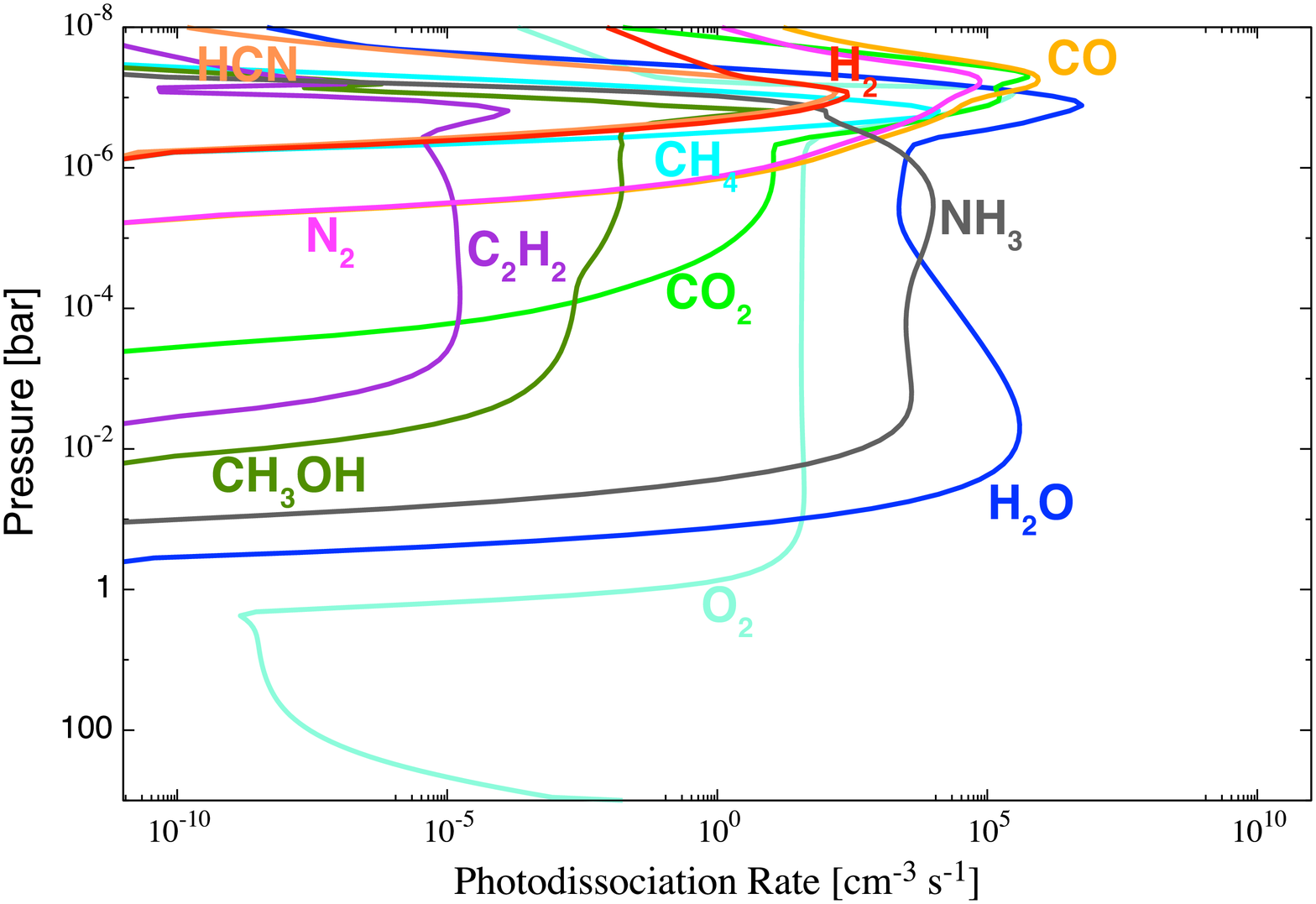}{0.5\textwidth}{(f)~1000 $\times$ Solar}
}
\caption{Same as Fig.~\ref{fig-photo} but for the cases of (a, b)~10 $\times$ Solar, (c, d)~100 $\times$ Solar, and (e, f)~1000 $\times$ Solar metallicity atmospheres.
\label{fig-photo_metallicity}}
\end{figure*}

Figure~\ref{fig-photo_metallicity} \ikomat{shows} the calculated vertical distributions \ikomat{of the volume mixing ratio{s} (left column) and photodissociation rate{s} (right column)} of gaseous species for the 10$\times$Solar, 100$\times$Solar, and 1000$\times$Solar cases \ikomat{(See also Fig.~\ref{fig-photo} for the solar case)}.

Apart from simple increases in the fractions of heavy elements,
the vertical profile of each species for the metal-rich cases is similar to that for the fiducial case (Fig.~\ref{fig-photo}~(a)).
A difference in vertical profile is that $\mathrm{H_2O}$, $\mathrm{CH_4}$, and $\mathrm{NH_3}$ are lost via photodissociation at slightly lower pressures for higher metallicity atmospheres.
This is because the optical depth at a certain pressure is larger for higher metallicity due to the larger abundance of the UV absorbers.
This trend can be more clearly seen in \ikomat{the right-column panels of} Figure~\ref{fig-photo_metallicity}.

{It is notable that in spite of their increased abundances for higher metallicity,} 
the {integrated} photodissociation rates of all the haze precursors, $\mathrm{CH_4}$, HCN, and $\mathrm{C_2H_2}$, are lower {except for that of $\mathrm{CH_4}$ in the 1000$\times$Solar case} (see Table~\ref{tab-photo}).
This is \ikomat{due to} increases of the major photon absorbers, $\mathrm{H_2O}$, $\mathrm{CO}$, $\mathrm{CO_2}$, and $\mathrm{O_2}$, exiting at higher altitudes than the hydrocarbons, the effect of which is larger than the effect of the decrease in the number of photons absorbed by $\mathrm{H_2}$.
\kawashima{Recent laboratory experiments, however, implied that not only the photodissociation of hydrocarbons, but also that of $\mathrm{CO}$, $\mathrm{CO_2}$, and $\mathrm{H_2O}$ can lead to the formation of haze, implying the existence of multiple formation pathways \citep{2018NatAs...2..303H, doi:10.1021/acsearthspacechem.8b00133}.
A brief discussion is made in \S~\ref{experiments}.}

Compared to the \ikomat{low-metallicity} cases, the atmospheric composition for the 1000$\times$Solar case (Fig.~\ref{fig-photo_metallicity}~(e)) is obviously different: $\mathrm{H_2O}$ is the most abundant species over the almost entire region.
Among the carbon-bearing species, $\mathrm{CO_2}$ is the most abundant rather than $\mathrm{CH_4}$ and the abundance of $\mathrm{C_2H_2}$ is small because of the high O/H ratio of 0.5812.
From Figure~\ref{fig-photo_metallicity}~(d) and (f), 
it is found that the photodissociation rate of $\mathrm{CH_4}$ is higher due to its larger abundance in the 1000$\times$Solar case than in the 100$\times$Solar case, while that of $\mathrm{C_2H_2}$ is smaller due to its smaller abundance in the former case than in the latter. 
In total, the total photodissociation rates of haze precursors for the 100$\times$Solar and 1000$\times$Solar cases are almost similar (see Table~\ref{tab-photo}).

\subsubsection{Particle Growth} \label{metallicity_growth}
\begin{figure*}
\plotone{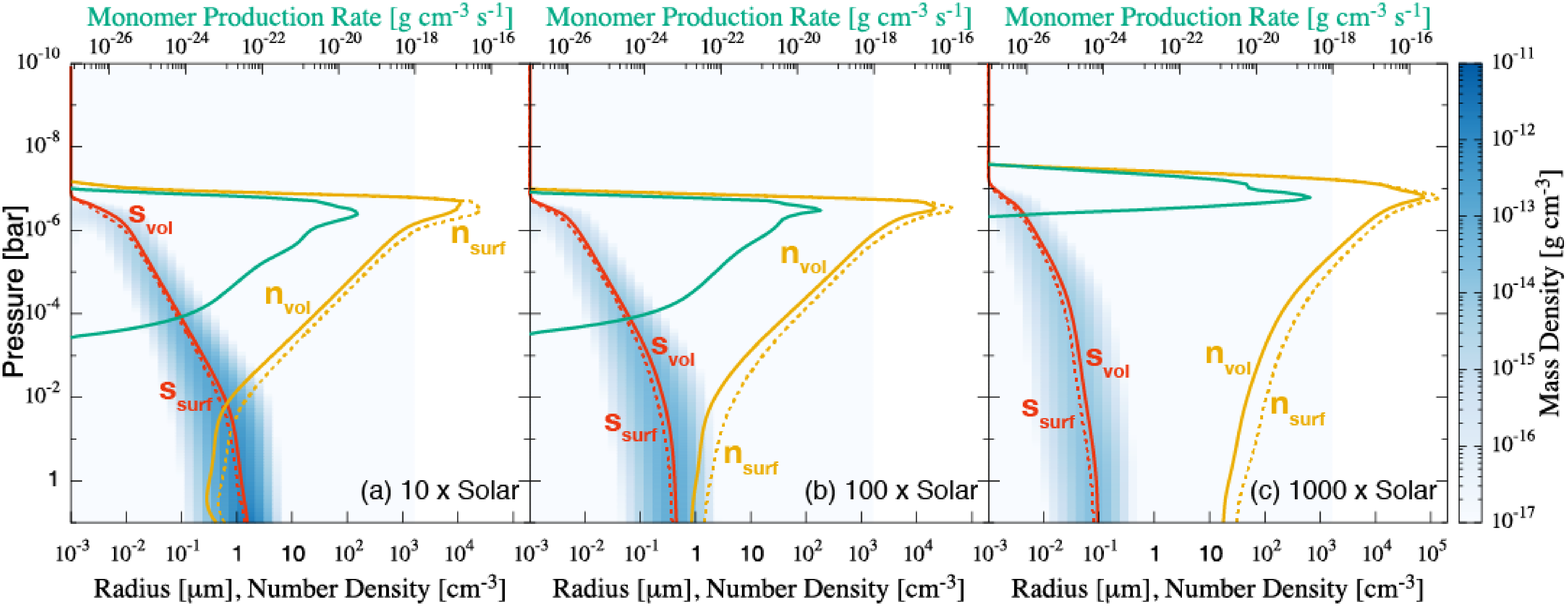}
\caption{Same as Fig.~\ref{fig-growth} but for the cases of (a)~10 $\times$ Solar, (b)~100 $\times$ Solar, and (c)~1000 $\times$ Solar metallicity atmospheres.
\label{fig-growth_metallicity}}
\end{figure*}

Figure~\ref{fig-growth_metallicity} shows the vertical distribution of haze particles for \ikomat{such} three \ikomat{super-solar} metallicities.
Both the mass density and average radii of the haze particles in the lower atmosphere are found to be smaller for higher metallicities (see also Fig.~\ref{fig-growth} for the 1$\times$Solar metallicity).
This is partly because the {integrated} monomer production rate is smaller for the higher metallicity atmosphere, as described in the last subsection.
Also, the particles settle down from the upper to lower boundary on a shorter timescale for the higher metallicity atmosphere. 
This is because the pressure scale height $H$ becomes small with the atmospheric mean molecular weight $\mu$ more greatly than the particle sedimentation velocity $v_\mathrm{sed}$ does (i.e., $H \propto \mu^{-1}$, while $v_\mathrm{sed} \propto \mu^{-1/2}$; see Eq.~(13) of Paper~I).
Thus, in the higher metallicity cases, there is less time for the particles to grow before reaching the lower boundary.
This effect can be seen clearly by comparison between the 100$\times$Solar and 1000$\times$Solar metallicity cases.
Although the {integrated} monomer production rates are comparable to each other, the mass density and averaged radii of the haze particles are smaller in the 1000$\times$Solar case.

%% file: result_co.tex
\subsection{Dependence on C/O ratio} \label{co}
\begin{figure}
\plotone{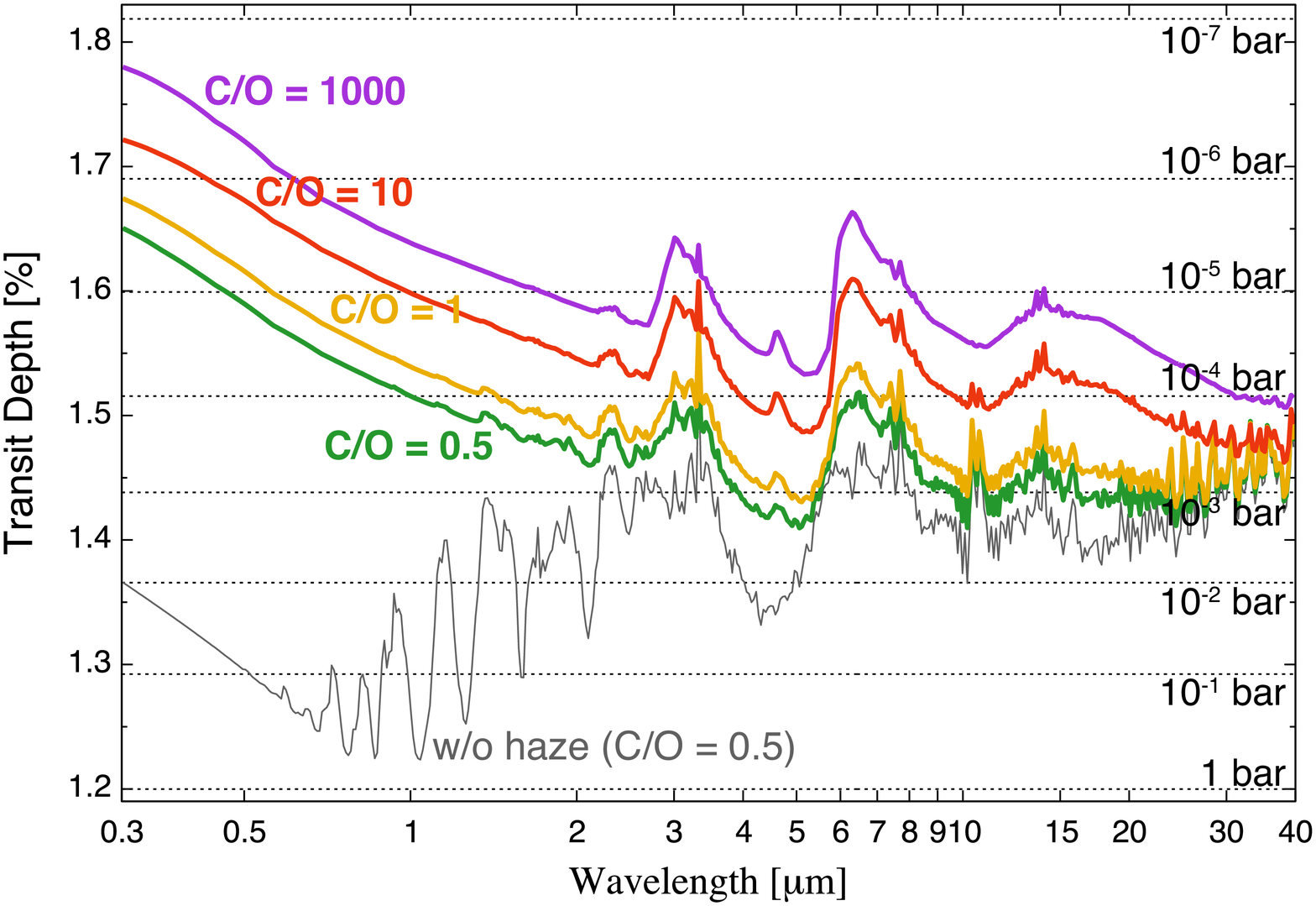}
\caption{Transmission spectrum models for the atmosphere with haze for the four different cases where $\mathrm{C/O} =$~0.5 (green line, same as the green line in Fig.~\ref{fig-spectra}), 1 (orange line), 10 (red line), and 1000 (purple line).
The transmission spectrum for the atmosphere without haze for the case of $\mathrm{C/O} =$~0.5 is also plotted (black line, same as the black line in Fig.~\ref{fig-spectra}).
As in Fig.~\ref{fig-spectra}, horizontal dotted lines represent the transit depths corresponding to the pressure levels from $1 \times 10^{-7}$~bar to 1~bar for the atmosphere in the case of $\mathrm{C/O} =$~0.5.
Note that the transmission spectrum models are smoothed for clarity.
\label{fig-spectra_co}
}
\end{figure}

Figure~\ref{fig-spectra_co} shows the transmission spectrum models for the atmosphere with haze for four different values of $\mathrm{C/O} =$~0.5 (green), 1 (orange), 10 (red), and 1000 (purple).
The transmission spectrum for the haze-free atmosphere for $\mathrm{C/O} =$~0.5 is also plotted (black line). 
Here we consider large values of C/O, because 
the effect of hydrocarbon haze on transmission spectra is of special interest in this study.

The transit depth increases with increasing C/O at each wavelength. 
This is simply because larger amounts of haze are produced for higher values of C/O (see \S~\ref{co_growth}). 
In particular, in the case of $\mathrm{C/O} =$~1000, {almost} all the molecular absorption features are hidden by the abundant haze and, instead, the absorption features of the haze particles still remain at 3.0, 4.6, and 6.3~$\mu$m.
We have confirmed that even in the extreme case of $\mathrm{C/O} = 10^{10}$, the spectrum for the atmosphere with haze shows such haze features and never becomes flat, because the total {photodissociation} rates of hydrocarbons are limited not by the amount of carbon, but by the incoming photon flux.

\subsubsection{Photochemistry {and Haze Precursor Production}} \label{co_photo}
\begin{figure*}
\gridline{
\fig{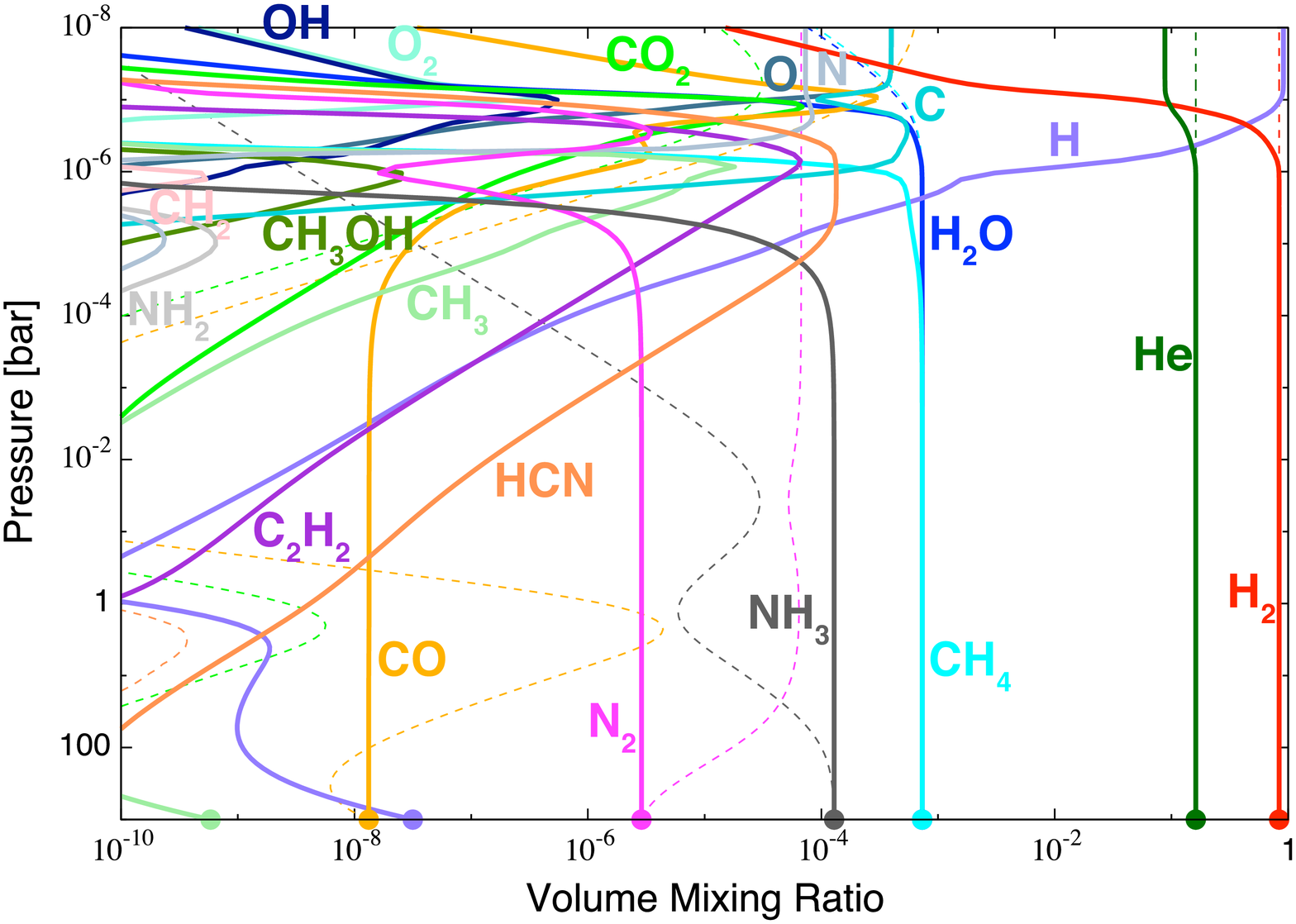}{0.5\textwidth}{(a)~$\mathrm{C/O} = 1$}
\fig{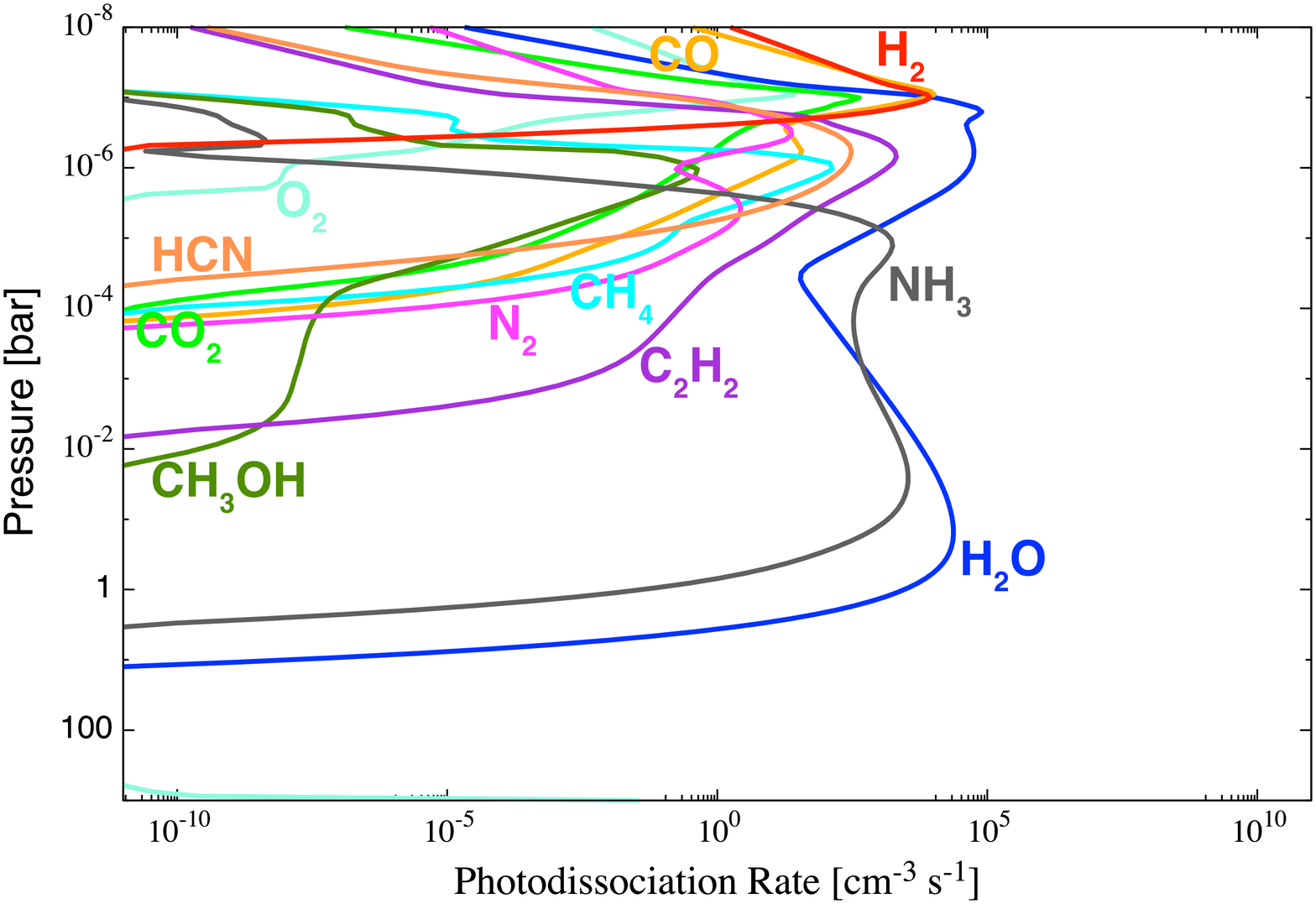}{0.5\textwidth}{(b)~$\mathrm{C/O} = 1$}
}
\gridline{
\fig{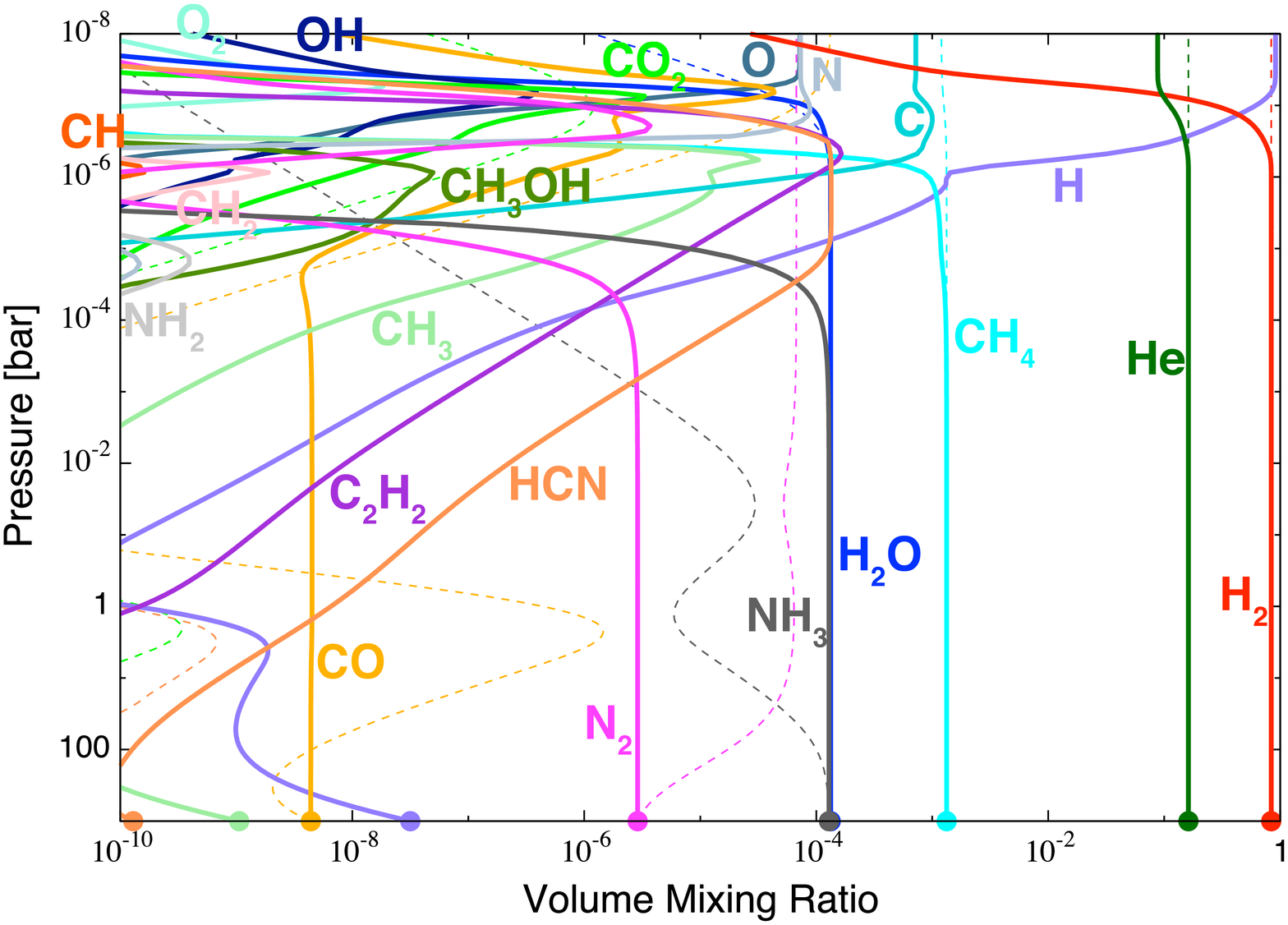}{0.5\textwidth}{(c)~$\mathrm{C/O} = 10$}
\fig{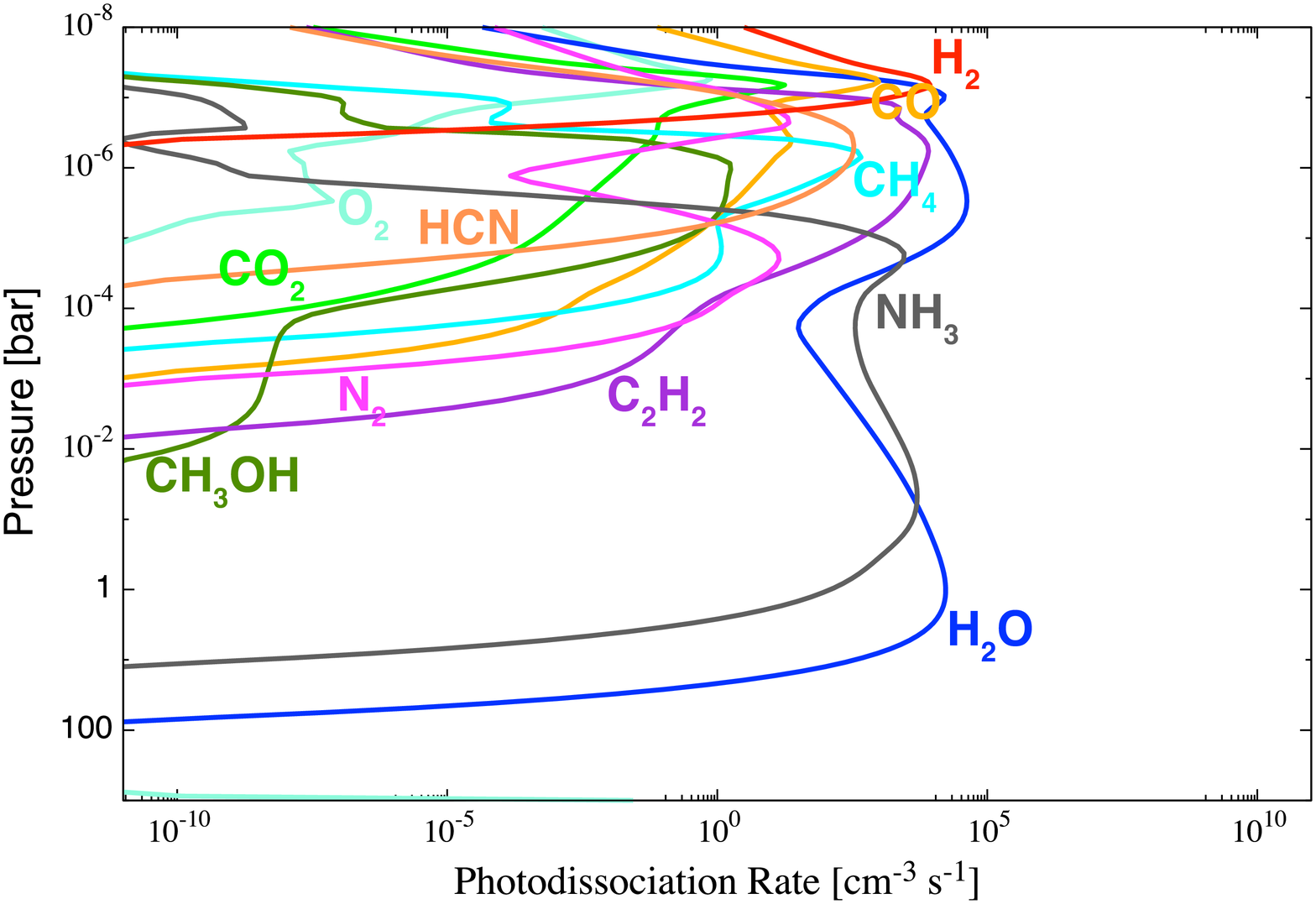}{0.5\textwidth}{(d)~$\mathrm{C/O} = 10$}
}
\gridline{
\fig{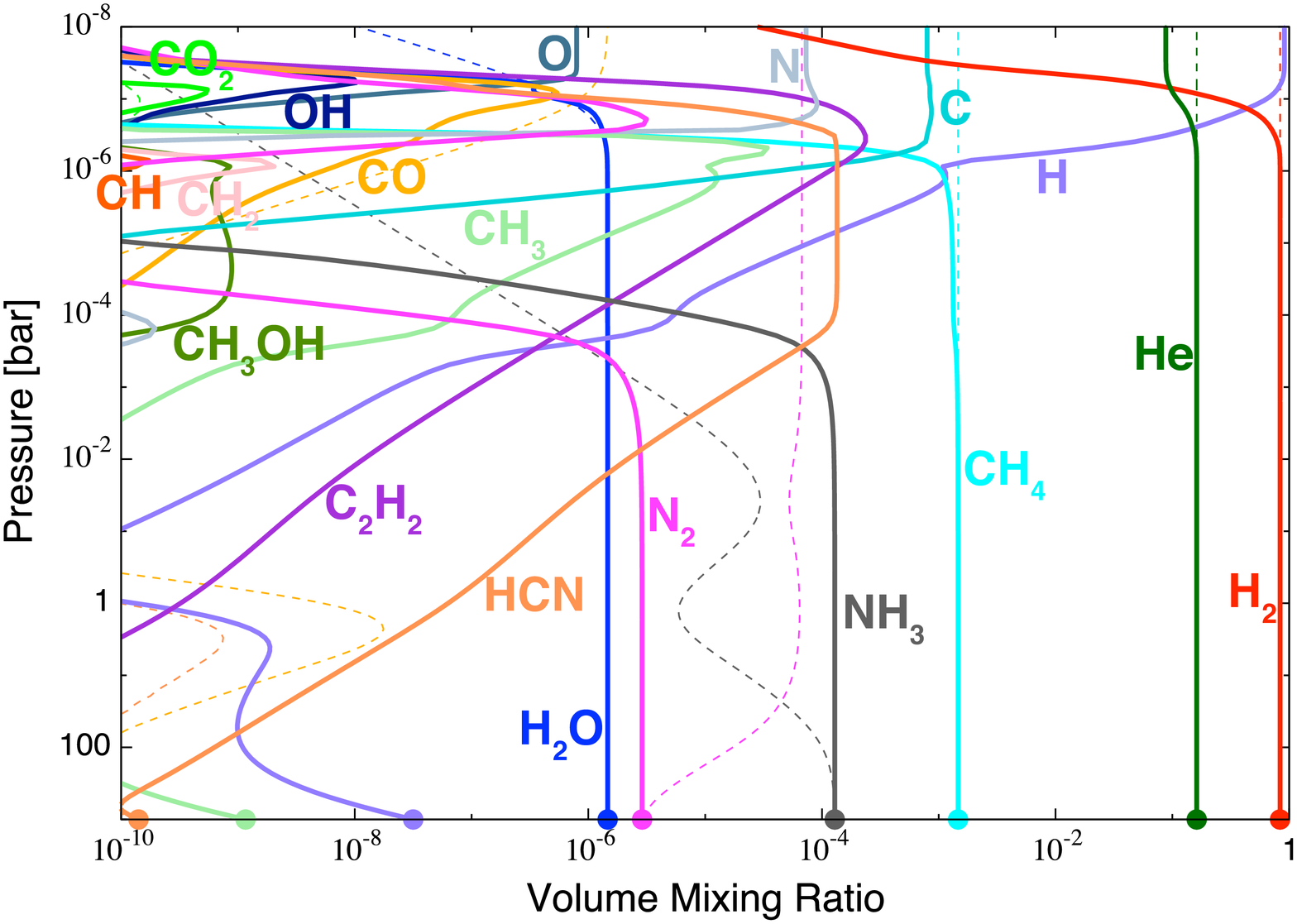}{0.5\textwidth}{(e)~$\mathrm{C/O} = 1000$}
\fig{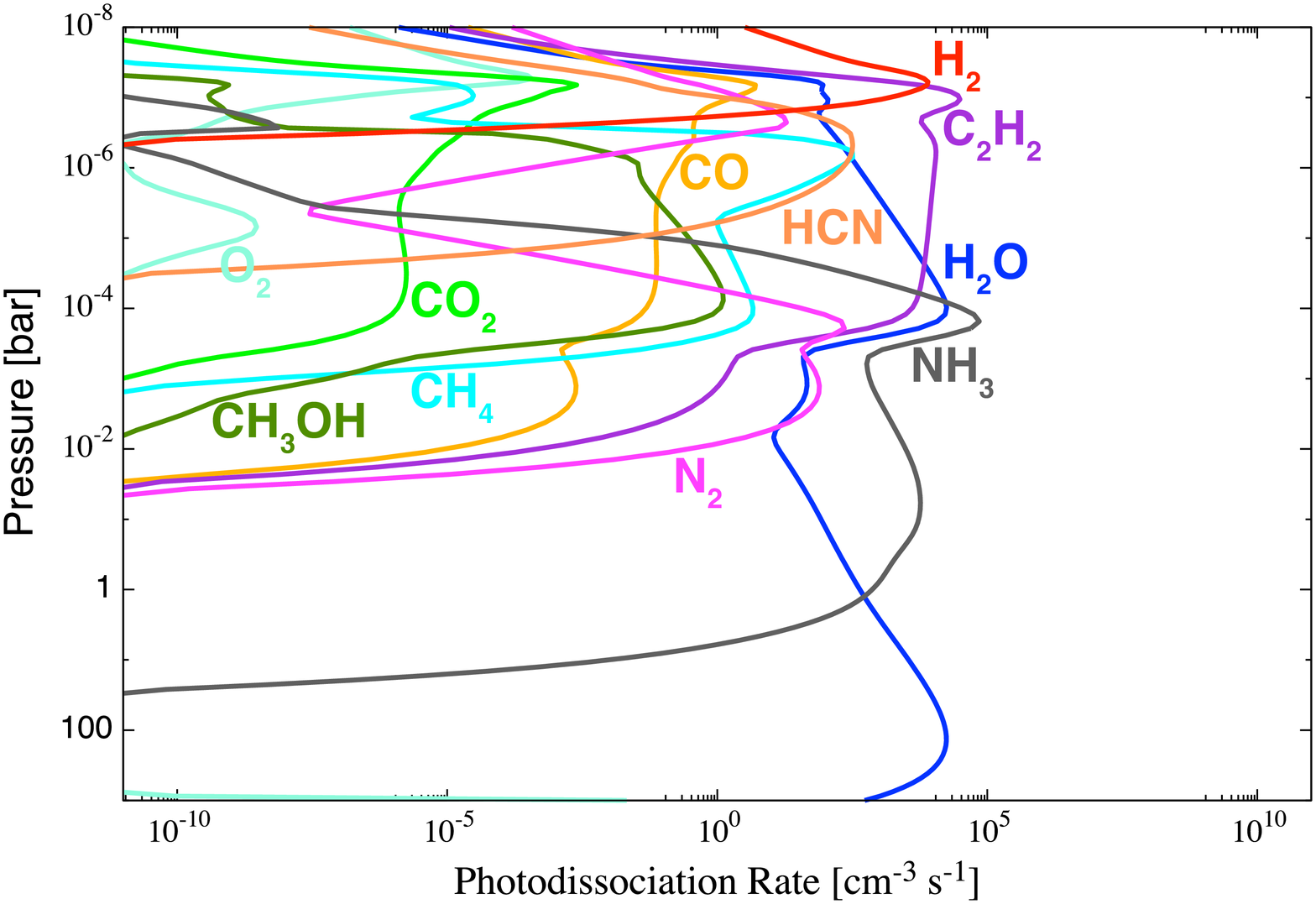}{0.5\textwidth}{(f)~$\mathrm{C/O} = 1000$}
}
\caption{Same as Fig.~\ref{fig-photo} but for the cases of the atmospheres with $\mathrm{C/O} =$~(a, b)~1, (c, d)~10, and (e, f)~1000.
\label{fig-photo_co}}
\end{figure*}

Figure~\ref{fig-photo_co}
shows the calculated vertical distributions of \ikomat{volume mixing ratio{s} (left column) and photodissociation rate{s} (right column) of} gaseous species for $\mathrm{C/O} =$~1, 10, and 1000.
In the case of $\mathrm{C/O} =$~1, 
the abundance of $\mathrm{CH_4}$ is almost equal to that of $\mathrm{H_2O}$ in the lower atmosphere ($P \gtrsim 10^{-4}$~bar), 
because almost all of the C and O are in the form of $\mathrm{CH_4}$ and $\mathrm{H_2O}$, respectively.
As can be expected, the abundances of oxygen-bearing species such as $\mathrm{H_2O}$, O, CO, $\mathrm{CO_2}$, OH, $\mathrm{O_2}$, and $\mathrm{CH_3OH}$ decrease with increasing C/O, while the other trends are similar to those in the fiducial case (\ikomat{C/O = 0.5;} Fig.~\ref{fig-photo}~(a)).

The \ikomat{integrated} 
photodissociation rates of the haze precursors, $\mathrm{CH_4}$, HCN, and $\mathrm{C_2H_2}$, become higher with increasing C/O ratio,
{except for $\mathrm{CH_4}$ and HCN \ikomat{for} $\mathrm{C/O} = 1000$} (see Table~\ref{tab-photo}). 
This is \ikomat{due to} two effects:
One is the increased abundances of those haze precursors, and another is the decreased abundances of $\mathrm{H_2O}$ and CO, which are the major photon absorbers in the fiducial case.
{Note that the {integrated} photodissociation rates of $\mathrm{CH_4}$ and HCN \ikomat{for} $\mathrm{C/O} = 1000$ becomes smaller than those \ikomat{for} $\mathrm{C/O} = 10$, because $\mathrm{C_2H_2}$ existing at higher altitudes absorbs more photons and inhibits them from absorbing the photons.}
Thus, the sum of the {integrated} photodissociation rate\ikomat{s} of the three haze precursors is higher for higher value{s} of C/O.
{Also, the photodissociation region of the three haze precursors become\ikomat{s} broader for higher C/O. This is because the atmosphere becomes less optically-thick due to the decrease of the major photon absorbers such as CO and $\mathrm{H_2O}$.}

\subsubsection{Particle Growth} \label{co_growth}
\begin{figure*}
\plotone{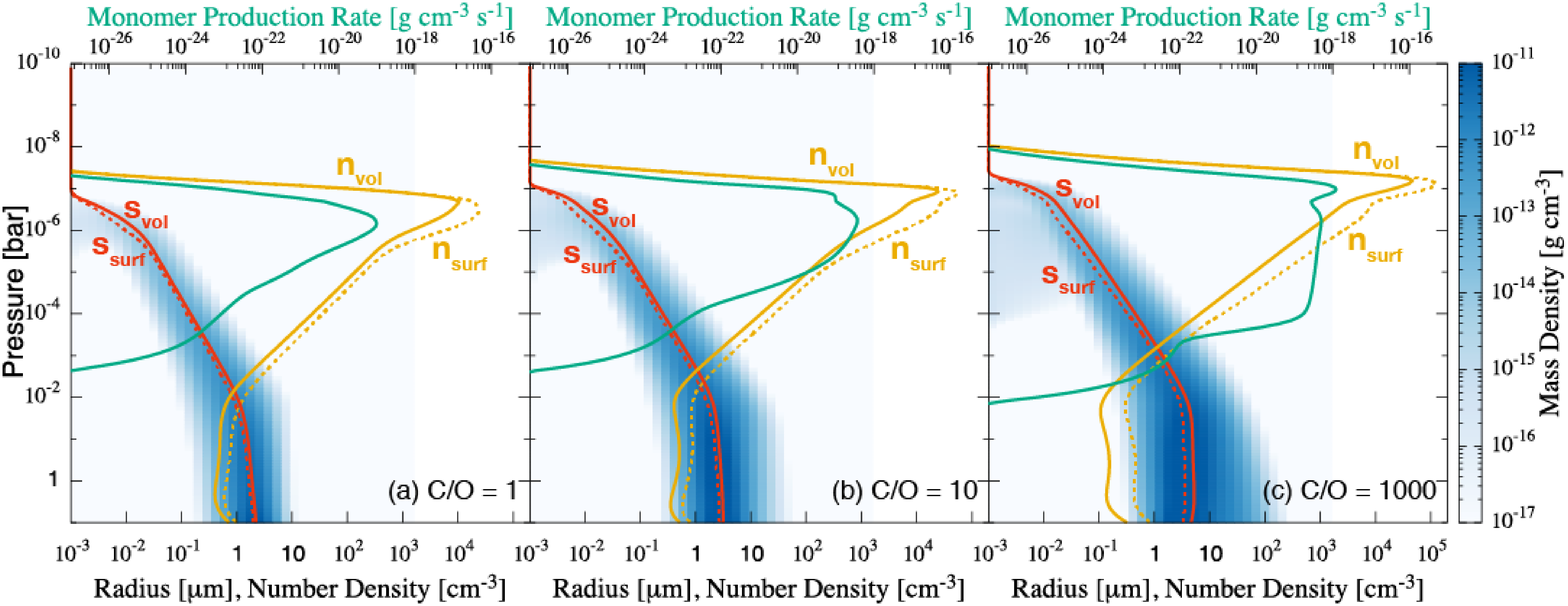}
\caption{Same as Fig.~\ref{fig-growth} but for the cases of atmospheres with $\mathrm{C/O} =$~(a)~1, (b)~10, and (c)~1000.
\label{fig-growth_co}}
\end{figure*}

Figure~\ref{fig-growth_co} shows the vertical distribution of the haze particles for \ikomat{such} three different values of C/O.
As explained in \S~\ref{co_photo}, the {integrated} monomer production rate is higher for higher C/O.
{
As a result, both the mass density and average radii of the haze particles in the lower atmosphere are found to be larger for higher C/O} 
(see also Fig.~\ref{fig-growth} for $\mathrm{C/O} = 0.5$)\ikomat{, which has a significant effect on the transit depths (see Fig.~\ref{fig-spectra_co})}.
{Also, the size\ikomat{s} of the haze particles are more diverse for higher C/O because of the broader monomer production region.}

%% file: result_eddy.tex
\subsection{Dependence on Eddy Diffusion Coefficient} \label{eddy}
\begin{figure}
\plotone{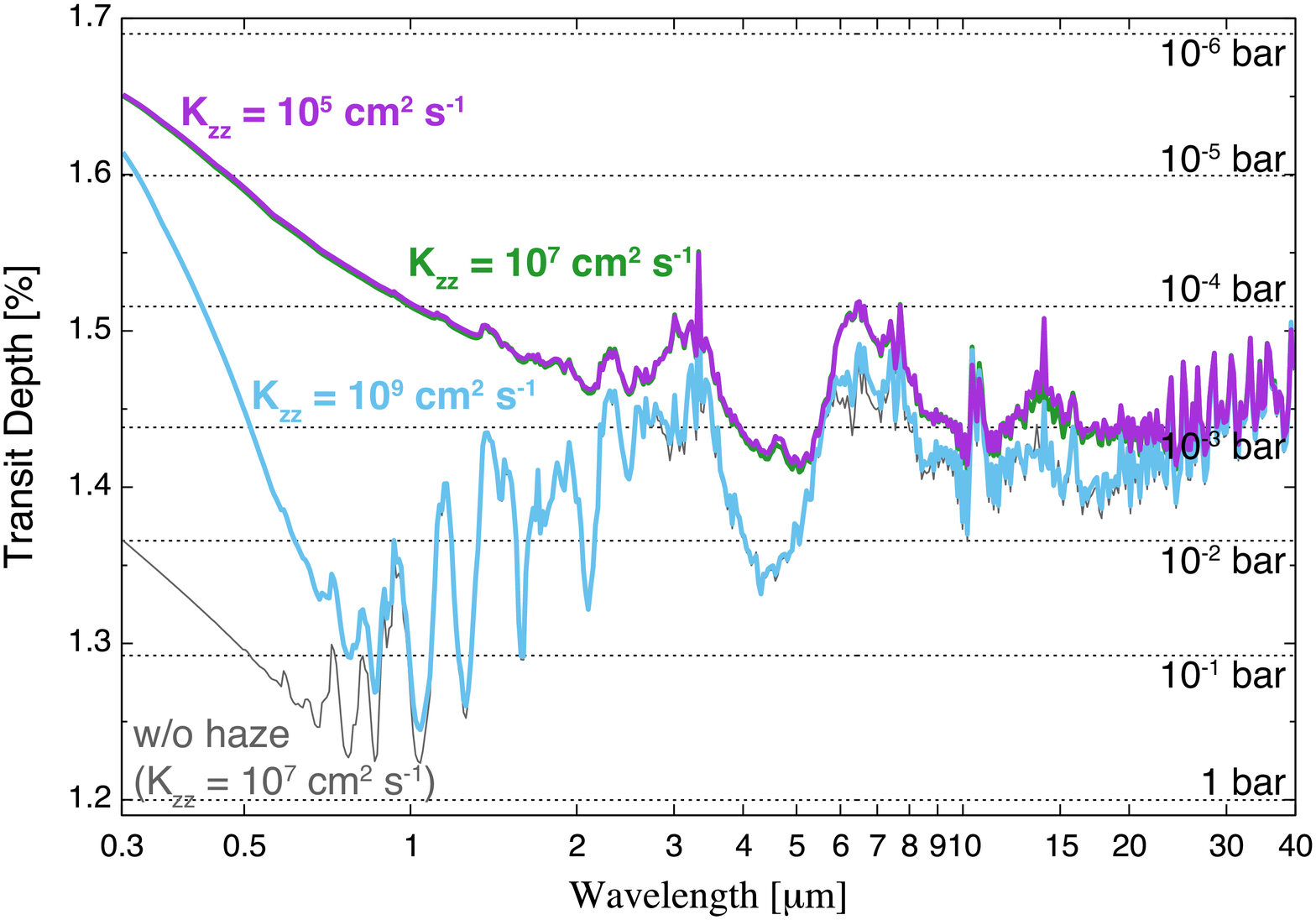}
\caption{Transmission spectrum models for the atmosphere with haze for the three different cases where $K_{zz} = 1 \times 10^9$~$\mathrm{cm}^2$~$\mathrm{s}^{-1}$ (light-blue line), $K_{zz} = 1 \times 10^7$~$\mathrm{cm}^2$~$\mathrm{s}^{-1}$ (green line, same as the green line in Fig.~\ref{fig-spectra}), and $K_{zz} = 1 \times 10^5$~$\mathrm{cm}^2$~$\mathrm{s}^{-1}$ (purple line).
The transmission spectrum for the atmosphere without haze for the case of $K_{zz} = 1 \times 10^7$~$\mathrm{cm}^2$~$\mathrm{s}^{-1}$ is also plotted (black line, same as the black line in Fig.~\ref{fig-spectra}).
As in Fig.~\ref{fig-spectra}, horizontal dotted lines represent the transit depths corresponding to the pressure levels from $1 \times 10^{-6}$~bar to 1~bar for the atmosphere in the case of $K_{zz} = 1 \times 10^7$~$\mathrm{cm}^2$~$\mathrm{s}^{-1}$.
Note that the transmission spectrum models are smoothed for clarity.
\label{fig-spectra_eddy}}
\end{figure}


Figure~\ref{fig-spectra_eddy} shows the transmission spectrum models for the atmosphere with haze for three different values of the eddy diffusion coefficient, $K_{zz}$, $1 \times 10^9$ (light blue; high-$K_{zz}$), $1 \times 10^7$ (green; fiducial), and $1 \times 10^5$~$\mathrm{cm}^2$~$\mathrm{s}^{-1}$ (purple; low-$K_{zz}$). 
The green and gray lines are the fiducial models with and without haze, respectively, shown in Fig.~\ref{fig-spectra}.

In the high-$K_{zz}$ model, one finds a steep Rayleigh-scattering slope in the optical and more prominent molecular-absorption features in the infrared, relative to the fiducial and low-$K_{zz}$ models. 
As explained in detail in \S~\ref{eddy_growth}, this is because efficient eddy diffusion removes haze particles from the upper atmosphere, making the atmosphere optically thinner.

The low-$K_{zz}$ model is almost the same as the fiducial one. 
This is due to similar distributions of haze particles (also see \S~\ref{eddy_growth}).
%
Slight differences in absorption feature are found at 11 and 14~$\mu$m. 
The former is due to a different distribution of NH$_3$, while the latter due to that of NH$_3$ + HCN, as shown in \S~\ref{eddy_chem}.

\subsubsection{Photochemistry {and Haze Precursor Production}} \label{eddy_chem}
\begin{figure*}
\gridline{
\fig{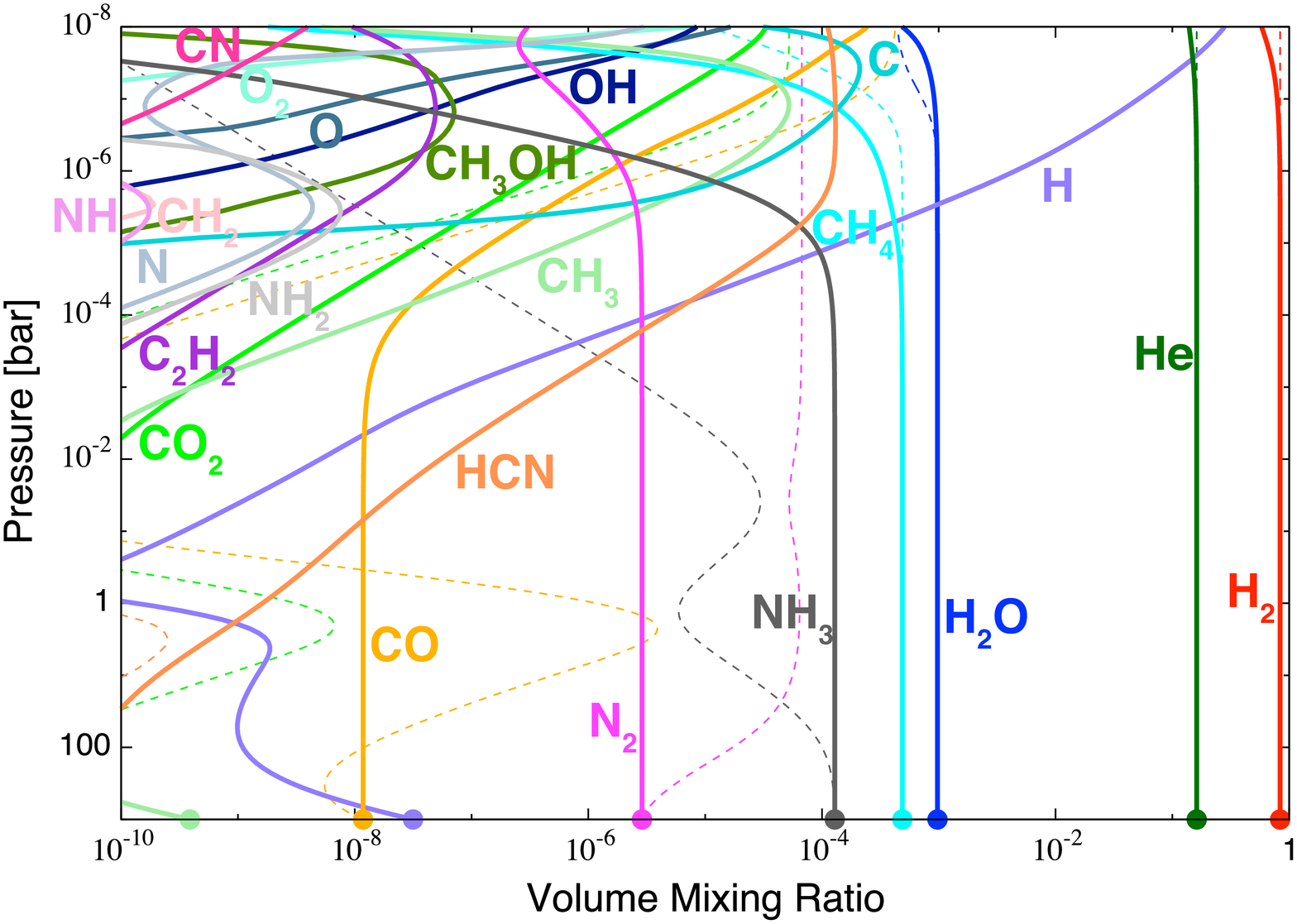}{0.5\textwidth}{(a)~$K_{zz} = 1 \times 10^9$~$\mathrm{cm}^2$~$\mathrm{s}^{-1}$}
\fig{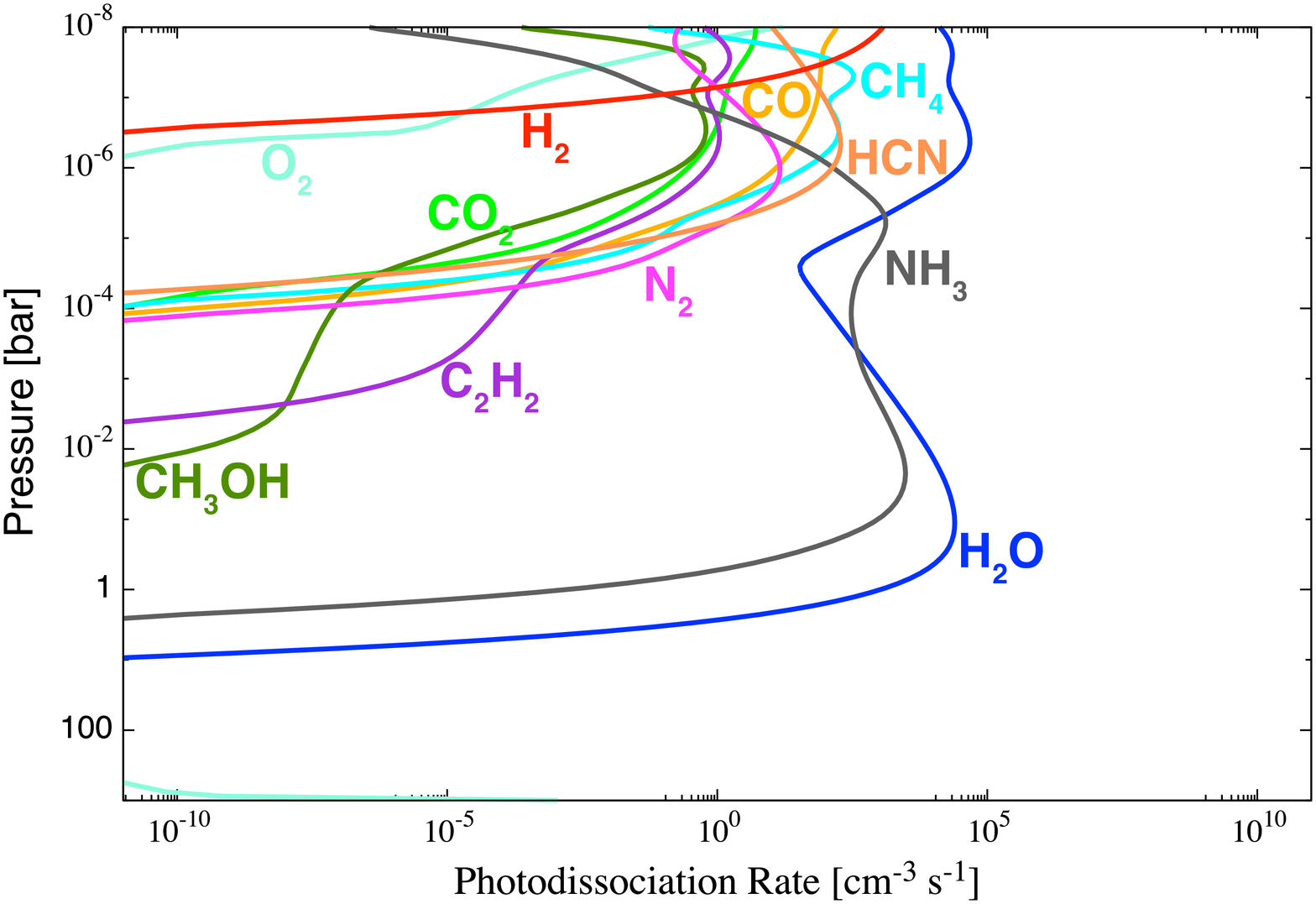}{0.5\textwidth}{(b)~$K_{zz} = 1 \times 10^9$~$\mathrm{cm}^2$~$\mathrm{s}^{-1}$}
}
\gridline{
\fig{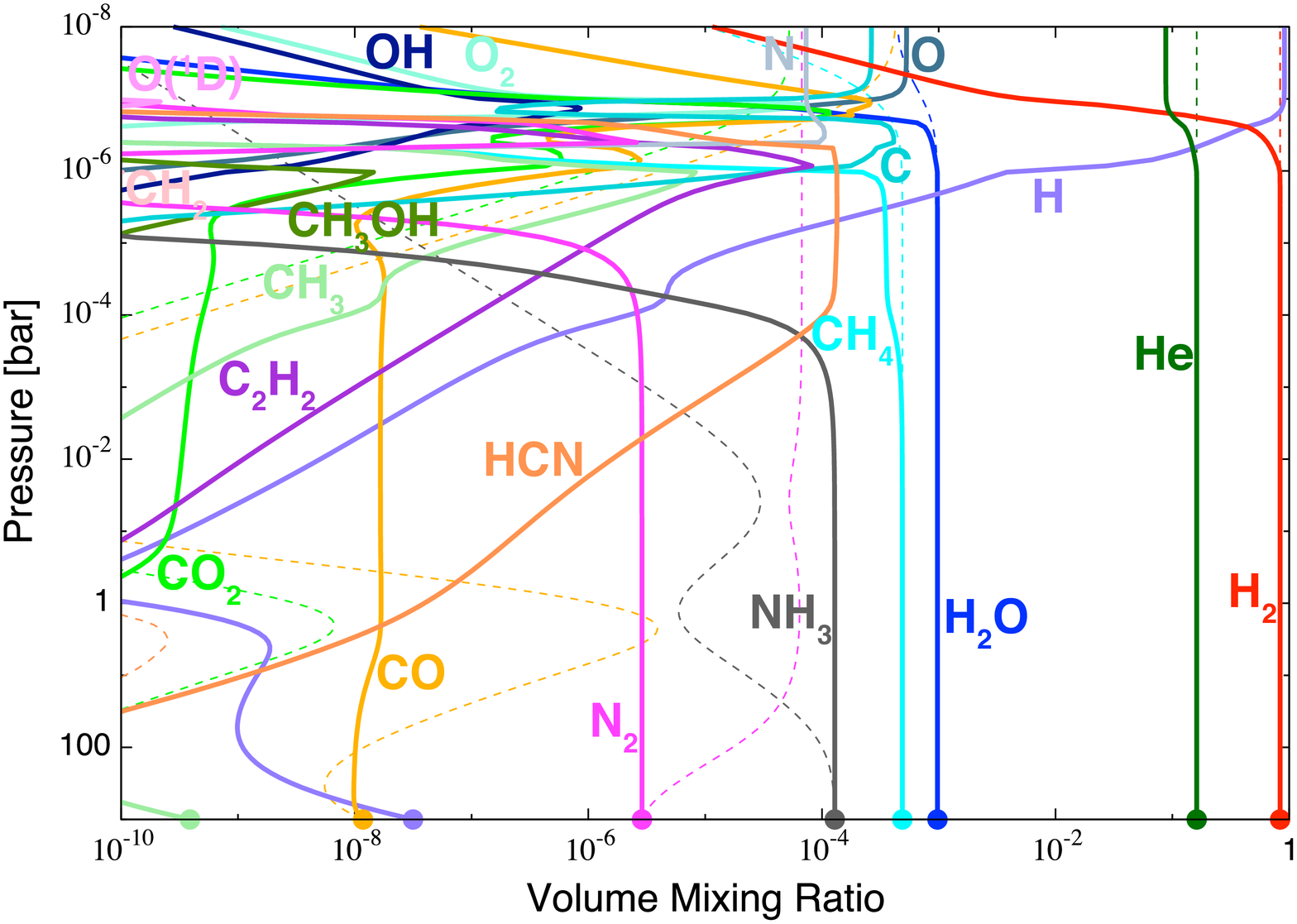}{0.5\textwidth}{(c)~$K_{zz} = 1 \times 10^5$~$\mathrm{cm}^2$~$\mathrm{s}^{-1}$}
\fig{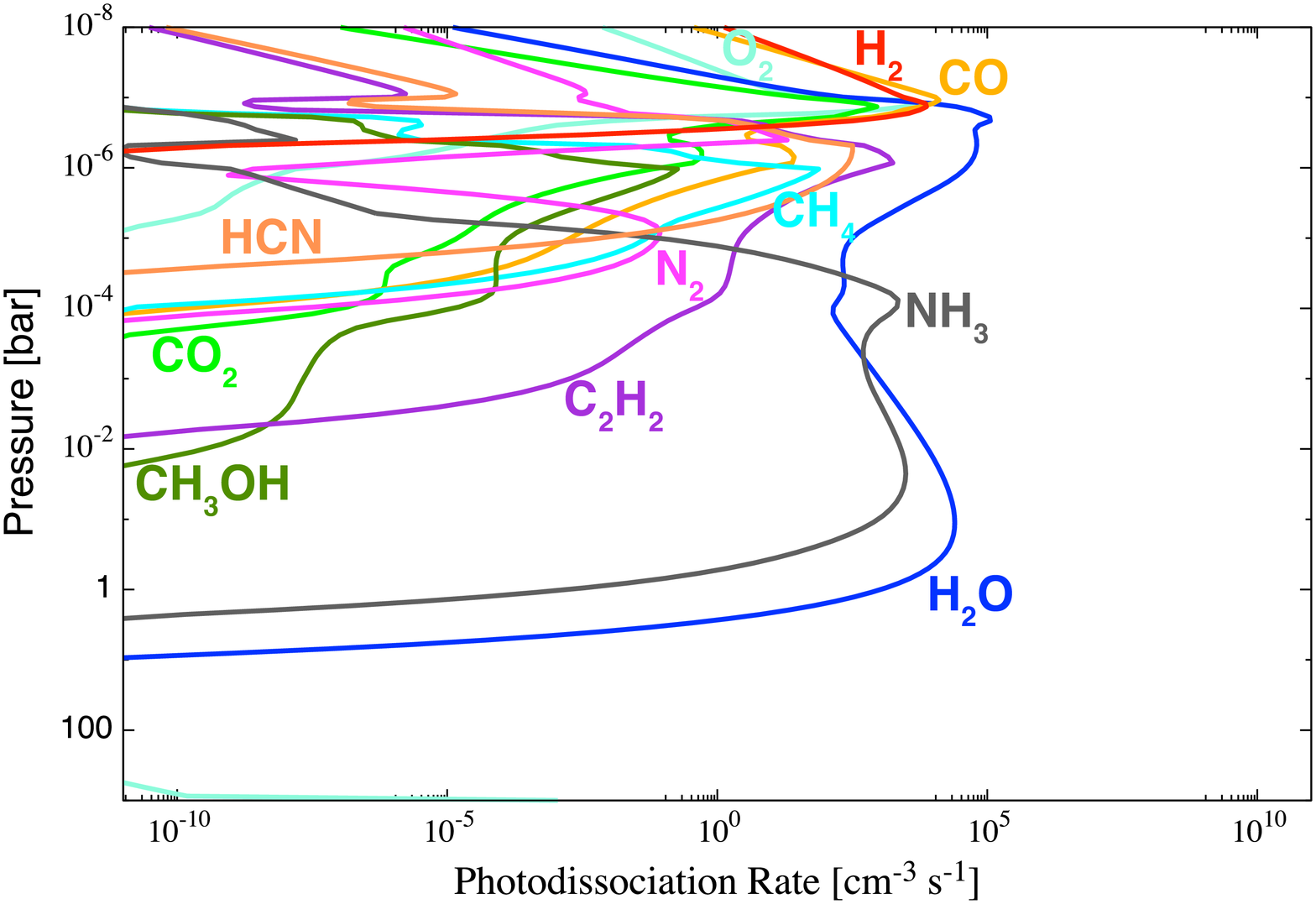}{0.5\textwidth}{(d)~$K_{zz} = 1 \times 10^5$~$\mathrm{cm}^2$~$\mathrm{s}^{-1}$}
}
\caption{Same as Fig.~\ref{fig-photo}, but for \ikomat{two different values} of eddy diffusion coefficient, $K_{zz}$: (a, b)~$K_{zz} = 1 \times 10^9$~$\mathrm{cm}^2$~$\mathrm{s}^{-1}$ and (c, d)~$K_{zz} = 1 \times 10^5$~$\mathrm{cm}^2$~$\mathrm{s}^{-1}$.
\label{fig-photo_eddy}}
\end{figure*}

Figure \ref{fig-photo_eddy}~(a) and (c) show the calculated vertical distributions  of gaseous species for $K_{zz} = 1 \times 10^9$ and $1 \times 10^5$~$\mathrm{cm}^2$~$\mathrm{s}^{-1}$, respectively (also see Fig.~\ref{fig-photo} for $K_{zz} = 1 \times 10^7$~$\mathrm{cm^2}$ $\mathrm{s^{-1}}$).
\ikomat{As found in each panel}, below a certain pressure (or above an altitude),
the volume mixing ratios of $\mathrm{H_2O}$, $\mathrm{CH_4}$, and $\mathrm{NH_3}$ are found to deviate from the constant values{, which are almost equal to those} at the lower boundary. 
Such a threshold pressure (or altitude) decreases (or increases) with increasing $K_{zz}$,
because the transport by eddy diffusion occurs more efficiently to compensate for the loss of those molecules via photodissociation for the higher value of $K_{zz}$. 
As for NH$_3$, such an effect appears as its 11~$\mu$m absorption feature in the transmission spectrum, as described above (see Fig.~\ref{fig-spectra_eddy}). 
By contrast, photochemical products such as C$_2$H$_2$ become more abundant, as $K_{zz}$ decreases, because photochemistry dominates over the transport of thermochemical products.

The distribution of HCN\ikomat{, which contributes to the 14~$\mu$m feature in Fig.~\ref{fig-spectra_eddy},} depends on eddy diffusion coefficient in \ikomat{a} somewhat complicated way.
\ikomat{As seen in Figure~\ref{fig-photo_eddy}~(a) and (c),} the higher the value of $K_{zz}$, the higher the altitude above which $\mathrm{HCN}$ outnumbers $\mathrm{NH_3}$ as the dominant nitrogen-bearing species.
Also, $\mathrm{HCN}$ remains as the dominant nitrogen-bearing species against $\mathrm{N}$ up to higher altitudes for higher values of $K_{zz}$. 
This is because the efficient eddy diffusion transport prevents N atoms from being produced by photodissociation of $\mathrm{HCN}$.
In short, the region where $\mathrm{HCN}$ exists as the dominant nitrogen-bearing species is the narrowest in the fiducial case among the three cases.
{This is one of the effects responsible for the \ikomat{stronger} absorption feature at 14~$\mu$m in the low-$K_{zz}$ model than in the fiducial case.
In the  low-$K_{zz}$ model, in addition to $\mathrm{NH_3}$, HCN also contribute\ikomat{s} to the \ikomat{absorption} feature. 
\ikomat{To be exact,} the existence of HCN in the broader pressure range \ikomat{compensates for} the loss of $\mathrm{NH_3}$ from the lower altitude and\ikomat{, thus,} results in the increase of absorption feature at 14~$\mu$m in the low-$K_{zz}$ model.
In the high-$K_{zz}$ model, the 14 $\mu$m feature is almost the same as that in the haze-free model \ikomat{with the fiducial value of $K_{zz}$} because the \ikomat{haze} abundance is small and also, HCN exists only \ikomat{at quite high altitudes, so} that it has negligible contribution to the feature in this case.}

Figure~\ref{fig-photo_eddy}~(b) and (d) plot the vertical profile of the photodissociation rate of each species for \ikomat{such two values of $K_{zz}$}.
In higher-$K_{zz}$ cases, photodissociation occurs at lower pressures (\ikomat{higher} altitudes) because molecules are transported up to lower pressures (\ikomat{higher} altitudes) for higher values of $K_{zz}$.
As for haze precursors, the {integrated} photodissociation rates of $\mathrm{CH_4}$ and HCN are higher for higher values of $K_{zz}$, because they are lifted up to higher altitudes by the efficient transport (see Table~\ref{tab-photo}).
By contrast, 
the {integrated} photodissociation rate of $\mathrm{C_2H_2}$ becomes smaller, as $K_{zz}$ becomes large.
This is because of its lower abundance for the higher value of $K_{zz}$.
Thus, in contrast to the fiducial and low-$K_{zz}$ cases, in the high-$K_{zz}$ case, $\mathrm{C_2H_2}$ is the last precursors to photodissociate among the three.
Putting all those effects together, the sum of the photodissociation rate of the three haze precursors is lower for higher values of $K_{zz}$.

\subsubsection{Particle Growth} \label{eddy_growth}
\begin{figure*}
\plotone{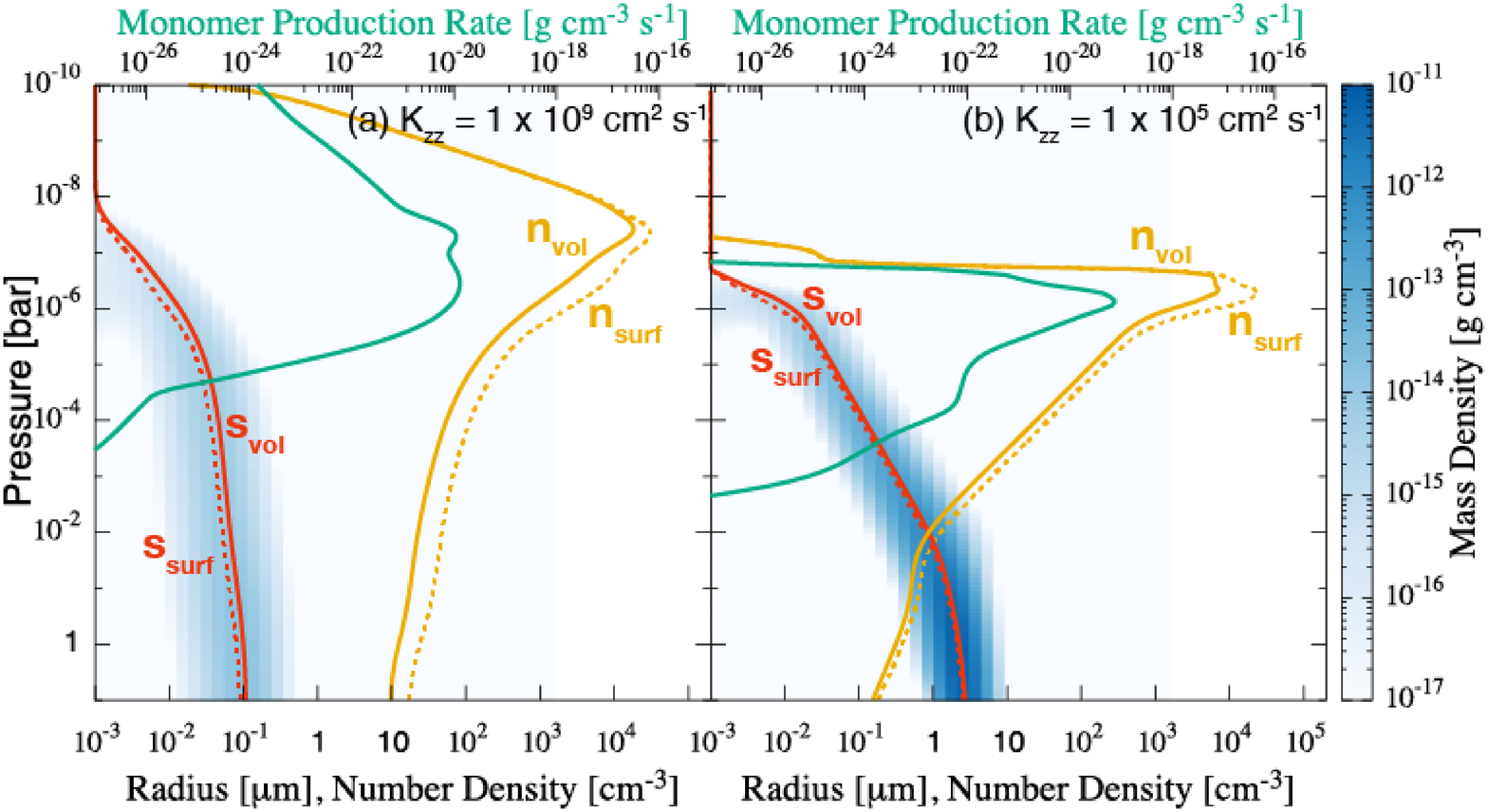}
\caption{Same as Fig.~\ref{fig-growth}, but for two different values of eddy diffusion coefficient, $K_{zz}$: (a)~$K_{zz} = 1 \times 10^9$~$\mathrm{cm}^2$~$\mathrm{s}^{-1}$ and (b)~$K_{zz} = 1 \times 10^5$~$\mathrm{cm}^2$~$\mathrm{s}^{-1}$.
\label{fig-growth_eddy}}
\end{figure*}

Figure~\ref{fig-growth_eddy} shows the vertical profiles of 
{the haze particles} for $K_{zz}$ = (a)~$1 \times 10^9$ and (b)~$1 \times 10^5$~$\mathrm{cm}^2$~$\mathrm{s}^{-1}$,
%
First, the particles are found to grow only to 0.1~$\mu$m in the high-$K_{zz}$ case in contrast to the fiducial and low-$K_{zz}$ cases where the particles grow to $\gtrsim$ 1~$\mu$m.
This is because {over the almost entire region} 
the eddy diffusion with such high $K_{zz}$ brings about descent of the particles and its velocity is higher
than the sedimentation velocity, resulting in a rapid downward transport of the particles.
Such an effect is negligible in the fiducial and low-$K_{zz}$ cases.
This is why the haze particles hardly obscure molecular features in the infrared transmission spectrum in the high-$K_{zz}$ case (see Fig.~\ref{fig-spectra_eddy}).
{\ikomat{Instead, in the high-$K_{zz}$ case, such small} haze particles \ikomat{($\lesssim 0.1$~$\mu$m), which exist} even in the lower atmosphere\ikomat{,} produce the steep Rayleigh-scattering slope in the optical.}
Note that the slightly low {integrated} monomer production rate for the high-$K_{zz}$ 
also inhibits particle growth. 
We have, however, confirmed that such a slighly low production rate has a minor effect, but the efficient eddy diffusion is responsible for the prohibition of the particle growth in the high-$K_{zz}$ case.

The distribution of haze particles for the low-$K_{zz}$ is almost similar to that for the fiducial case (see Fig.~\ref{fig-growth}).
This is because eddy-diffusive mixing hardly contributes to the particle transport and the monomer production rates are almost similar in both cases.
Thus, the transmission spectra are also similar in both cases, as shown in Fig.~\ref{fig-spectra_eddy}.

%% file: result_temp.tex
\subsection{Dependence on Temperature 
} \label{temp}
\begin{figure}
\plotone{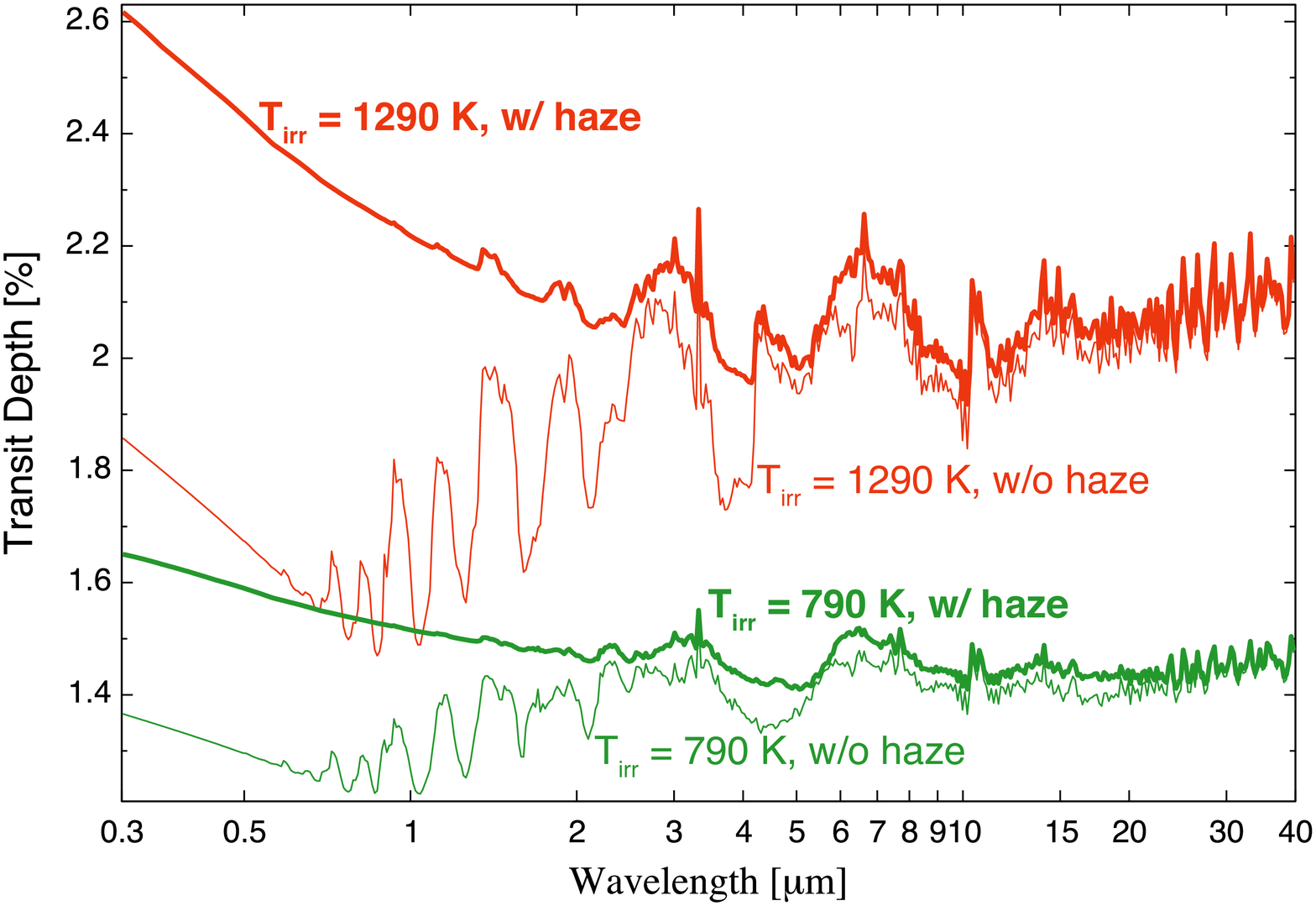}
\caption{Transmission spectrum models for the atmospheres with haze for the cases of $T_\mathrm{irr} = 1290$~K (red thick line) and $T_\mathrm{irr} = 790$~K (green thick line, same as the green line in Fig.~\ref{fig-spectra}).
The transmission spectrum for the atmosphere without haze for the cases of $T_\mathrm{irr} = 1290$~K (red thin line) and $T_\mathrm{irr} = 790$~K (green thin line, same as the black line in Fig.~\ref{fig-spectra}) are also plotted.
Note that the transmission spectrum models are smoothed for clarity.
\label{fig-spectra_temp}}
\end{figure}

Figure~\ref{fig-spectra_temp} shows the transmission spectrum models for the higher irradiation temperature $T_\mathrm{irr}$ of 1290~K (red), which are compared with the fiducial models with $T_\mathrm{irr} = 790$~K (green). 
Regardless of haze, the transit depths are larger and the spectral features are more pronounced for the higher-$T_\mathrm{irr}$ atmosphere, because of larger atmospheric scale height. 
Comparing the spectra with (thick line) and without (thin line) haze for each temperature case, one finds that the presence of the haze has a smaller effect on the spectrum in the higher-$T_\mathrm{irr}$ case {when taking the scale-height difference into account}. 
This is because of the smaller mass density of the haze particles, as explained below.

\begin{figure*}
\gridline{
\fig{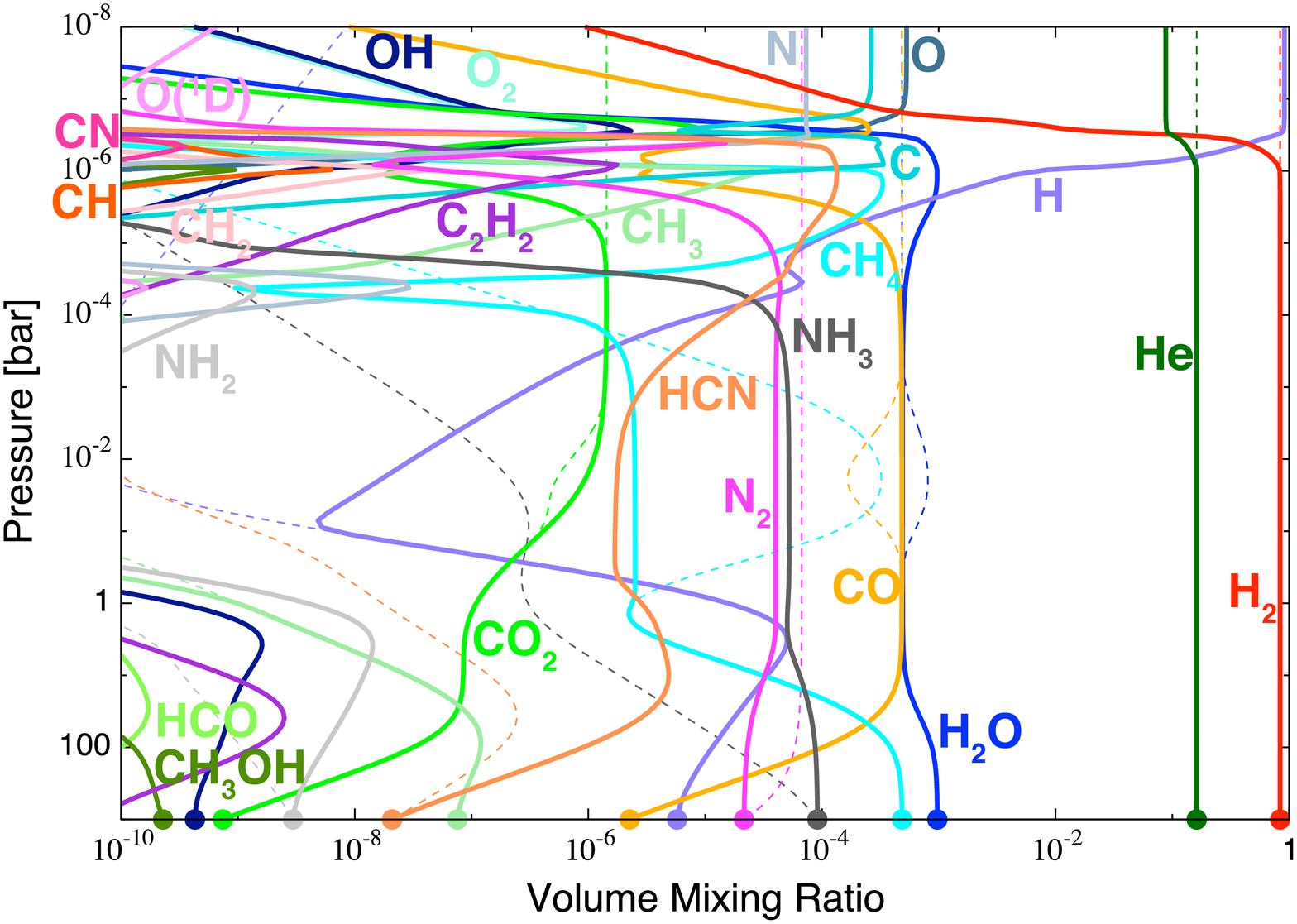}{0.5\textwidth}{(a)~$T_\mathrm{irr} = 1290$~K}
\fig{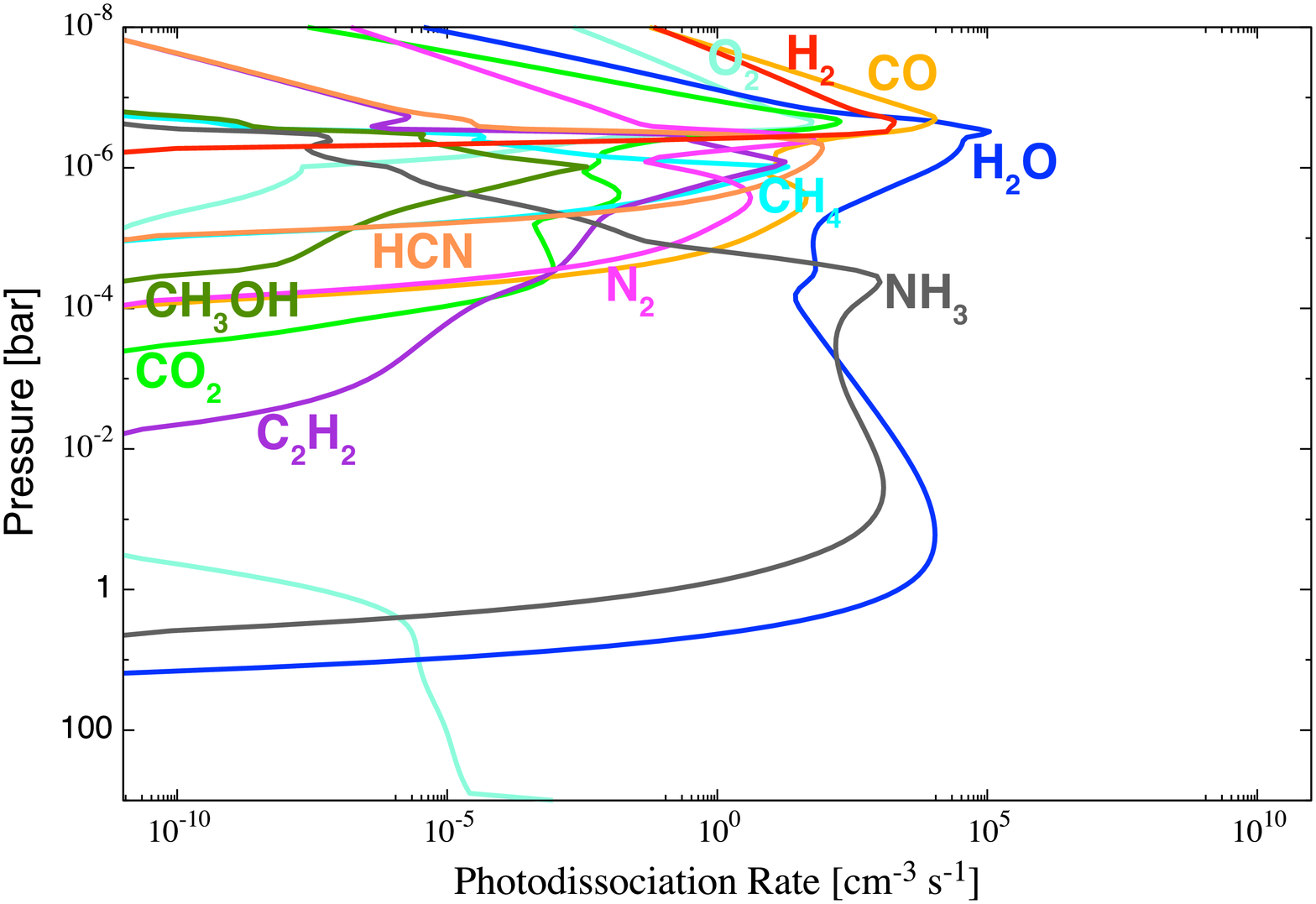}{0.5\textwidth}{(b)~$T_\mathrm{irr} = 1290$~K}
}
\caption{Same as Fig.~\ref{fig-photo} but for the cases of the irradiation temperature of $T_\mathrm{irr} = 1290$~K.
\label{fig-photo_temp}}
\end{figure*}

In Figure~\ref{fig-photo_temp}, we show (a) the calculated vertical distributions and (b) photodissociation rate profiles of gaseous species for $T_\mathrm{irr} = 1290$~K.
Unlike in the fiducial case with $T_\mathrm{irr}$ = 790~K (Fig.~\ref{fig-photo}~(a)), not CH$_4$ but CO is the dominant carbon-bearing species and is equal in abundance to $\mathrm{H_2O}$ in the middle atmosphere ($10^{-4}$~bar $\lesssim P \lesssim$ 10~bar), because CO is more thermodynamically stable than $\mathrm{CH_4}$ at high temperatures. 
Therefore, the photodissociation rate of CO is also larger than in the fiducial case. 
Likewise, $\mathrm{N_2}$, H, and $\mathrm{CO_2}$ are more abundant than in the fiducial case because of stability at high temperatures, while $\mathrm{CH_4}$ and $\mathrm{NH_3}$ are less abundant.

The haze precursors are {all} less abundant {in the photodissociation region ($P \sim 10^{-6}$~bar)} because carbon exists in the form of CO over the almost entire region. 
HCN also exists even in the lower atmosphere unlike in the fiducial case{, where it is thermochemically produced due to the high temperatures}. 
The {integrated} photodissociation rates of $\mathrm{CH_4}$, HCN, and  $\mathrm{C_2H_2}$ are all smaller compared to those in the fiducial case because of their decreased abundances (see Table~\ref{tab-photo})

\begin{figure}
\plotone{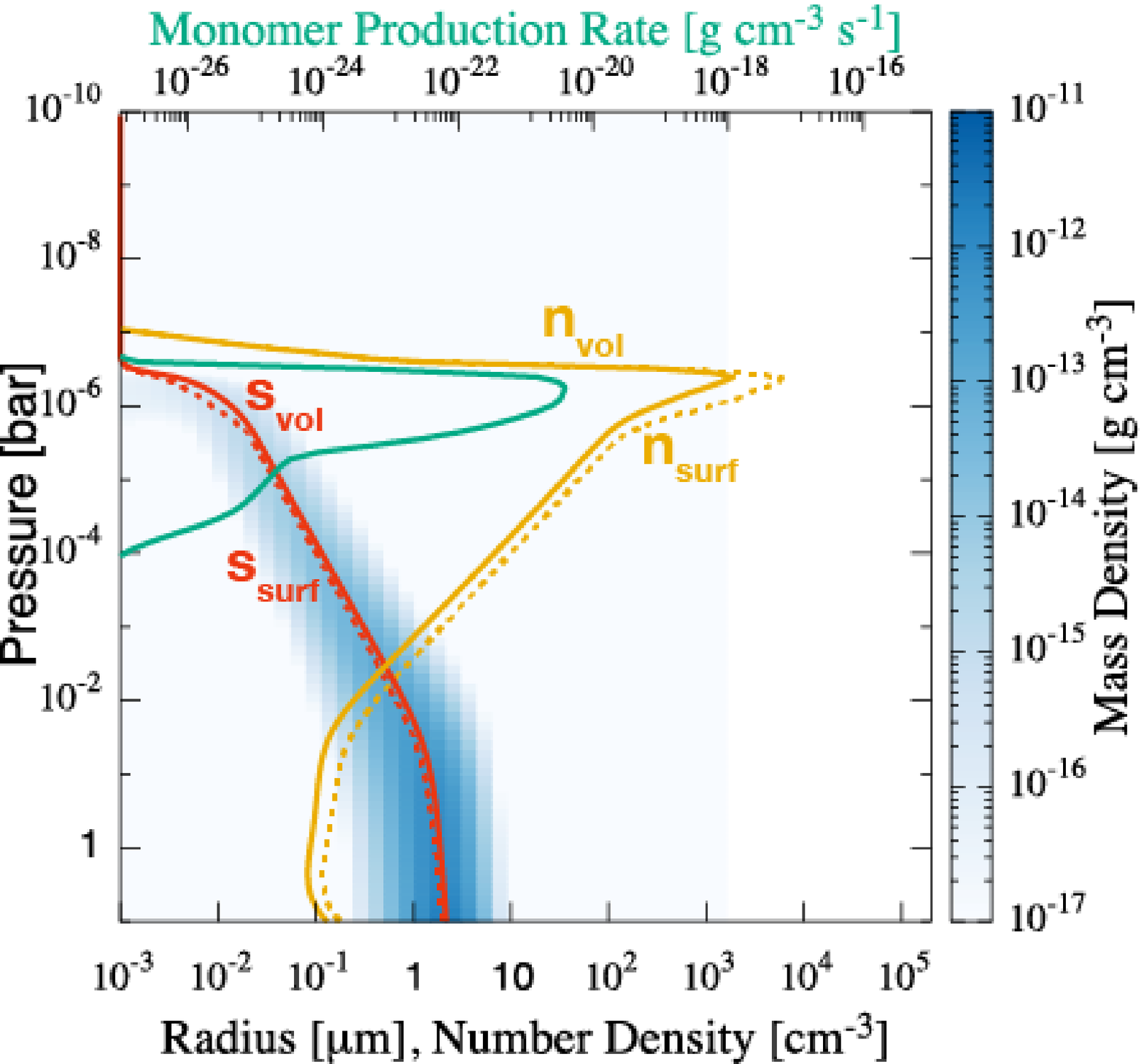}
\caption{Same as Fig.~\ref{fig-growth} but for the case of $T_\mathrm{irr} = 1290$~K.
\label{fig-growth_temp}}
\end{figure}

Figure~\ref{fig-growth_temp} shows the vertical distribution of the haze particles for $T_\mathrm{irr} = 1290$~K.
The total mass density of the haze particles is found to be slightly smaller, because as seen above, the sum of the {integrated} photodissociation rates of haze precursors is smaller and, thus, the assumed monomer production rate is smaller than in the fiducial case.

%% file: discussion.tex
\section{Discussions} \label{discussion}

\subsection{Implications for Observations} \label{obs}


{
In this paper, \ikomat{over} broad ranges of poorly-constrained parameters, we have \ikomat{explored which} combinations of those parameters result in larger or smaller absorption features in transmission spectra \ikomat{of warm {super-Earth} 
atmospheres with hydrocarbon haze. We have found that}
atmospheres with lower \ikomat{UV irradiation, lower} C/O ratio, higher eddy diffusion coefficient, and \ikomat{higher} temperature result in \ikomat{more pronounced molecular-}absorption features compared to the fiducial case. 
\ikomat{We have also found that transmission spectra depend on metallicity in a somewhat complicated way:} \ikomat{At relatively short wavelengths ($\lesssim$~2-3~$\mu$m),} moderate metallicities (such as 100 times the solar metallicity) result in \ikomat{strong} absorption features\ikomat{, because the effects of low monomer production rate and small scale height compete with each other for high metallicities;} 
\ikomat{At relatively long wavelengths ($\gtrsim$~2-3~$\mu$m), where haze has a small effect on transmission spectra}, lower metallicities result in \ikomat{more pronounced features}.
}

\ikomat{The idea that warm exoplanets have atmospheres covered with hydrocarbon haze seems to be consistent with an observationally suggested correlation between transmission spectra and atmospheric temperatures.}
{In \S~\ref{temp}, we have demonstrated that the higher \ikomat{the} temperature\ikomat{,} \ikomat{the} smaller \ikomat{the} photodissociation rates of the hydrocarbon \ikomat{precursors} \ikomat{and the} production rates of haze monomers \ikomat{are.} 
\ikomat{That is because} CO rather than $\mathrm{CH_4}$ becomes the dominant carbon-bearing species at high temperatures.
Thus, the resultant \ikomat{spectra} for higher atmospheric temperatures \ikomat{show} more distinct molecular absorption features. 
\ikomat{This} is consistent with the reported observational trend
\ikomat{such that stronger} absorption features \ikomat{are seen in transmission spectra of} hotter planets, although other planetary properties may affect such a correlation \citep{2016ApJ...817L..16S, 2016ApJ...826L..16H, 2017AJ....154..261C}.}
\kawashima{However, recent laboratory experiments implied that $\mathrm{CH_4}$ is not necessarily required for the haze formation, and instead, $\mathrm{CO}$ and $\mathrm{CO_2}$ can provide an alternative source of carbon \citep{2018NatAs...2..303H, doi:10.1021/acsearthspacechem.8b00133}.
A brief discussion on this is made in \S~\ref{experiments}.}

\begin{figure*}
\gridline{
\fig{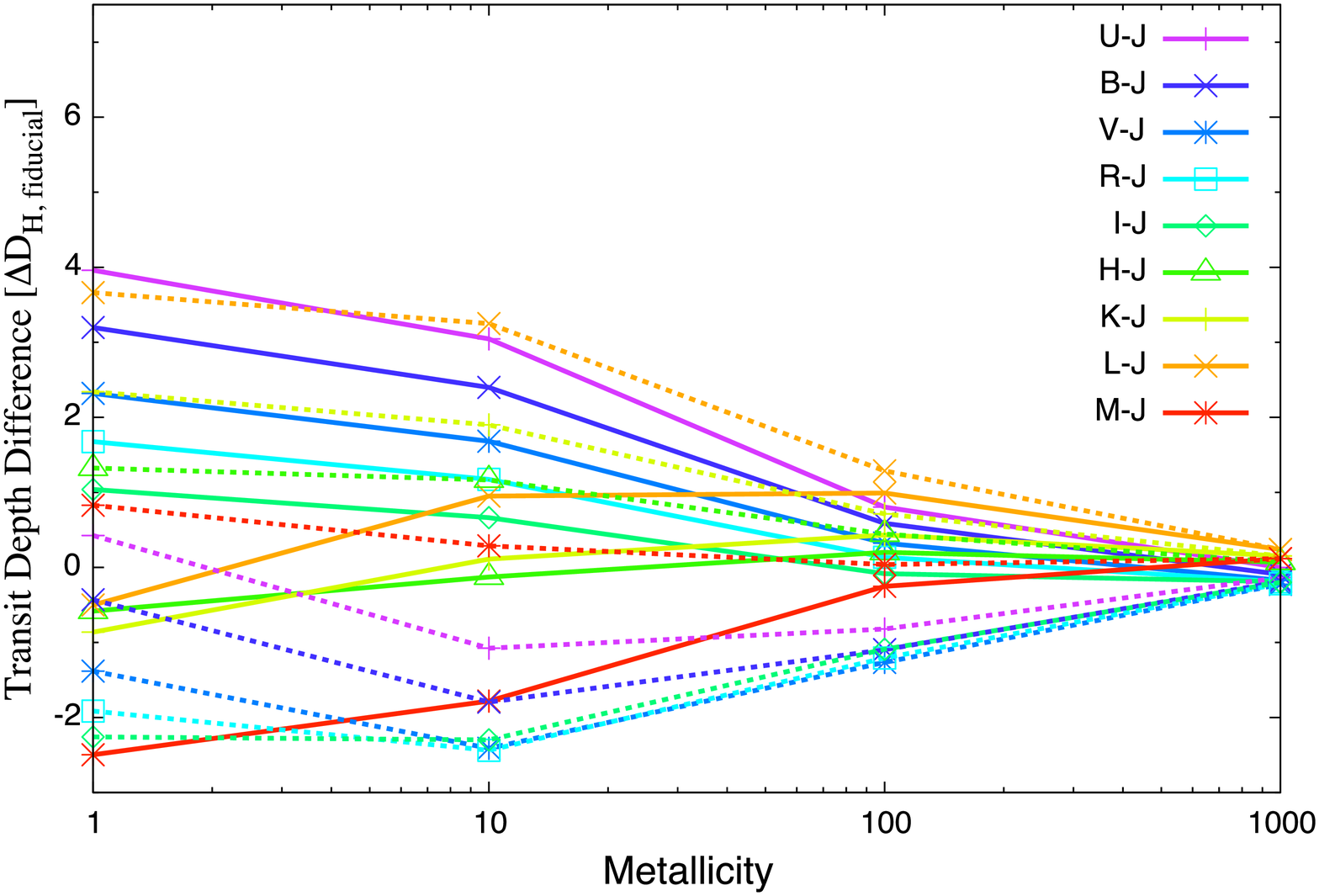}{0.5\textwidth}{(a)~$K_{zz} = 1 \times 10^7$~$\mathrm{cm}^2$~$\mathrm{s}^{-1}$}
\fig{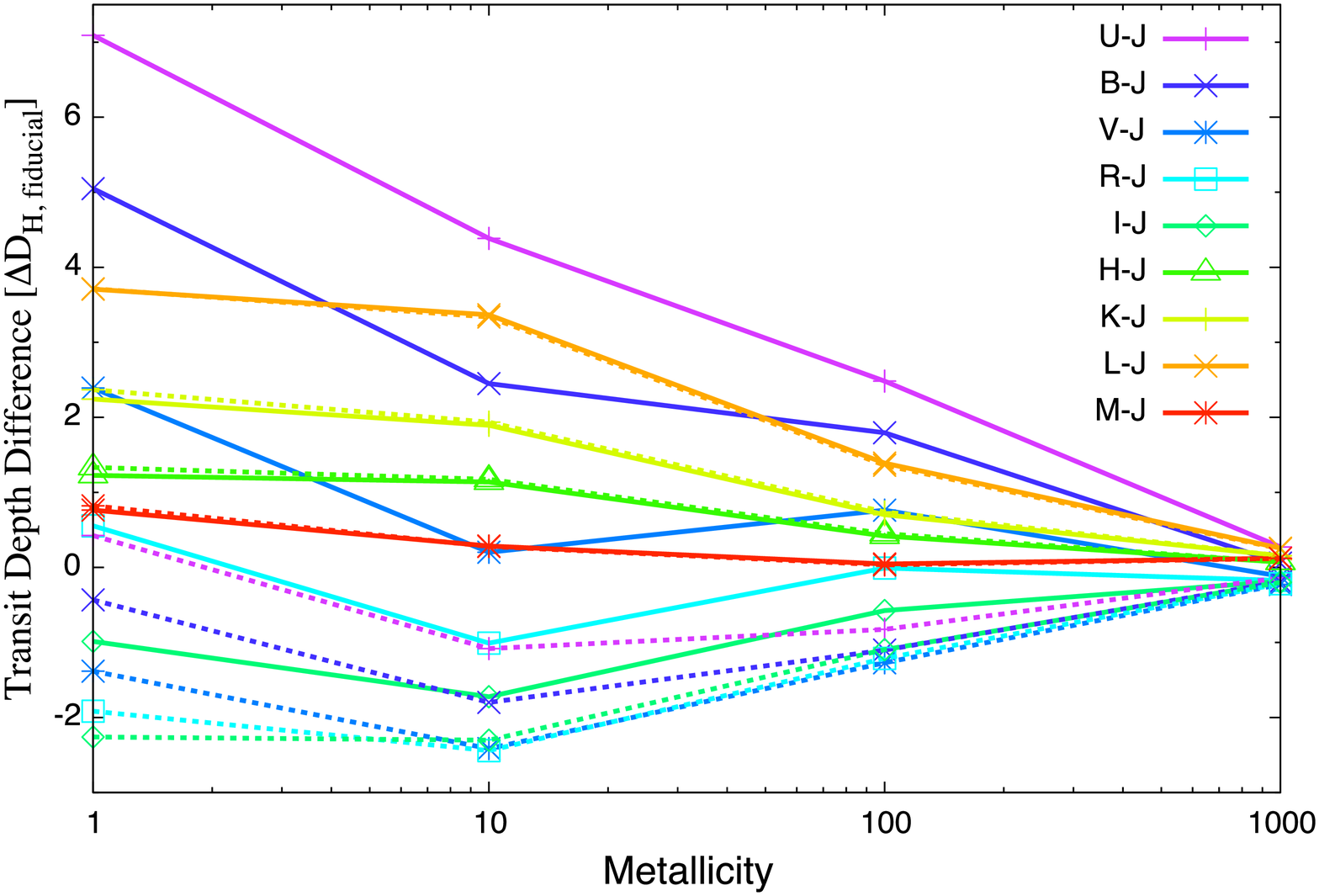}{0.5\textwidth}{(b)~$K_{zz} = 1 \times 10^9$~$\mathrm{cm}^2$~$\mathrm{s}^{-1}$}
}
\caption{{
\ikomat{Differences of the} band-integrated transit depth\ikomat{s} from that in \ikomat{the} $J$~band 
as a function of \ikomat{atmospheric} metallicity for two different values of the eddy diffusion coefficient, $K_{zz}$, (a)~$1 \times 10^7$ and (b)~$1 \times 10^9$~$\mathrm{cm}^2$~$\mathrm{s}^{-1}$.
{Transit depths for the hazy atmospheres are plotted with solid lines, while those for the haze-free atmospheres are plotted with dashed lines.} 
The values of the transit depths are normalized by \ikomat{the transit depth difference caused by one atmospheric scale height {in the fiducial case ($1 \times$Solar and $K_{zz} = 1 \times 10^7$~$\mathrm{cm}^2$~$\mathrm{s}^{-1}$)},} $\Delta D_{H, \mathrm{fiducial}} = 2 R_\mathrm{ref}H_\mathrm{ref}/R_s^2$\ikomat{, where $R_\mathrm{ref}$ is the planetocentric distance to a reference pressure level, $H_\mathrm{ref}$ is the pressure scale height at $R_\mathrm{ref}$ and $R_\mathrm{s}$ is the stellar radius (see text for the details)}.}
\label{fig-band}}
\end{figure*}

{Next, we explore} observational \ikomat{strategies for constraining} the composition of hazy atmospheres. 
In Figure~\ref{fig-band}, we plot \ikomat{the calculated differences in} band-integrated transit depth\ikomat{s} \ikomat{between several photometric bands and the} $J$~band as a function of metallicity for two different values of the eddy diffusion coefficient, $K_{zz}$, (a)~$1 \times 10^7$ and (b)~$1 \times 10^9$~$\mathrm{cm}^2$~$\mathrm{s}^{-1}$.
The values of the transit depths are normalized by $\Delta D_{H, \mathrm{fiducial}} = 2 R_\mathrm{ref}H_\mathrm{ref}/R_s^2$, which is the transit depth difference caused by one atmospheric scale height \ikomat{{for} $H_\mathrm{ref} \ll R_\mathrm{ref}$} {\citep{2001ApJ...553.1006B}} {in the fiducial case ($1 \times$Solar and $K_{zz} = 1 \times 10^7$~$\mathrm{cm}^2$~$\mathrm{s}^{-1}$)}. 
Here, $R_\mathrm{ref}$ is \ikomat{the} planetocentric radius at a reference pressure, \ikomat{for} which we assume $\ikomat{1 \times} 10^{-3}$~bar, $H_\mathrm{ref}$ is an atmospheric scale height at $R_\mathrm{ref}$, and $R_s$ is \ikomat{the} stellar radius.
We adopt the value of $\Delta D_{H, \mathrm{fiducial}}$ \ikomat{in} all the cases of the different metallicities and $K_{zz}$.
The values of the effective wavelength $\lambda_\mathrm{eff}$ and the band width $\Delta \lambda$ we adopt for each band are listed in Table~\ref{tab-band}, which we take from Table~2.1 of \cite{1998gaas.book.....B}.

\begin{deluxetable}{crr}
\tablecaption{Band \label{tab-band}}
\tablehead{
\colhead{Band} & \colhead{$\lambda_\mathrm{eff}$~[nm]} & \colhead{$\Delta \lambda$~[nm]} 
}
\startdata
$U$ & 365 & 66 \\
$B$ & 445 & 94 \\
$V$ & 551 & 88 \\
$R$ & 658 & 138 \\
$I$ & 806 & 149 \\
$J$ & 1220 & 213 \\
$H$ & 1630 & 307 \\
$K$ & 2190 & 390 \\
$L$ & 3450 & 472 \\
$M$ & 4750 & 460 \\
\enddata
\tablecomments{{Values of the effective wavelength $\lambda_\mathrm{eff}$ and the band width $\Delta \lambda$ we adopt for each band, which is taken from Table~2.1 of \cite{1998gaas.book.....B}.}}
\end{deluxetable}

\ikomat{As shown in Figure~\ref{fig-band},}
{the absolute value of the \ikomat{difference in} transit depth between each band and \ikomat{the} $J$ band generally decreases with increasing metallicity for both the hazy \ikomat{(solid lines)} and haze-free \ikomat{(dotted lines)} atmospheres because atmospheric scale height \ikomat{decreases}.
It \ikomat{is demonstrated} that \ikomat{simultaneous observations at shorter wavelengths are more} suitable for \ikomat{constraining} atmospheric metallicity, because of the larger dependence of the transit depth difference on metallicity. This is \ikomat{due to} the Rayleigh-scattering slope produced by haze particles.} 

{For the \ikomat{high-$K_{zz}$} case (\ikomat{panel} b), 
the transit depth differences \ikomat{for two shortest} wavelength bands ($U$ and $B$) are larger than those \ikomat{for} the fiducial $K_{zz}$ case due to the efficient downward transport of haze particles  
(\S~\ref{eddy}).
Also, because of the small monomer production rate, the transit depth differences \ikomat{for long} wavelength bands ($H$, $K$, $L$, and $M$) for the hazy atmospheres almost coincide with those for the haze-free atmospheres.}

{From these two figures, one finds that {in some cases,} from broad-band observations alone, {we suffer from a degeneracy between metallicity and eddy diffusion coefficient.
{For example, the transit depth difference of $U$-$J$ (purple line) for 1 $\times$ Solar metallicity atmosphere with the fiducial $K_{zz}$ value~(a) is almost the same as that for 10 $\times$ Solar metallicity atmosphere with the high $K_{zz}$ value~(b).}
\ikomat{Certainly} such broad-band observations especially \ikomat{in the optical} are useful for target selection aiming for detailed space-based observations in the sense that atmospheres showing steep spectral scope\ikomat{s} in the optical are likely to show strong absorption features in the infrared\ikomat{.} 
\ikomat{However, for obtaining precise constraints on metallicity, it is necessary to measure the} strength of absorption features in the infrared, where the contribution of haze is small.
}

{
\ikomat{Recent} 
{observational} precision\ikomat{s} {are already} \ikomat{high} enough to distinguish} transit depth difference\ikomat{s} \ikomat{smaller than} one atmospheric scale height for some {super-Earths} such as {for} GJ~1214b and HAT-P-11b \ikomat{even with ground-based telescopes such as the Okayama 188cm one equipped with} {the Multicolor Simultaneous Camera for studying Atmospheres of Transiting exoplanets (MuSCAT)} \citep{2015JATIS...1d5001N, Fukui:2016ky}. 
Although \ikomat{being} still limited due to lack of bright targets,
\ikomat{the number of such exoplanets is expected to greatly increase,}
thanks to the Transiting Exoplanet Survey Satellite \ikomat{\citep[TESS;][]{2014SPIE.9143E..20R}} launched in April 2018 and 
\ikomat{PLAnetary Transits and Oscillations of stars \ikomat{\citep[PLATO;][]{2014ExA....38..249R}} to be launched in 2026}.
{Also, the James Webb Space Telescope \ikomat{\citep[JWST;][]{2006SSRv..123..485G}} to be launched in 2021 {and Atmospheric Remote-sensing Exoplanet Large-survey \citep[ARIEL;][]{2018ExA....46..135T} to be launched in 2028} will enable us to {precisely measure the strength of absorption features at longer wavlengths, where the contribution of haze is small.}
{These facts indicate} the possibility of \ikomat{obtaining constraints as to atmospheric properties such as metallicity} for a number of exoplanets.}

\subsection{Implications from Experiments} \label{experiments}
A series of experiments have been recently conducted to measure the production rate of hydrocarbon haze, simulating the environments of warm, high-metallicity exoplanet atmospheres in a chamber \citep{2018ApJ...856L...3H, 2018NatAs...2..303H, 2018AJ....156...38H}.
For the two enriched gases, 100$\times$Solar and 1000$\times$Solar, with temperature of 400~K, the measured production rates 
are 0.25 and 10.00~$\mathrm{mg}$~$\mathrm{h}^{-1}$, respectively (cf. 
7.4~$\mathrm{mg}$~$\mathrm{h}^{-1}$ for the simulated Titan's atmosphere), indicating that the efficiency of photo-dissociative conversion from hydrocarbons to haze particles becomes higher with metallicity.
However, we have to keep in mind that results obtained in a chamber cannot be applied directly to the real atmosphere because photochemistry in atmospheres is controlled not only by local conditions unlike in chamber experiments. 
We must take into account the extinction of photo-dissociating radiation through the atmosphere from the host star to the altitude of interest. Indeed,
as we have shown in \S~\ref{metallicity}, the higher the atmospheric metallicity is, the lower the photodissociation rates of the low-order hydrocarbons such as $\mathrm{CH_4}$, HCN, and $\mathrm{C_2H_2}$ are basically (see Table~\ref{tab-photo}), 
since a rise in atmospheric metallicity leads to increases of $\mathrm{H_2O}$, $\mathrm{CO}$, $\mathrm{CO_2}$, and $\mathrm{O_2}$ exiting at higher altitudes than the hydrocarbons and those molecules absorb more photons, hampering the photodissociation of the hydrocarbons.

In this section, using the data of haze production rate from the above experiments, 
we explore the transmission spectra for the 100$\times$Solar and 1000$\times$Solar atmospheres. 
Unlike the assumption we have used so far in this paper, here we define the {integrated} monomer production rate throughout the atmosphere, $\dot{M}_\mathrm{exp}$, as
\begin{equation}
\dot{M}_\mathrm{exp} = \beta \frac{I_\mathrm{Ly\alpha}}{I_\mathrm{Ly\alpha, Titan}} \dot{M}_{\mathrm{Titan}},
\label{eq_prod}
\end{equation}
where \kawashima{$\beta$ is a tuning parameter described below}, $I_\mathrm{Ly\alpha}$ is the incident stellar $\mathrm{Ly\alpha}$ flux at the planet's orbital distance, and $I_\mathrm{Ly\alpha, Titan}$ and $\dot{M}_{\mathrm{Titan}}$ are the incident solar $\mathrm{Ly\alpha}$ flux and {integrated} monomer production rate in the atmosphere of Titan, respectively.
Note that this is the same equation as Eq.~(35) of Paper~I.
Namely, we assume that the {integrated} production rate is proportional to the incident stellar Ly$\alpha$ flux like in Paper~I.
{Note that although we have found that the dependence of the photodissociation rate{s} of the haze precursors on the UV irradiation intensity is weaker than the linear-relationship in \S~\ref{uv}, we adopt this assumption for simplicity.}
We adopt the parameter $\beta$ as the ratio of the experimental production rates of the 100$\times$Solar and 1000$\times$Solar gas to that for the simulated Titan's atmosphere, namely, $0.25 / 7.4 = 0.034$ and $10.00 / 7.4 = 1.4$, respectively.
For the value of $\dot{M}_{\mathrm{Titan}}$, we adopt $1 \times$~$10^{-14}$~g~$\mathrm{cm^{-2}}$~$\mathrm{s^{-1}}$ since microphysical models, photochemical models, and laboratory simulations all imply that the production rate of the monomers on Titan is in the range between $0.5 \times 10^{-14}$ and $2 \times 10^{-14}$~g~$\mathrm{cm^{-2}}$~$\mathrm{s^{-1}}$ \citep{2001P&SS...49...79M}.
Also, we use $6.2 \times 10^9$~photons~$\mathrm{cm^{-2}}$~$\mathrm{s^{-1}}$ for $I_\mathrm{Ly\alpha, Titan}$ \citep{2006PNAS..10318035T} and $3.3 \times 10^{13}$~photons~$\mathrm{cm^{-2}}$~$\mathrm{s^{-1}}$ for $I_\mathrm{Ly\alpha}$ using the values of the observed $\mathrm{Ly\alpha}$ flux of GJ~1214 \citep{Youngblood:2016ib} and GJ~1214b's semi-major axis \citep{2013AA...551A..48A}.

We assume that the monomer production rate is proportional to the sum of the photodissociation rates of $\mathrm{CH_4}$, $\mathrm{HCN}$, and $\mathrm{C_2H_2}$ and define the modified monomer production rate at each altitude, $p_\mathrm{exp} (v_1, z)$, as
\begin{equation}
p_\mathrm{exp} (v_1, z) = \frac{\dot{M}_\mathrm{exp}}{\dot{M}} p(v_1, z),
\end{equation}
where $\dot{M}$ is the {integrated} production rate throughout the atmosphere, which we have adopted in the previous part and is given by
\begin{equation}
\dot{M} = \int_0 ^{\infty} p(v_1, z) dz.
\end{equation}
From Eq.~(\ref{eq_prod}), the values of $\dot{M}_\mathrm{exp}$ come out to be $1.79 \times 10^{-12}$ and $7.18 \times 10^{-11}$~$\rm g \, cm^{-2} \, s^{-1}$ for the 100$\times$Solar and 1000$\times$Solar cases, respectively, which are 9.66 and 380 times larger than those of $\dot{M}$ that we used in \S~\ref{metallicity}.

Figure~\ref{fig-spectra_metallicity-Exp} shows the transmission spectrum models for the hazy 100$\times$Solar atmosphere calculated with $\dot{M}_\mathrm{exp}$ (yellow line) and $\dot{M}$ (red thick line, same as the red thick line in Fig.~\ref{fig-spectra_metallicity}) and the hazy 1000$\times$Solar atmosphere calculated with $\dot{M}_\mathrm{exp}$ (light-blue line) and $\dot{M}$ (purple thick line, same as the purple thick line in Fig.~\ref{fig-spectra_metallicity}). 
The transmission spectra for the corresponding haze-free atmospheres are also plotted with the red thin line (same as the red thin line in Fig.~\ref{fig-spectra_metallicity}) and purple thin line (same as the purple thin line in Fig.~\ref{fig-spectra_metallicity}), respectively.

The transmission spectra for $\dot{M}_\mathrm{exp}$ turn out to be more featureless relative to those for $\dot{M}$. This is because $\dot{M}_\mathrm{exp}$ is much higher than $\dot{M}$. 
Especially, the spectrum for the hazy 1000$\times$Solar atmosphere calculated with $\dot{M}_\mathrm{exp}$ is flat in almost the entire wavelength region. 
%
\kawashima{We consider that the several factors are responsible for the much higher values of $\dot{M}_\mathrm{exp}$ than those of $\dot{M}$.
First, the experiments include much more complex chemistry than our photochemical simulations, while we assume 100\% conversion efficiency of forming haze from the photodissociation of our relatively limited number of haze precursors to haze monomers.
Recent laboratory experiments identified some other potential key precursors such as $\mathrm{CH_2NH}$ and $\mathrm{HCHO}$ in addition to the species assumed in our simulations, $\mathrm{CH_4}$, HCN, and $\mathrm{C_2H_2}$ \citep{doi:10.1021/acsearthspacechem.8b00133}.
Our limited number of haze precursors can make $\dot{M}$ smaller than $\dot{M}_\mathrm{exp}$, while $\dot{M}$ can be overestimated since the 100\% conversion efficiency is obviously unlikely to be achievable.
On the other hand, the vertical photon-shielding effect by the other molecules, existing at higher altitudes than the hydrocarbons, is not considered in the experiments, while considered in our simulations. This can also make $\dot{M}_\mathrm{exp}$ higher. Moreover, although we assume the linear-dependence of the photodissociation rates of the haze precursors on the incident UV flux, our results in \S~\ref{uv} imply that the relationship is slightly weaker than the linear-one, while the dependence is not monotonic. This can be also responsible for the higher values of $\dot{M}_\mathrm{exp}$.}

\kawashima{As mentioned in \S~\ref{metallicity}, recent laboratory experiments implied that not only the photodissociation of hydrocarbons, but also that of $\mathrm{CO}$, $\mathrm{CO_2}$, and $\mathrm{H_2O}$ can lead to the formation of haze, implying the existence of multiple formation pathways \citep{2018NatAs...2..303H, doi:10.1021/acsearthspacechem.8b00133}.
If we also include the photodissociation rates of $\mathrm{CO}$, $\mathrm{CO_2}$, and $\mathrm{H_2O}$, the total monomer production rates become $1.93 \times 10^{-10}$, $3.08 \times 10^{-10}$, $3.50 \times 10^{-10}$, and $4.07 \times 10^{-10}$~$\rm g \, cm^{-2} \, s^{-1}$ for the 1, 10, 100, and 1000$\times$Solar cases, respectively, and thus larger for higher metallicities.
As for C/O ratio, those values become $1.93 \times 10^{-10}$, $1.81 \times 10^{-10}$, $1.53 \times 10^{-10}$, and $1.03 \times 10^{-10}$~$\rm g \, cm^{-2} \, s^{-1}$ for the cases of $\mathrm{C/O} = 0.5$, 1, 10, and 1000, respectively, and thus slightly smaller for higher values of C/O ratio.
Finally for temperature, those values become $1.93 \times 10^{-10}$ and $2.51 \times 10^{-10}$~$\rm g \, cm^{-2} \, s^{-1}$ for the cases of $T_\mathrm{irr} = 790$ and 1290~K, respectively, and thus larger for higher temperatures.
These opposite dependence from our results for all of the above three parameters, metallicity, C/O ratio, and temperature, come from the fact that $\mathrm{CO}$, $\mathrm{CO_2}$, and $\mathrm{H_2O}$ absorb much more photons than the hydrocarbons we have assumed as the precursors and also, their abundances are larger for the higher metallicities and temperatures, and smaller for the higher C/O ratios.
However, key haze precursors have not been fully understood and are still in debate. In addition, the conversion efficiency of forming haze from the photodissociation of each precursor is still quite uncertain and the 100\% conversion efficiency we have assumed here is obviously overestimated.}
In summary, to gain a deeper understanding of haze production, we need more data from laboratory experiments obviously, but, furthermore, incorporate them correctly in atmospheric models.

\begin{figure}
\plotone{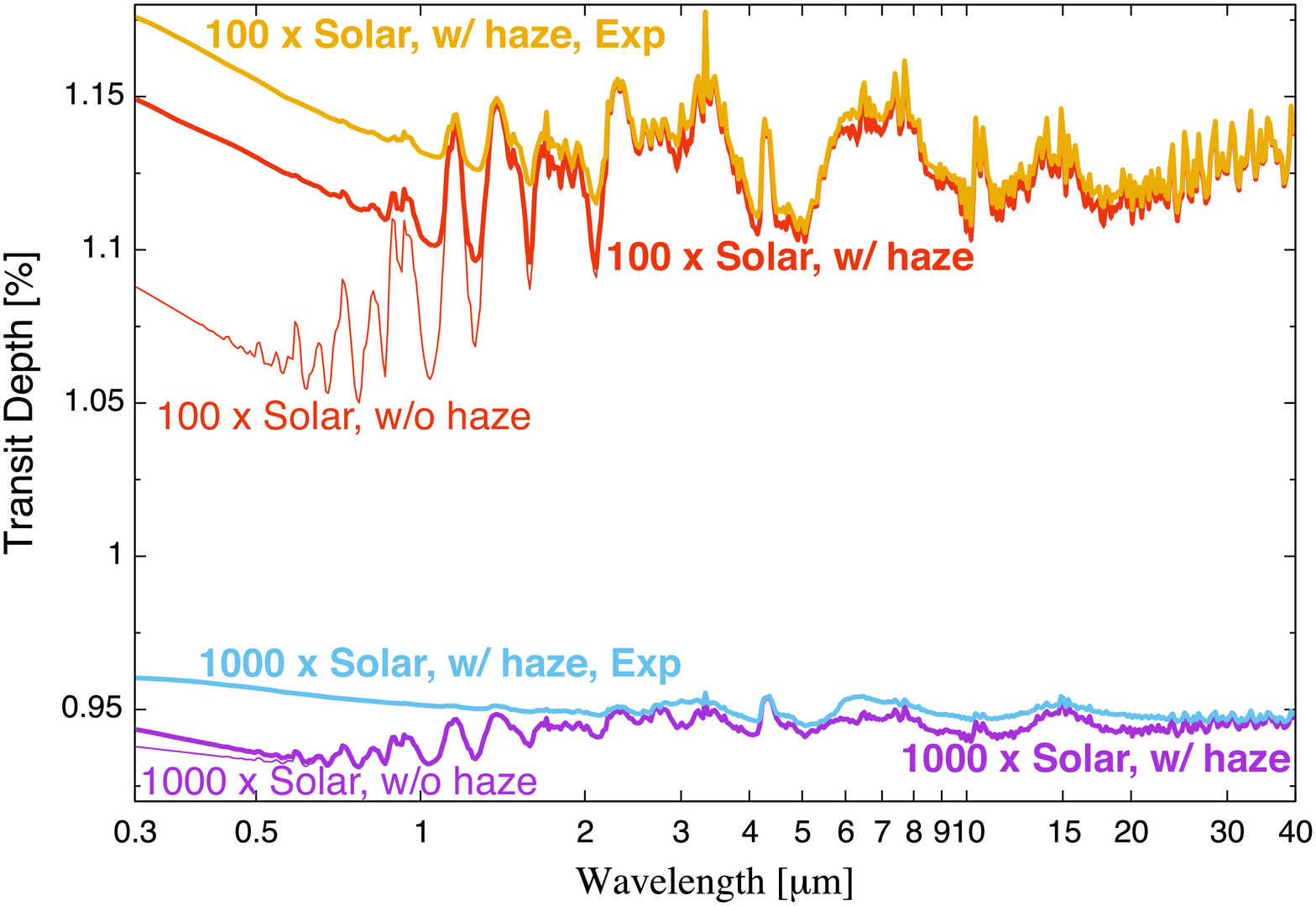}
\caption{Transmission spectrum models for the hazy 100 $\times$ Solar atmospheres calculated with $\dot{M}_\mathrm{exp}$ (yellow line) and $\dot{M}$ (red thick line, same as the red thick line in Fig.~\ref{fig-spectra_metallicity}) and 1000 $\times$ Solar atmospheres calculated with $\dot{M}_\mathrm{exp}$ (light-blue line) and $\dot{M}$ (purple thick line, same as the purple thick line in Fig.~\ref{fig-spectra_metallicity}) .
See the text for the definition of each quantity.
The transmission spectrum for the 100 $\times$ Solar and 1000 $\times$ Solar atmospheres without haze are also plotted in red thin line (same as the red thin line in Fig.~\ref{fig-spectra_metallicity}) and purple thin line (same as the purple thin line in Fig.~\ref{fig-spectra_metallicity}), respectively.
Note that the transmission spectrum models are smoothed for clarity.
\label{fig-spectra_metallicity-Exp}}
\end{figure}

\subsection{\ikomat{Comparison with Previous Studies}}

There are some parameter studies that explored transmission spectra of hydrogen-rich atmospheres with focus on the effects of haze or cloud particles.

First, while we have focused on super-Earths with moderate temperatures in this study, considering hot Jupiters such as HD~209458b and HD 189733b, 
\cite{2017ApJ...847...32L} calculated photochemistry and microphysics of haze particles, modeled the transmission spectra, and compared them with the observed spectra.
They also explored the dependence of the transmission spectra on the \ikomat{poorly constrained} parameters including the monomer production rate, eddy diffusion, and temperature-pressure profile.
They found that higher monomer production rates led to the formation of larger particles, yielding flatter transmission spectra, which is consistent with our finding in Paper~I.
In addition, they found that efficient eddy diffusion hampers collision between particles, making the atmosphere optically thin, which is also consistent with our results in \S~\ref{eddy}.
Finally, as for the temperature-pressure profile, they showed that the resultant transmission spectra for \ikomat{the} hotter and cooler temperature-pressure profiles were almost similar 
\ikomat{to each other, in contrast to our finding that the transmission spectrum for the hotter atmosphere is less affected by haze and has more prominent absorption features (see \S~\ref{temp}). 
Such a difference in temperature dependence of transmission spectrum comes from the fact that they assumed the fixed value of the monomer production rate regardless of temperature, whereas we have determined it from the temperature-dependent photodissociation rates of the hydrocarbons.}
%

Regarding condensation clouds, 
the dependence of atmospheric transmission spectra on eddy diffusion have been investigated by \cite{2018ApJ...863..165G} and \cite{2019A&A...622A.121O}. 
Whereas the atmosphere with the photochemical haze becomes less thick and the resultant transmission spectrum has more prominent molecular-absorption features for higher values of eddy diffusion coefficient (see \S~\ref{eddy}), 
the amount of condensation clouds increases with increasing eddy diffusion coefficient \citep{2018ApJ...863..165G, 2019A&A...622A.121O}.
\ikomat{Such a} difference comes from the fact that haze particles are formed in upper atmospheric regions and transported downward, while condensation clouds are formed in lower atmospheric regions and transported upward. 
{This implies that both haze and condensation clouds \ikomat{being considered}, efficient eddy diffusion does not necessarily \ikomat{reduce} the optical thickness of the atmosphere as we have shown in \S\ref{eddy}. Taking their coexistence in the atmosphere into account is one of our important future stud\ikomat{ies}.
}

\subsection{Caveats}
\subsubsection{Monomer \kawashima{Radius}} \label{monomers}

\begin{figure}
\plotone{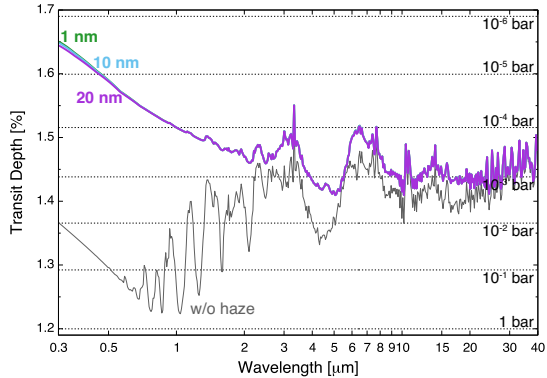}
\caption{\kawa{Transmission spectrum models for the atmosphere with haze for the three different values of the monomer radius, $1$~nm (green line, same as the green line in Fig.~\ref{fig-spectra}), $10$~nm (light-blue line), and $20$~nm (purple line).
The transmission spectrum for the atmosphere without haze is also plotted (black line, same as the black line in Fig.~\ref{fig-spectra}).
As in Fig.~\ref{fig-spectra}, horizontal dotted lines represent the transit depths corresponding to the pressure levels from $1 \times 10^{-6}$~bar to 1~bar.
Note that the transmission spectrum models are smoothed for clarity.}
\label{fig-spectra_monomer}
}
\end{figure}

While we have adopted 1~nm for the monomer \kawashima{radius} \kawa{following some previous studies \citep[e.g.,][]{1992Icar...95...24T, 2017ApJ...847...32L}} in this paper, we have confirmed that it has a little effect on the transmission spectrum.
Indeed, calculating the vertical distribution\kawa{s} of haze particles with monomer \kawa{radii} of 10 \kawa{and 20}~nm, we have found that all the average radii, number densities, and total mass density agree fairly well with those in the fiducial 1--nm case in the lower atmosphere of $P \gtrsim 10^{-6}$~bar, so that the resultant transmission spectra are almost similar \kawa{as shown in Figure~\ref{fig-spectra_monomer}}.
In the upper atmosphere ($P \lesssim 10^{-6}$~bar), where the particle sizes are still small and affected by the initial monomer size, the total mass densit\kawa{ies} for the 10 \kawa{and 20}--nm case\kawa{s} \kawa{are} smaller by \kawa{about} an order of magnitude because the larger particles fall more rapidly.
However, this difference occurs high enough in the upper atmosphere that it has little effect on the resultant transmission spectrum.

\subsubsection{\kawashima{Material Density}} \label{density}

\begin{figure}
\plotone{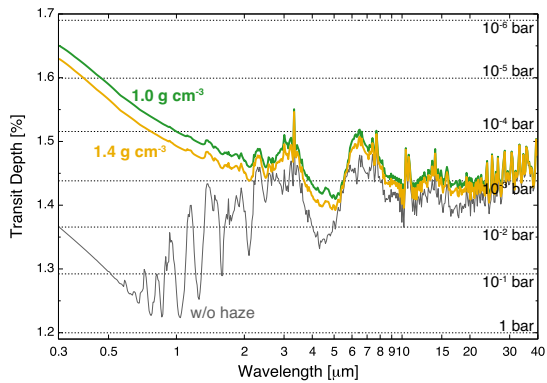}
\caption{\kawashima{Transmission spectrum models for the atmosphere with haze for the two different values of the material density, $1.0~\mathrm{g}~\mathrm{cm}^{-3}$ (green line, same as the green line in Fig.~\ref{fig-spectra}) and $1.4~\mathrm{g}~\mathrm{cm}^{-3}$ (orange line).
The transmission spectrum for the atmosphere without haze is also plotted (black line, same as the black line in Fig.~\ref{fig-spectra}).
As in Fig.~\ref{fig-spectra}, horizontal dotted lines represent the transit depths corresponding to the pressure levels from $1 \times 10^{-6}$~bar to 1~bar.
Note that the transmission spectrum models are smoothed for clarity.}
\label{fig-spectra_density}
}
\end{figure}

\kawashima{
In this study, we have adopted 1.0~g~$\mathrm{cm^{-3}}$ for the value of the material density of haze particles as the widely used value for the microphysical models of Titan's haze \citep[e.g.,][]{1992Icar...95...24T, 2011ApJ...728...80L}.
However, recent laboratory experiments of Titan's haze analogues measured the slightly higher material density of 1.3--1.4~g~$\mathrm{cm^{-3}}$ \citep{2012Icar..218..247I, 2017ApJ...841L..31H}.}
\kawashima{Thus, we simulate the transmission spectrum model for the atmosphere with haze, adopting the value of $1.4~\mathrm{g}~\mathrm{cm}^{-3}$ for the material density, which is shown in Figure~\ref{fig-spectra_density}.
Since the particle sedimentation velocity is linearly proportional to the material density (see Eq.~(13) of Paper I), haze particles fall more rapidly for the larger material density case and thus the atmosphere becomes slightly less optically-thick.}

\subsubsection{Fluffy Particles}
In this study, we have modeled particle growth assuming compact spherical particles.
For the Titan's atmosphere, observation suggests that the haze particles have a fractal structure with a fractal dimension of $\sim 2$ \citep{1992Icar...96..176C, 1993P&SS...41..257C, 1995Icar..118..355R, 1997JGR...10210997R}.
Since the sedimentation velocity of such fluffy particles is small \kawashima{and their collisional cross section is large} compared to that of spherical compact ones with the same mass, 
atmospheric transmission spectra with the former would be flatter than those with the latter \kawashima{as recently shown by \cite{2019ApJ...874...61A}}. 
This effect will be explored in detail in our forthcoming papers.

%% file: summary.tex
\section{Summary and Conclusions} \label{summary}
{In this study, we have 
\ikomat{investigated} 
transmission spectra 
\ikomat{of atmospheres of close-in warm ($\lesssim$~1000~K) exoplanets with hydrocarbon haze} 
\ikomat{for wide ranges of the} model parameters, namely, UV irradiation intensity, metallicity, carbon-to-oxygen ratio (C/O), eddy diffusion coefficient, and \ikomat{atmospheric} temperature.
\ikomat{We have focused on the vertical distributions of the haze particles and gaseous species. In particular, in contrast to previous studies including Paper~I,}
we have \ikomat{made a more realistic assumption} that the monomer production rate is equal to be the sum of the photodissociation rates of the hydrocarbons, $\mathrm{CH_4}$, HCN, and $\mathrm{C_2H_2}$.}

We have found that differences in UV irradiation intensity yields \ikomat{a} diversity of transmission spectra\ikomat{, which are} observationally suggested \ikomat{(see \S~\ref{uv}).}
The photodissociation rates of the hydrocarbons \ikomat{depend relatively weakly} on the UV irradiation intensity {with} the proportion of the incoming photons used for the photodissociation of the haze precursors {decreasing} with increasing UV flux.
{This is due to an enhanced photon-shielding effect by CO and $\mathrm{O_2}$, which exist at higher altitudes than the hydrocarbons.}

{As for metallicity, we have demonstrated that the photodissociation rates of the hydrocarbons \ikomat{(and thus the monomer production rates)} are basically smaller for higher metallicities in spite of their increased abundances \ikomat{(see \S~\ref{metallicity})}.
This is \ikomat{due to an} enhanced photon-shielding effect by the major photon absorbers, $\mathrm{H_2O}$, $\mathrm{CO}$, $\mathrm{CO_2}$, and $\mathrm{O_2}$, existing at higher altitudes than the hydrocarbons.
{However, since the atmospheric scale height is also smaller for higher metallicities, the metallicity \ikomat{affects strengths of} absorption features in the transmission spectra \ikomat{in a somewhat} complicated \ikomat{way}: Moderate metallicities (such as 100 times the solar metallicity) result in large absorption features at short wavelengths ($\lesssim$ 2-3~$\mu$m), while lower metallicities result in larger ones at longer wavelengths ($\gtrsim$ 2-3~$\mu$m).}
While recent chamber-experiments for production of hydrocarbon\ikomat{s} demonstrated higher production rate\ikomat{s} \ikomat{in} higher metallicity \ikomat{gases} \citep{2018ApJ...856L...3H, 2018NatAs...2..303H, 2018AJ....156...38H}, we warn that when applying to the real atmospheres, we also need to consider \ikomat{such a} shielding effect by the other molecules, which we have found \ikomat{is} larger for higher metallicity.}

{\ikomat{Regarding} carbon-to-oxygen ratio \ikomat{(see \S~\ref{co})}, higher values of C/O generally yield larger photodissociation rates of the hydrocarbons, because of the decreased abundances of the major \ikomat{photon-shielding molecules} $\mathrm{H_2O}$, $\mathrm{CO}$, $\mathrm{CO_2}$, and $\mathrm{O_2}$.
As a result, the molecular absorption features in the transmission spectrum become \ikomat{less prominent} with increasing C/O.
However, \ikomat{the transmission spectrum never becomes completely flat, because the absorption features of haze particles remain} 
even \ikomat{for extremely high} $\mathrm{C/O}$ \ikomat{(= $10^{10}$).} 
\ikomat{This is} because the total photodissociation rates of hydrocarbons are limited not by the amount of carbon, but by the incoming photon flux.}

We have demonstrated that the efficient eddy diffusion yields a steep Rayleigh-scattering slope in the optical and more prominent molecular-absorption features \ikomat{(see \S~\ref{eddy})}. 
\ikomat{Such a dependence is opposite to that for the case of} condensation clouds \citep{2018ApJ...863..165G, 2019A&A...622A.121O}.
This is because haze particles are formed \ikomat{at high altitudes} and transported downward, while condensation clouds are formed \ikomat{at relatively low altitudes} and transported upward.

{Finally, \ikomat{we have found that} higher temperature results in smaller photodissociation rates of the hydrocarbons since CO rather than $\mathrm{CH_4}$ becomes the dominant carbon-bearing species at high temperatures.
Thus, the resultant spectrum for the higher atmospheric temperature has more distinct molecular absorption features.}

In conclusion,
\ikomat{detection of} molecular absorption features in hazy atmospheres \ikomat{favors} 
planets with lower incoming UV flux and {higher} temperature.
{As for metallicity, we have revealed \ikomat{its} somewhat complicated effect on transmission spectra. From this, we warn that the featureless spectra recently observed for some exoplanets do not necessarily indicate high atmospheric metallicit\ikomat{ies}.
\ikomat{As} an observational strategy for \ikomat{obtaining} constraint\ikomat{s} on atmospheric metallicit\ikomat{ies} \ikomat{of many} exoplanets expected to be discovered by TESS \citep[][]{2014SPIE.9143E..20R} and PLATO \citep[][]{2014ExA....38..249R}, 
\ikomat{multi-color} broad-band observations especially at optical wavelengths \ikomat{with} ground-based telescopes such as MuSCATs {\citep{2015JATIS...1d5001N, narita2018muscat2}} are useful \ikomat{to select} targets for further detailed observations\ikomat{,} since atmospheres showing steep spectral scope\ikomat{s} in the optical are likely to show strong absorption features in the infrared.
{However,} in order to \ikomat{put more precise} constrain\ikomat{s on} atmospheric metallicity by breaking the degeneracy {of metallicity with eddy diffusion coefficient, we claim that \ikomat{it is necessary to measure the} strength of absorption features in the infrared, where the contribution of haze is small.
For this purpose, \ikomat{of great importance are} space-based telescopes feasible for observations at mid-infrared wavelengths \ikomat{such as JWST \citep[][]{2006SSRv..123..485G}} and those dedicated to exoplanet \ikomat{chemical} characterization such as ARIEL \citep[][]{2018ExA....46..135T}.
}